\documentclass[twoside,11pt]{article}

\usepackage{jmlr2e}
\usepackage{stfloats}
\usepackage{graphicx}
\usepackage{subfigure}
\usepackage{amsmath}
\usepackage{amsfonts,amssymb}
\usepackage{color}
\usepackage{booktabs}
\usepackage{threeparttable} 
\usepackage{paracol}
\usepackage{algorithm}
\usepackage{algorithmic}
\jmlrheading{}{}{}{}{}{Huihui Wang et al.}

\ShortHeadings{dgm for eit reconstruction}{dgm for eit reconstruction}
\firstpageno{1}

\begin{document}

\title{A Comparative Study of Variational Autoencoders, Normalizing Flows, and Score-based Diffusion Models for Electrical Impedance Tomography}

\author{\name Huihui Wang \email huihuiwang@csu.edu.cn \\
       \addr School of Mathematics and Statistics\\
       Central South University
       \AND
       \name Guixian Xu \email xuguixian@csu.edu.cn \\
       \addr School of Mathematics and Statistics\\
       Central South University
       \AND
       \name Qingping Zhou \email
       qpzhou@csu.edu.cn\\
       \addr School of Mathematics and Statistics\\
       Central South University
}


\maketitle

\begin{abstract}
Electrical Impedance Tomography (EIT) is a widely employed imaging technique in industrial inspection, geophysical prospecting, and medical imaging. However, the inherent nonlinearity and ill-posedness of EIT image reconstruction present challenges for classical regularization techniques, such as the critical selection of regularization terms and the lack of prior knowledge. Deep generative models (DGMs) have been shown to play a crucial role in learning implicit regularizers and prior knowledge. This study aims to investigate the potential of three DGMs—variational autoencoder networks, normalizing flow, and score-based diffusion model—to learn implicit regularizers in learning-based EIT imaging. We first introduce background information on EIT imaging and its inverse problem formulation. Next, we propose three algorithms for performing EIT inverse problems based on corresponding DGMs. Finally, we present numerical and visual experiments, which reveal that (1) no single method consistently outperforms the others across all settings, and (2) when reconstructing an object with 2 anomalies using a well-trained model based on a training dataset containing 4 anomalies, the conditional normalizing flow (CNF) model exhibits the best generalization in low-level noise, while the conditional score-based diffusion model (CSD*) demonstrates the best generalization in high-level noise settings. We hope our preliminary efforts will encourage other researchers to assess their DGMs in EIT and other nonlinear inverse problems.
\end{abstract}


\section{Introduction}

Electrical impedance tomography (EIT) is a radiation-free imaging technique that enables repetitive and non-invasive measurement of regional changes within an object, making it useful for various applications~\cite{adler2021electrical}, such as medical diagnosis, industrial inspection, and environmental monitoring. 
However, the intrinsic nonlinearity and ill-posedness of the image reconstruction in EIT pose a significant challenge, especially when the measured data are noisy~\cite{gehre2012sparsity}. 
Classical iterative approaches, such as Gauss-Newton~\cite{colibazzi2022learning} and Landweber iterative algorithm~\cite{1315989}, face challenges such as the critical choice of regularization terms and the lack of prior knowledge. To enhance EIT image reconstruction, researchers have enlisted the help of deep generative models (DGM) that benefit the prior knowledge in big data and the regularizer term representation learning of the deep convolutional networks~\cite{zhang2020supervised}. 

Combining the strengths of probabilistic modeling and data-driven deep learning solutions, DGMs have become popular in many areas of nonlinear inverse problems due to their state-of-the-art performance. Generative adversarial networks~\cite{goodfellow2020generative}, variational autoencoders~\cite{kingma2014autoencoding}, and normalizing flows~\cite{DBLP:journals/corr/DinhKB14} are common typical generative models. 
When these generative models are used to solve nonlinear inverse problems,  \cite{bohra2022bayesian} presents a Bayesian reconstruction framework and develops a tractable posterior sampling scheme based on the Metropolis-adjusted Langevin algorithm where the prior knowledge about the image is specified by a deep generative model to solve optical diffraction tomography. \cite{9495275} presents a unique ProjectionGAN that fixes the missing cone issues with the same issue as the former by utilizing a novel unsupervised learning method.
When combined with variational autoencoder, \cite{seo2019learning} converts the ill-posed inverse problem to the well-posed one. \cite{li2021differentiable} proposes to use novel differentiable data-dependent layers, where the custom operators are defined by solving optimization problems, to re-parameterize and gaussian the latent tensors in generative models such as StyleGAN2, which can therefore produce high-fidelity in-distribution solutions and be valid in solving nonlinear problems, such as eikonal tomography.
Score-based Diffusion models are a class of likelihood-based generative models and have achieved state-of-the-art results in density estimation as well as in sample quality~\cite{song2021maximum}, which learn to match a data distribution by reversing a multi-step, gradual noising process. 

In this paper, we systematically survey the recent applications of deep generative models to inverse problems and find that they are solved by constructing conditional generative models. When using conditional generative models, such as the one in~\cite{saharia2022image}, to solve inverse problems, they can generate high-quality samples but require a large number of training iterations and time-consuming inference. A similar situation arises in~\cite{denker2021conditional}. Moreover, a common issue with conditional generative models is the need to retrain the model when the dataset changes, resulting in low efficiency. Consequently, an unconditional model may be a better choice, as it can avoid repetitive training and allow for inference using the same trained model. Recent studies~\cite{song2022solving,singh2022conditioning} have demonstrated that utilizing pre-trained score-based diffusion networks and incorporating distorted images with the corresponding reverse diffusion process yields comparable visual quality to traditional conditional learning techniques.
However, only a few studies employing this model have focused on nonlinear inverse problems thus far. By utilizing posterior sampling approximations, \cite{chung2023diffusion} extends diffusion solvers to effectively address various noisy nonlinear inverse problems. Similarly, $\Pi$GDM~\cite{song2023pseudoinverseguided}, like~\cite{kawar2022jpeg}, calculates conditional scores from the measurement model of inverse problems without the need for additional training, making it applicable for solving nonlinear problems as well. Based on~\cite{song2022solving}, we here propose our own conditional score-based diffusion model to solve EIT reconstruction.

The main contributions of our work are:
    
\begin{itemize} 
    \item Developing a novel methodology of nonlinear inverse problems that takes advantage of the deep generative models (DGMs) in learning implicit regularizers and prior knowledge. As an example, we demonstrate the EIT reconstruction problem using variational autoencoder networks, normalizing flow, and score-based diffusion models.

    \item Proposing a new conditional score-based diffusion model ($\mathrm{CSD}^*$) for reconstruction.
    It treats the score-based diffusion as a post-processing operator of the solution of the Gauss-Newton method, leading to a significant speed-up in sampling compared to the original score-based diffusion model.

    \item Illustrating the power of the proposed methodology in a variety of numerical and visual experiments, including reconstruction estimation, robustness to different noisy levels, and generalization capability.
\end{itemize}

This paper is organized as follows: Section~\ref{sec:eit} provides an overview of EIT forward and inverse problems, along with a classic reconstruction method. In Section~\ref{sec:method}, we review several widely-used deep learning models and briefly describe their applications in solving the EIT reconstruction problem. Section~\ref{sec:experiments} presents the main experimental results, including performance evaluations under higher noise levels and generalization studies. Finally, Section~\ref{sec:conclusion} offers a conclusion and future outlook.

\section{EIT} \label{sec:eit}

Focusing on medical demands, EIT is applied to obtain tomographic images of the conductivity distribution to figure out the lesion area. Essentially, by changing the electrical current on the electrodes and measuring the boundary voltage difference of the object to reconstruct the conductivity distribution of inclusions~\cite{ guo2021construct}.

The EIT forward problem, given the conductivity distribution and the injected current, calculates the voltage across the electrodes through the forward model, which can be depicted through the following elliptic partial differential equation~\eqref{eq:EIT}~\cite{somersalo2004statistical}. Specifically, in the bounded domain $\Omega \in \mathbb{R}^n, n=2,3$, let $\Omega$ be simply connected and have smooth boundary $\partial \Omega$. Assuming there are $L$ electrodes, they can be modeled as strictly disjoint surface patches $e_{\ell} \subset \partial \Omega $, $1 \leq \ell \leq L$, when $\ell \neq k$, $\overline{e_{\ell}} \cap \overline{e_{k}} = \emptyset $. When different currents $I_{\ell} (1 \leq \ell \leq L)$ are injected to the boundary $\partial\Omega$, the target electrical potential $ V_{\ell} (1 \leq \ell \leq L) $ with different boundary settings can be calculated, in which $z_{\ell}$ represents the contact impedance at the $\ell_{th}$ electrode. Supposing the background consists of a homogeneous material with the same conductivity $\sigma_0$. The whole area can be seen as $\sigma \in L^{\infty}(\Omega)$, then the inhomogeneous inclusions can be detected by $\sigma - \sigma_0$. In practice, it is usually solved by the finite element method (FEM). 

\begin{figure}[ht]	
    \centering
    \includegraphics[width =0.7 \linewidth ]{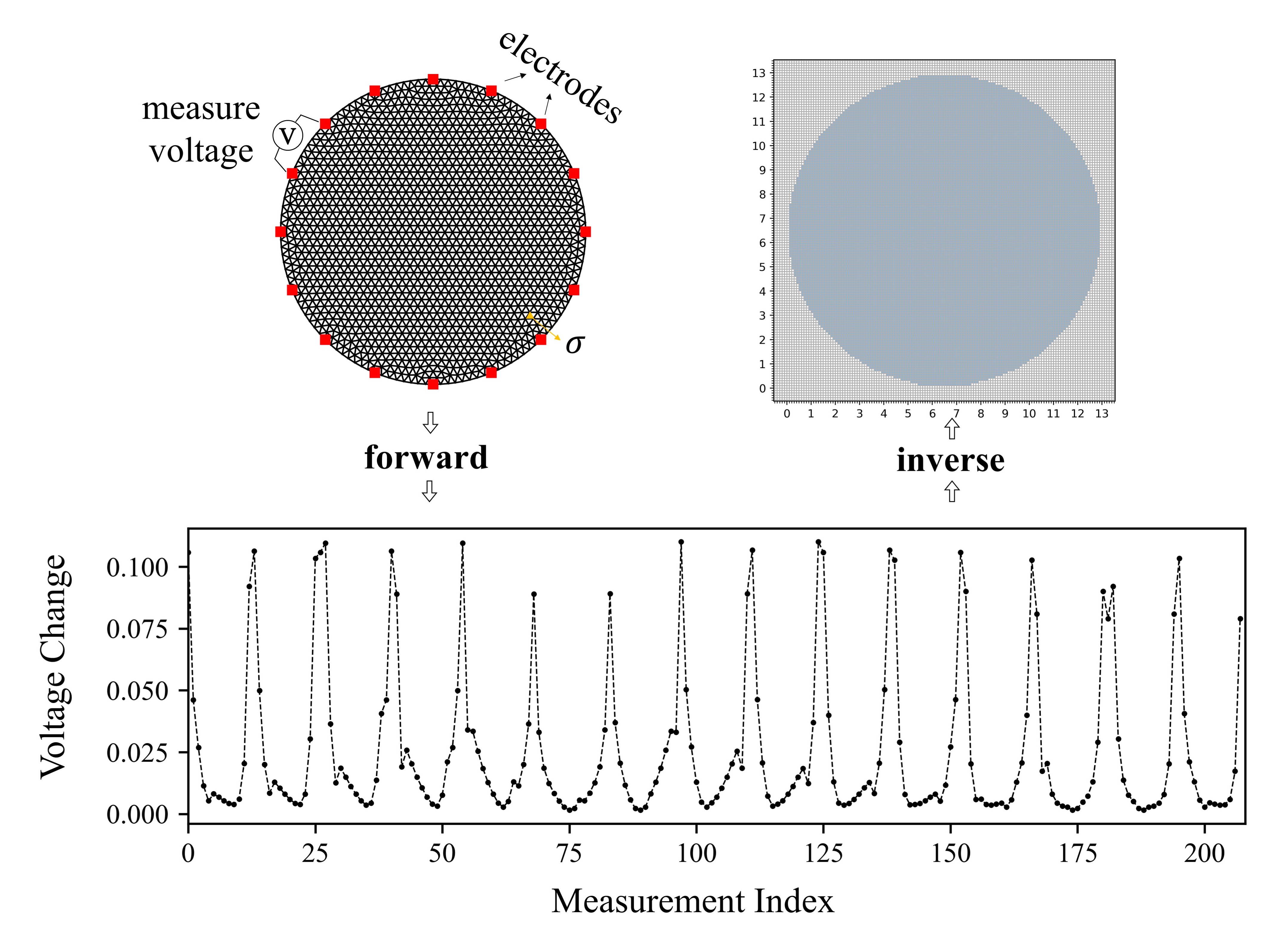}
    \caption{\textbf{EIT forward and inverse.} In the circular boundary 16 electrodes are equally spaced located, and the adjacent injection-adjacent measurement protocol is used to compute the voltage difference~\cite{liu2018pyeit}. The forward process uses FEM to transform the mesh data $\sigma$ into the measured voltage difference $v_{208 \times 1}$; the inverse one aims to reconstruct the ground truth in the form of $128\times128$ pixels from this measured voltage difference, which can avoid inverse crime.}\label{fig:EIT}
\end{figure}


\begin{subequations}
\begin{align}
\nabla \cdot (\sigma \nabla u) = 0 \ \quad &\text{in} \quad \Omega,\label{ell1} \\
\int_{e_{\ell}}\sigma \frac{\partial u}{\partial n}dS = I_{\ell} \ \quad &\text{on} \quad \partial \Omega, \quad\ell = 1, 2, \cdots, L, \label{ell2} \\
\sigma \frac{\partial u}{\partial n}\bigg|_{\partial \Omega \backslash \cup e_{\ell}} = 0 \ \quad &\text{on} \quad \partial \Omega, \quad\ell = 1, 2, \cdots, L,\label{ell3}\\
u + z_{\ell} \sigma \frac{\partial u}{\partial n} \vert_{e_{\ell}} = U_{\ell}\quad &\text{on} \quad \partial \Omega, \quad \ell = 1, 2, \cdots, L, \label{ell4} 
\end{align}
\label{eq:EIT}
\end{subequations}

In addition, the injected current and the measured potential satisfy the following two unique constraints
\begin{equation}
\sum_{\ell=1}^L I_{\ell} = 0, \qquad \sum_{\ell=1}^L U_{\ell} = 0,
\end{equation}

Considering the presence of additive noise in measured data, We abbreviate the noisy nonlinear observation model as follows
\begin{equation}
v=F(\sigma)+\eta,
\label{eq:eit_nonlinear}
\end{equation}
where $v \in \mathbb{R}^{m}$ denotes the vector of all the measured electrode potentials, whose dimension $m$ depends on the choice of the measurement protocol. Moreover, $\eta \in \mathbb{R}^{m}$ represents a zero-mean Gaussian-distributed measurement noise vector. According to the forward operator $F$, the inverse problem, inner conductivity distribution reconstruction, is to estimate the conductivity $\sigma$ in $\Omega$, both for the position and shape of the inclusions by solving the following nonlinear least squares problem
\begin{equation}
{\sigma}^{\prime}= \underset{\sigma}{\arg \min }f(\sigma) \triangleq  
 \underset{\sigma}{\arg \min } \frac{1}{2}\|v-F(\sigma)\|_2^2+\lambda R(\sigma),
\label{eq:eit-inverse}
\end{equation}
where $\frac{1}{2}\|v-F(\sigma)\|_2^2$ is the data fidelity term, $R(\sigma)$ is the regularization term, and $\lambda$ is the trade-off parameter.  

In EIT, however, both forward solving by the FEM and inverse problem solving are widely recognized as challenging. For the former, the voltage differences are calculated at discrete points, the solving therefore suffers from a high condition number, which is dependent on the discretization~\cite{bar2021strong}. In EIT reconstruction, one of the primary issues is that the number of variables to be solved is significantly greater than the number of observations. While the voltage data is sensitive to errors in the forward model, it is not as sensitive to slight perturbations in conductivity. Similarly, the boundary voltage measurements are highly sensitive to changes in impedance near the electrodes, but not as sensitive to changes in central impedance. These factors contribute to EIT reconstruction being a highly ill-posed and challenging problem. In this paper, we mainly focus on its inverse problem.

\noindent \textbf{The Gauss-Newton Method.}
The Gauss-Newton method is one of the widely spread methods in solving the EIT inverse problem, which performs a line search strategy in a specific descent direction and ignores all of the second-order terms in Taylor's expansion approximation of the function. Specifically
\begin{equation}
f(\sigma+d) \approx f(\sigma)+\nabla f(\sigma)^T d+\frac{1}{2} d^T \nabla^2 f(\sigma) d, 
\label{GN0}
\end{equation}
where the gradient and the Hessian of $f(\sigma)$ are given respectively by
\begin{equation}
\nabla f(\sigma)=J(\sigma)^T\left(F(\sigma)-v\right), \quad \nabla^2 f(\sigma)=J(\sigma)^T J(\sigma)+\sum_{s=1}^m r_s(\sigma) \nabla^2 r_s(\sigma),
\end{equation}
with $J(\sigma)$ being the Jacobian matrix of $r(\sigma):=F(\sigma)-v$ and $s$ being the voltage difference index. The final search direction $d^{G N}$ can be obtained by basically ignoring all the second-order terms from $\nabla^2 f(\sigma)$
\begin{equation}
    J(\sigma)^T J(\sigma) d^{G N} = -J(\sigma)^T\left(F(\sigma)-v\right),
\label{eq:ori-gn}    
\end{equation}

Starting from an initial guess $\sigma_0$, the Gauss-Newton algorithm (see Algorithm \ref{alg:gn}) computes $\sigma_{k+1}$ based on the previous estimate $\sigma_{k}$ and the direction $d_k^{GN}$ 

\begin{equation}
\sigma_{k+1}=\sigma_k+d_k^{G N},
\label{eq:gn}
\end{equation}

\begin{algorithm}[ht]
\setcounter{algorithm}{0}
\caption{The Gauss-Newton method (GN).}
\label{alg:gn}
\begin{algorithmic}[1]
    \REQUIRE $\sigma_0$, $v$, $\lambda > 0$  ($\sigma_0$ follows the default setting in~\cite{liu2018pyeit})
    \ENSURE $\sigma_{k}$    
    
    \FOR{$k=1, \cdots, $}
    \STATE \ \, compute Jacobian matrix $J(\sigma)$
    \STATE \ \, compute direction $d_k^{\mathrm{LM}}$ by \eqref{eq:lm}
    \STATE \ \, compute $\sigma_{k} = \sigma_{k-1} + d_k^{\mathrm{LM}}$
    \ENDFOR
    \RETURN $\sigma_{k}$
\end{algorithmic}
\end{algorithm}

However, due to the Jacobian matrix's compact structure, the coefficient matrix possesses an unbounded (discontinuous) inverse. Applying the Gauss-Newton method will therefore yield inaccurate solutions. One could instead employ some form of regularization on the sought solution $\sigma$. The Levenberg-Marquardt algorithm provides direct regularization on the ill-conditioned system by incorporating a diagonal matrix whose values are the diagonal elements of the matrix $J(\sigma)^TJ(\sigma)$~\cite{liu2018pyeit}. This method is also given as a baseline in experiments shown in Section~\ref{sec:experiments}. Referring to the objective function in \eqref{eq:eit-inverse}, the descent direction of the EIT nonlinear problem can be approximated by solving the following system, and its algorithm is similar to Algorithm \ref{alg:gn} but replacing \eqref{eq:ori-gn} by \eqref{eq:lm}

    \begin{equation}
    \left(J\left(\sigma_k\right)^TJ\left(\sigma_k\right)
    +
    \lambda~\mathrm{diag}\left(J\left(\sigma_k\right)^T J\left(\sigma_k\right)\right)
    \right) d_k^{\mathrm{LM}}=J\left(\sigma_k\right)^T\left(v-F\left(\sigma_k\right)\right).
    \label{eq:lm}
    \end{equation}

Many different regularization methods have been proposed. However, the effectiveness of these kinds of methods hinges on the crucial selection of the regularization parameter $\lambda$, which process typically involves exhaustive research that can be arduous and time-consuming, making it unfeasible for practical applications.

\section{Methods} \label{sec:method}
This section will provide a concise review of several widely used deep generative models in solving the EIT reconstruction problem.

\begin{figure}[ht] 	
    \centering
    \includegraphics[width = 0.8\linewidth]{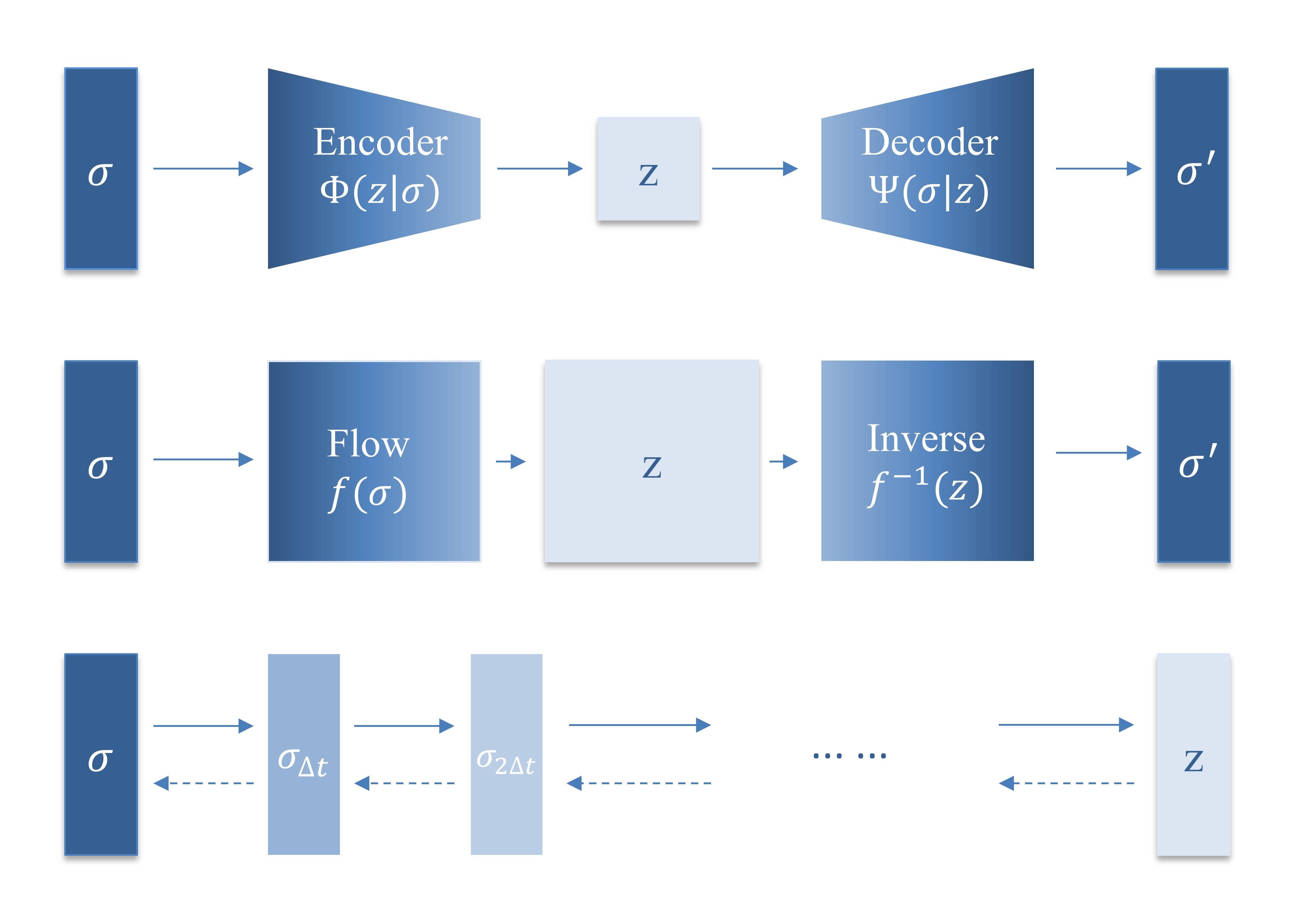}
    \caption{\textbf{Brief structures of three generative models.} The first row is the VAE model, the middle is normalizing flow, and the bottom is the score-based diffusion model. {\color{black}The symbol $\sigma$ denotes the ground truth, while $z$ represents the latent representation following a transformation, and ${\sigma}^{\prime}$ signifies the reconstructed image. In the context of the diffusion process, $\sigma_{\Delta t}$ represents the intermediate state of $\sigma$ at a specific time step in the process.}}\label{fig:com}
\end{figure}

\subsection{Conditional variational autoencoder} \label{subsection:cvae}
\begin{figure}[ht] 	
    \centering
    \includegraphics[width = 0.8\linewidth]{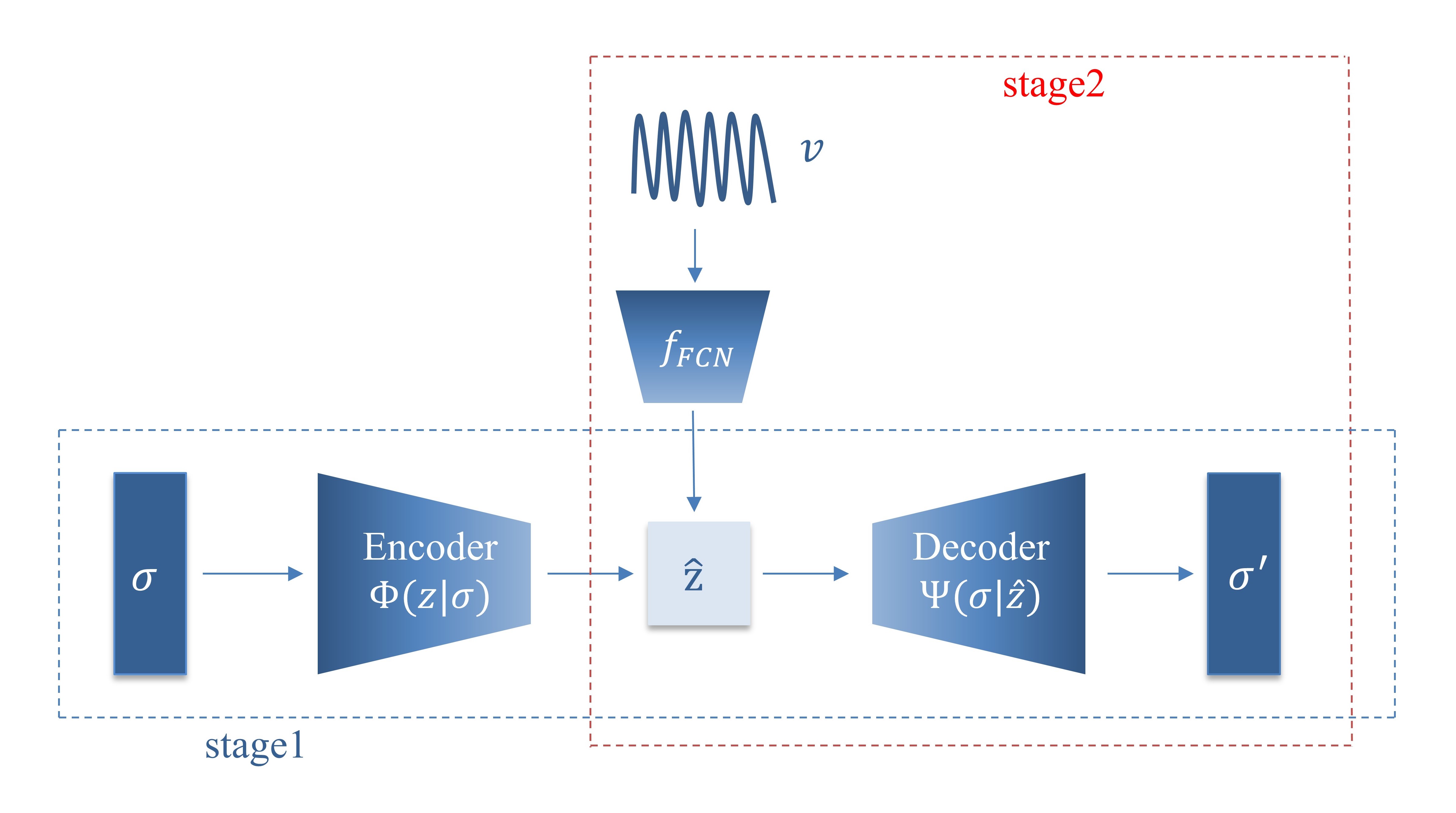}
    \caption{\textbf{Model structure of CVAE mentioned above}. The blue frame represents the first stage in CVAE also the original structure of VAE, and the red is the second stage to incorporate condition $v$ through $f_{FCN}$.}
    \label{fig:cvae}
\end{figure}

Variational autoencoder (VAE)~\cite{kingma2014autoencoding} is an effective learning model with the encoder-decoder architecture as shown in the first row in Figure \ref{fig:com}. The encoder part tries to encode data into a hidden representation or learn it hidden while the decoder tries to decode the hidden representation to input space.

\begin{algorithm}[htbp]
\setcounter{algorithm}{1}
\caption{Conditional variational autoencoder(CVAE) algorithm.}
\label{alg:cvae}
\begin{algorithmic}[1]
    \STATE \textbf{Stage1. Training VAE to find a low-dimensional representation}
    \FOR{number of training steps} 
    \STATE \ \, Sample the minibatch of $m$ images $\left\{\sigma_1, \cdots, \sigma_m\right\}$ from training data.
    \STATE \ \, Sample the minibatch of $m$ auxiliary noise {\color{black}$\left\{\varepsilon_{\text {noise}, 1}, \cdots, \varepsilon_{\text {noise}, m}\right\}$} from standard normal $\mathcal{N}(0, I)$.
    \STATE \ \, Update the parameters of VAE using the gradient of the loss in \eqref{vae}.
    \ENDFOR
    \vspace{3 mm}
    \STATE \textbf{Stage2. Training the nonlinear regression map $f_{\mathrm{FCN}}$ }
    \FOR{number of training steps}
    \STATE \ \, Sample the minibatch of m images $\left\{\sigma_1, \cdots, \sigma_m\right\}$ from training data and encode them to $\left\{z_1, \cdots, z_m\right\}$.
    \STATE \ \, Sample the minibatch of $m$ paired voltage difference data $\left\{v_1, \cdots, v_m\right\}$.
    \STATE \ \, Update the parameters of $f_{\mathrm{FCN}}$ using gradient of loss in \eqref{fcn}.
    \ENDFOR
    \vspace{3 mm}
    \STATE \textbf{Stage3. EIT image reconstruction}
    \STATE Combining the trained nonlinear regression map $f_{\mathrm{FCN}}$ with VAE decoder $\Psi$.
    \vspace{1 mm}
\end{algorithmic}
\end{algorithm}

Specifically, the output of encoder $\Phi$ can be described as 
\begin{equation}
z_i = \Phi(\sigma_i)=\mu_i+\sigma_i^{std} \odot \mathbf{\varepsilon}_{\text{noise }},
\end{equation}
where $\sigma_i$ is the input, $\mu_i=(\mu_i(1), \cdots, \mu_i(k)) \in \mathbb{R}^k$ and $\sigma_i^{std}=(\sigma_i^{std}(1), \cdots, \sigma_i^{std}(k)) \in \mathbb{R}^k$ are the mean and the standard deviation of encoder respectively(To distinguish from input $\sigma_i$, we here label the standard deviation of encoder as $\sigma_i^{std}$). $\mathbf{\varepsilon}_{\text {noise }}$ is an auxiliary noise variable sampled from standard normal distribution $\mathcal{N}(0, I)$, which means $z$ is a nondeterministic variable. By introducing noise, $\sigma_i$ now can be encoded as a range of perturbations, $z_i \sim \mathcal{N}(\mu_i, \Sigma_i)$, in the latent space. Then the decoder $\Psi$ is used to map these perturbations to useful images of $\sigma^{\prime}$. An additional restriction is added to ensure variation in the latent space by penalizing the Kullback-Leibler divergence loss between $\mathcal{N}(\mu_i, \Sigma_i)$ and $\mathcal{N}(0, I)$.
\begin{equation}
D_{K L}\left(\mathcal{N}\left(\mu_i, \Sigma_i\right) \| \mathcal{N}(0, I)\right)=\frac{1}{2} \sum_{j=1}^k\left[\left(\mu_i(j)^2+\sigma^{std}_i(j)^2-\log \sigma^{std}_i(j)-1\right]\right.,
\end{equation}
Then the whole objective function can be described as 
\begin{equation}
(\Psi, \Phi)=\underset{(\Psi, \Phi) \in \mathbb{VAE}}{\operatorname{argmin}} \frac{1}{N} \sum_{i=1}^N\left[\left\|\Psi \circ \Phi\left(\sigma_i\right)- \sigma_i\right\|^2+D_{K L}\left(\mathcal{N}\left(\mu_i, \Sigma_i\right) \| \mathcal{N}(0, I)\right)\right],
\label{vae}
\end{equation}

When it comes to the nonlinear inverse problem EIT, to solve the highly under-determined problem, the author~\cite{seo2019learning} follows a paradigm that realistic lung images lie on a nonlinear manifold that is much lower dimensional than the space of all possible images. That means by finding a suitable set which is hoped to be a low dimensional manifold of images displaying lung ventilation, the original problem can be converted into the approximately well-posed one. Consequently, they take advantage of the encoder-decoder architecture of VAE and then design a two-stage conditional variational autoencoder (CVAE) (see Algorithm \ref{alg:cvae} and Figure \ref{fig:cvae}).

Given the paired training dataset $\left\{\left(\sigma_i, v_i \right)\right\}_{i=1}^N$, where $N$ is the total number of training samples, by conditioning the latent space on the voltage difference $v_i$, CVAE attempts to reconstruct the ground truth $\sigma_i$.
The VAE part is still trained according to the original VAE objective function in \eqref{vae}. As for the fully connected encoder, it is trained by
\begin{equation}
\mathcal{L}_2=\frac{1}{N} \sum_{i=1}^N\left\|f_{\mathrm{FCN}}\left(v_i\right)-z_i\right\|^2,
\label{fcn}
\end{equation}
It is worth noting that we have replaced the original $\mu_i$ used in~\cite{seo2019learning} as $z_i$ in \eqref{fcn}, which improves the performance, and all subsequent CVAE results are derived from this modified version.

Overall, this two-stage model can be described as
\begin{equation}
\sigma^{\prime} = \Psi \circ f_{\mathrm{FCN}}(v),
\end{equation}

\subsection{Conditional normalizing flow} \label{subsection:cnf}
Normalizing flow is a powerful tool that enables the transformation of simple probability distributions into highly complex ones, with broad applications in generative models, reinforcement learning, variational inference, and more. Figure \ref{fig:com} depicts the structure of normalizing flow in the middle row. Let $z$ be a random variable with a known probability density function $p_z$. This distribution is called the base distribution and should be simple to evaluate and sample from. It induces a distribution via the invertible transformation $T_\theta$ on the image space $\sigma=T_\theta(z)$. Using the change-of-variable theorem, it is possible to evaluate the likelihood of this induced distribution:
\begin{equation}
p_\theta(\sigma)=p_{z}\left(T_\theta^{-1}(\sigma)\right)\left|\operatorname{det} J_{T_\theta}\left(T_\theta^{-1}(\sigma)\right)\right|^{-1},
\end{equation}
in which $T_\theta: z \rightarrow \sigma$, which is parameterized by $\theta$. This transformation has to be invertible, and both $T_\theta$ and $T_{\theta}^{-1}$ have to be differentiable, $J_{T_\theta}$ denotes the Jacobian of $T_\theta$. 
When applied to solve inverse problems, the conditional normalizing flow is used to build a probabilistic model $p_\theta(\sigma \mid v)$ to approximate the unknown conditional distribution $p(\sigma \mid v)$, where $v$ is the corresponding response of $\sigma$. Similar to normalizing flow, but the transformation has been replaced by $T_\theta: z \times v \rightarrow \sigma$, which is the crucial point that has to be invertible with respect to the first argument, and both $T_\theta(\cdot ; v)$ and $T_\theta^{-1}(\cdot ; v)$ have to be differentiable for $v$.
 \begin{equation}
p_\theta(\sigma \mid v)=p_{z}\left(T_\theta^{-1}(\sigma ; v)\right)\left|\operatorname{det}\left(\frac{\partial T_\theta^{-1}(\sigma ; v)}{\partial \sigma}\right)\right|,
\end{equation}

Given the independent and identically distributed paired training dataset $\left\{\left(\sigma_i, v_i \right)\right\}_{i=1}^N$, the maximum likelihood loss is used to fit the parameter $\theta$, and the Jacobian matrix is denoted by the abbreviation $J_{T_\theta^{-1}}(\sigma ; v)=\frac{\partial T_\theta^{-1}(\sigma; v)}{\partial \sigma}$

\begin{equation}
\begin{aligned}
\max _\theta \mathcal{L}(\theta) & =\sum_{i=1}^N \log \left(p_\theta\left(\sigma_i \mid v_i \right)\right) \\
& =\sum_{i=1}^N\left(\log p\left(T_\theta^{-1}\left(\sigma_i ; v_i\right)\right)+\log \left(\left|\operatorname{det} J_{T_\theta^{-1}}\left(\sigma_i ; v_i\right)\right|\right)\right),
\end{aligned}
\end{equation}

\begin{figure}[ht]	
    \centering
    \includegraphics[width =0.8 \linewidth ]{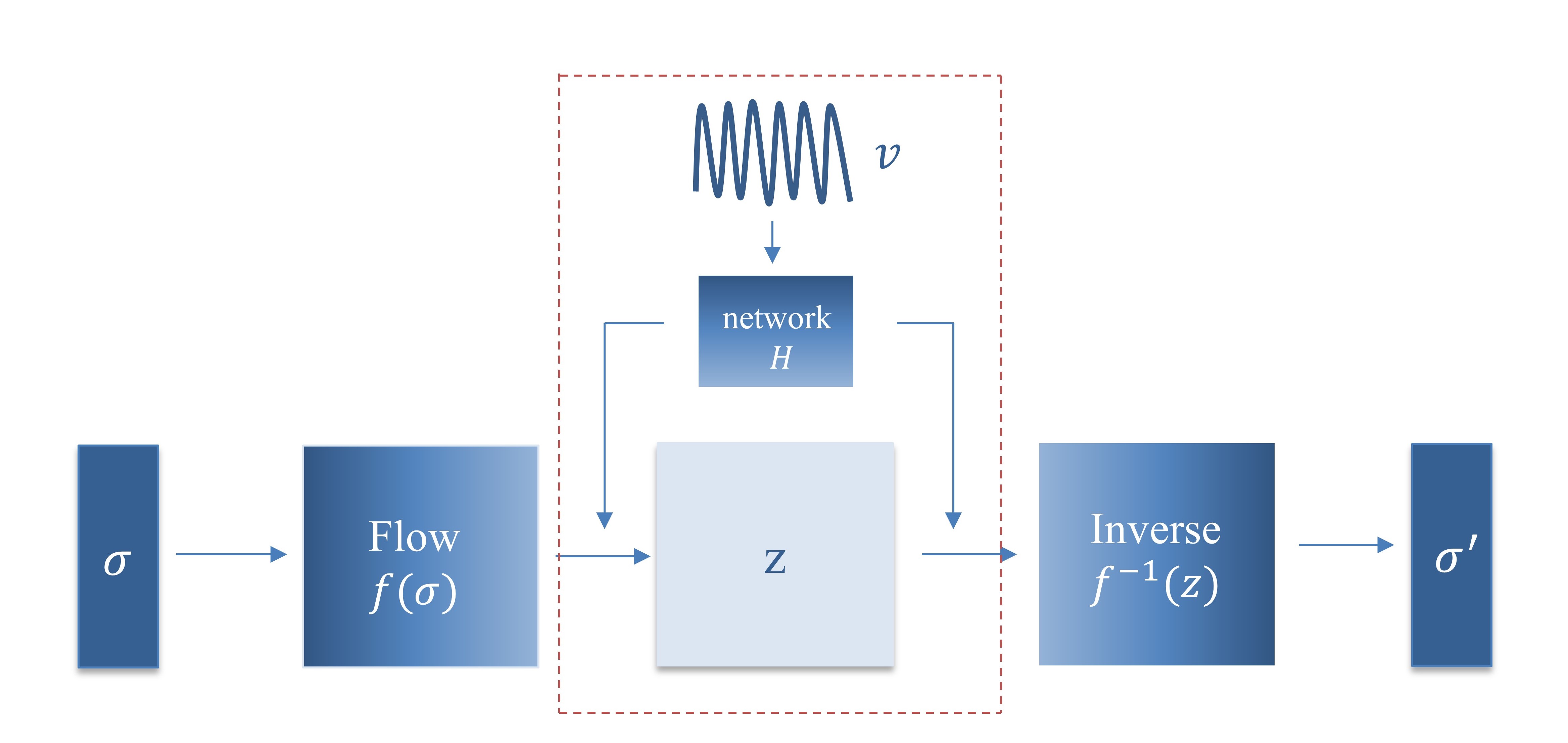}
    \caption{Structure of the CNF model. $v$ is input to the conditional coupling layer via an additional conditional network $H$: $h=H\left(v\right)$. }\label{fig:cnf}
\end{figure}

We here focus on the specific conditional normalizing flow (CNF) in~\cite{denker2021conditional} with multi-scale architecture combined by coupling blocks, downsampling, and splitting operations for conditional normalizing flow which follows the NICE and Real-NVP framework~\cite{DBLP:journals/corr/DinhKB14,dinh2017density} to solve EIT reconstruction and denote it as $f_{\theta}$. Expressive CNF models are primarily built using conditional coupling layers, which were introduced in~\cite{ardizzone2019guided} for modeling conditional image densities and can be thought of as an extension of the original coupling layers. For the conditional coupling layer, the author extends the coupling function $M$ to take the measurements $v$ as an additional input (see Figure \ref{fig:cnf}). Let $\sigma \in \mathbb{R}^n$ be the input, $v \in \mathbb{R}^m$ the measurements, and $I_1, I_2$ disjoint partitions of $\{1, \ldots, n\}$ with $\left|I_1\right|=d$ and $\left|I_2\right|=n-d$. Then, a conditional coupling layer is defined by

\begin{equation}
\begin{aligned}
& v^{I_1}=\sigma^{I_1}, \\
& v^{I_2}=G\left(\sigma^{I_2}, M\left(\sigma^{I_1}, v\right)\right),
\end{aligned} \label{coupling}
\end{equation}
where $G: \mathbb{R}^{n-d} \times \mathbb{R}^{n-d} \rightarrow \mathbb{R}^{n-d}$ is called the coupling law, which has to be invertible concerning the first argument. Conditional coupling function $M: \mathbb{R}^d \times \mathbb{R}^m \rightarrow \mathbb{R}^{n-d}$ does not need to be invertible and can be implemented as an arbitrary neural network. Here we adopt the affine coupling functions, the equation \eqref{coupling} can then be rewritten as
\begin{equation}
\begin{aligned}
& v^{I_1}=\sigma^{I_1} \\
& v^{I_2}=\sigma^{I_2} \odot \exp \left(s\left(\sigma^{I_1},v\right)\right)+t\left(\sigma^{I_1},v\right)
\end{aligned} \Leftrightarrow \begin{aligned}
\sigma^{I_1} & =v^{I_1} \\
\sigma^{I_2} & =\exp \left(-s\left(v^{I_1},v\right)\right) \odot\left(v^{I_2}-t\left(v^{I_1},v\right)\right)
\end{aligned}
\end{equation}
where $M(\cdot)=[s(\cdot),t(\cdot)]$, the scaling function $s(\cdot)$ and the translation $t(\cdot)$ are learned.

The measurements $v$ are transformed to $h=H\left(v\right)$ by a conditioning network $H$~\cite{ardizzone2019guided,winkler2020learning}, which can extract key features, feature extraction and density modeling therefore will be separated. It is possible to train the conditional network parallel to CNF or use a fixed, pre-trained network $H$. To fit the medical issue, the author implemented this conditioning network as a model-based inversion layer that maps from the measurement space to the image space, initial reconstruction, 
\begin{equation}
    \mathcal{L}_{cond}=\frac{1}{N} \sum_{i=1}^N\left\|    H\left(v_i\right)-\sigma_i\right\|^2
\end{equation}
and then concatenated with a convolutional neural network for post-processing to extract features.
The multi-scale architecture with $L$ scales can therefore be described by
\begin{equation}
\begin{aligned}
\sigma^{(0)} & =\sigma, \\
\left(z^{(i+1)}, \sigma^{(i+1)}\right) & =f^{(i+1)}\left(\sigma^{(i)}, H^{(i)}\left(v\right)\right), \\
z^{(L)} & =f^{(L)}\left(\sigma^{(L-1)}, H^{(L-1)}\left(v\right)\right), \\
z & =\left(z^{(1)}, \ldots, z^{(L)}\right),
\end{aligned}
\end{equation}
Each $f^i$ consists of a coupling $\rightarrow$ downsampling $\rightarrow$ coupling $\rightarrow$ splitting operation. The algorithm is given in Algorithm \ref{alg:cnf}.

\begin{algorithm}[htbp]
\setcounter{algorithm}{2}
\caption{Conditional normalizing flow(CNF) algorithm.}
\label{alg:cnf}
\begin{algorithmic}[1]
    \REQUIRE{The paired training dataset $\left\{\left(\sigma_i, v_i \right)\right\}_{i=1}^N$}\\
    \ENSURE{trained invertible neural network $f$}\\
    \FOR{number of training steps} 
    \STATE \ \, Sample the minibatch of m pairs $\left\{\left(\sigma_i, v_i \right)\right\}$ from training data.
    \STATE \ \, Update the parameters $\theta$ of $f$ using gradient of loss \begin{equation*}
        \min _\theta-\mathcal{L}(\theta)+\alpha \mathcal{L}_{cond}, \quad
        \text{where $\alpha>0$ is a weighting factor}
    \end{equation*}
    \ENDFOR
\end{algorithmic}
\end{algorithm}

\subsection{Conditional score-based diffusion model}
\subsubsection{General conditional score-based diffusion model (CSD)} 
Recently, score-based diffusion models have outperformed many other state-of-the-art generative models. It can be regarded as a process of adding noise forward and removing noise in reverse, and its basic model structure can be seen in the bottom line of Figure~\ref{fig:com}. When it is used to solve inverse problems, it can be divided into the following parts.

\noindent \textbf{The forward process.} Supposing continuously give Gaussian noise to data from an unknown data distribution $p_0(\sigma)$ through the following SDE
\begin{equation}
\mathrm{d} \sigma_t=f(\sigma_t, t)  \mathrm{~d} t+g(t) \mathrm{d} w_t, \quad t \in[0,1],
\label{eq:csd_f}
\end{equation}
where $f(\cdot \, ,t): \mathbb{R}^n \to \mathbb{R}^n$, $g(t) \in \mathbb{R}$ are called the drift and diffusion coefficients, $\lbrace w_t \in \mathbb{R}^n \rbrace_{t\in[0,1]}$ denotes a standard Wiener process, and $\lbrace \sigma_t \rbrace_{t\in[0,1]}$ represents the trajectory of random variables in the stochastic process. The marginal probability distribution of $\sigma_t$ is denoted as $p_t(\sigma_t)$, and the transition distribution from $\sigma_0$ to $\sigma_t$ as $p_{0t}(\sigma_t \mid \sigma_0)$. $f(\cdot \, ,t)$ and $g(t)$ will be specifically chosen according to $p_0(\sigma_0)$ to ensure $p_1(\sigma_1)$ approximates the final standard Gaussian distribution after the above noise-adding process.

\noindent \textbf{The reverse process.}
It could be seen as going back along the forward process to produce samples from $p_t(\sigma_t \mid v)$ by starting from $p_1(\sigma_1 \mid v)$ through the following conditional reverse-time SDE. Here we suppose $v$ does not have a separate diffusion process.
\begin{equation}
\mathrm{d} \sigma_t=\left[f(\sigma_t, t) -g(t)^2 \nabla_{\sigma_t} \log p_t\left(\sigma_t \mid v\right)\right] \mathrm{d} t+g(t) \mathrm{d} \overline{w}_t, \quad t \in[0,1], \label{ell10} 
\end{equation}
where $\overline{w}_{t\in[0,1]}$ denotes a standard Wiener process. In this direction, $dt$ is the infinitesimal negative time step, since the time is decreasing from $t=1$ to $t=0$.
\begin{figure}[ht]	
    \centering
    \includegraphics[width = \linewidth]{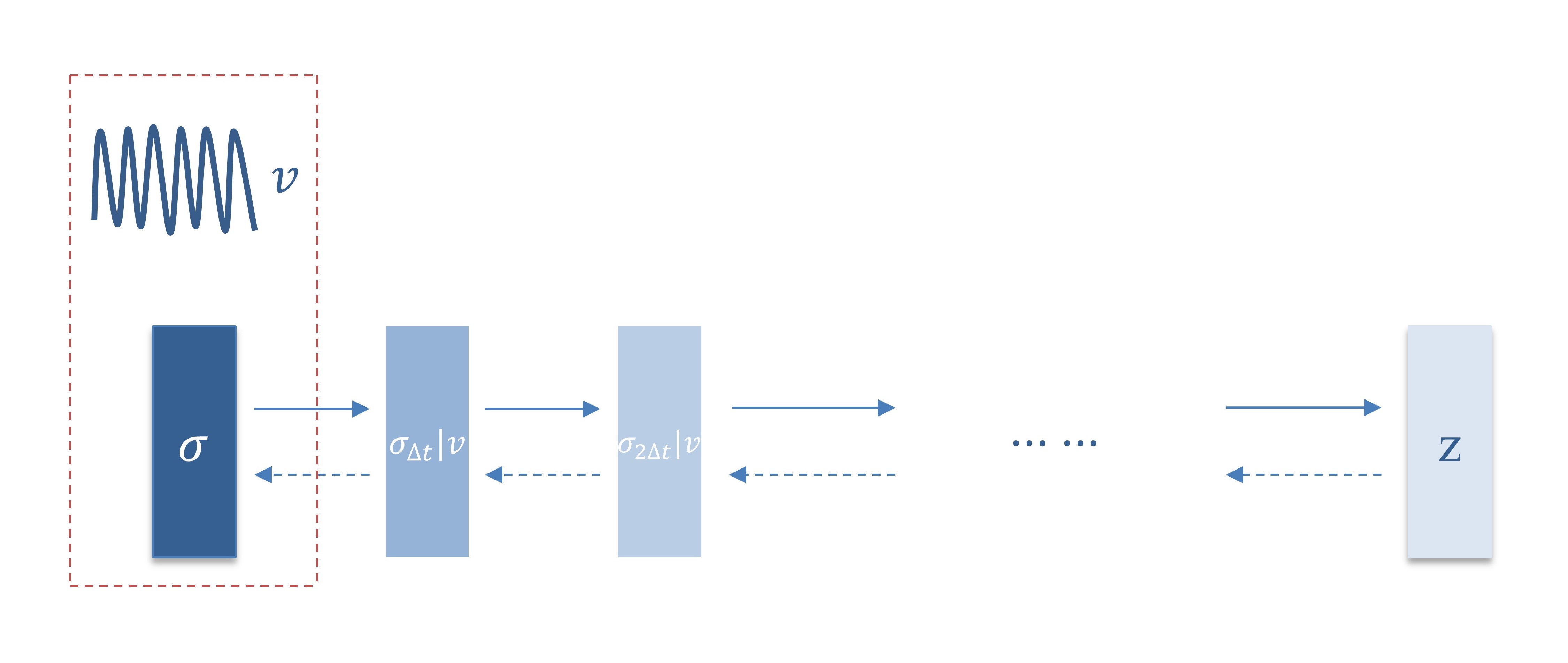}
    \caption{\textbf{Structure of conditional score-based diffusion model.} $v$ is usually incorporated by designing the conditional score function $\nabla_{\sigma_t} \log p_t\left(\sigma_t \mid v\right)$, shown in \eqref{eq:csd}.} \label{fig:conScore}
\end{figure}
This neural network is trained with denoising score matching~\cite{vincent2011connection} on training sample pairs $\left\{\left(\sigma_i, v_i \right)\right\}_{i=1}^N$ with the following objective function
\begin{equation}
\theta^*=\underset{\theta}{\arg \min } \mathbb{E}_t\left\{\lambda(t) \mathbb{E}_{\sigma_0,v} \mathbb{E}_{\sigma_t \mid \sigma_0}\left[\left\|s_{\theta}(\sigma_t,v, t)-\nabla_{\sigma_t} \log p_{t}(\sigma_t \mid v)\right\|_2^2\right]\right\},
\label{eq:csd}
\end{equation}
Where $\lambda:[0, 1] \rightarrow \mathbb{R}$ is a positive weighting function, $t$ is uniformly sampled over $[0, 1]$, $\sigma_0 \sim p_0(\sigma_0\mid v)$ and $\sigma_t \sim p_{t}(\sigma_t \mid v)$. Given enough data and model capacity, we can draw the optimal solution $s_{\theta}^*(\sigma, v,t)$, and then we can roll back to the initial data distribution along the following sample process. 

\noindent \textbf{Predictor-Corrector Samplers.} When the information of $\nabla_{\sigma_t} \log p_t(\sigma_t \mid v)$ from the objective loss function is available, the reconstruction result can be drawn by sampling. ~\cite{song2021scorebased} proposes employing score-based MCMC approaches such as Langevin MCMC~\cite{parisi1981correlation,grenander1994representations} to sample from $p_t$ directly and correcting the solution of numerical SDE solver. Sampling algorithms can be seen in Algorithm~\ref{alg:uncon}.

Specifically, in the inference stage, if we further write the conditional score function in \eqref{ell10} as 
\begin{equation}
\begin{aligned}
\nabla_{\sigma_t} \log p_t\left(\sigma_t \mid v\right) & =\nabla_{\sigma_t} \log p_t\left(v \mid \sigma_t\right)+\nabla_{\sigma_t} \log p_t\left(\sigma_t\right) \\
& \simeq \nabla_{\sigma_t} \log p_t\left(v \mid \sigma_t\right)+s_{\theta^*}\left(\sigma_t, t\right),
\label{eq:con-score}
\end{aligned}
\end{equation}
in which $s_{\theta^*}\left(\sigma_t, t\right)$ is the score function of unconditional score-based diffusion model. It means that if we can calculate $\nabla_{\sigma_t} \log p_t\left(v \mid \sigma_t\right)$ directly, we can combine it with an unconditional score function for inference, which therefore can avoid repetitive training for different datasets. However, we notice that $\nabla_{\sigma_t} \log p_t\left(v \mid \sigma_t\right)$ can be relatively easily calculated in linear inverse problem, while when the problem is nonlinear, the likelihood term $\log p_t\left(v \mid \sigma_t\right)$ is nontrivial~\cite{chung2023diffusion}.

\subsubsection{\texorpdfstring{The proposed method ($\mathrm{CSD^*}$)}{The proposed method (CSD*)}} \label{selfscore}

Rather than facing the nontrivial problem above directly, we incorporate $v$ in a totally different way. We mainly get our motivation from~\cite{chung2022come,guo2023physics}, that is we can generate the sample using the score-based diffusion model not from the pre-defined start point but a relatively clear state which is close to the ground truth, or a roughly estimated image using classical qualitative or quantitative methods, and then connecting score-based diffusion model as a post-processing operator.  We abbreviate this proposed conditional score-based diffusion model as $\mathrm{CSD}^*$.

\begin{figure}[htbp]	
    \centering
    \includegraphics[width = \linewidth]{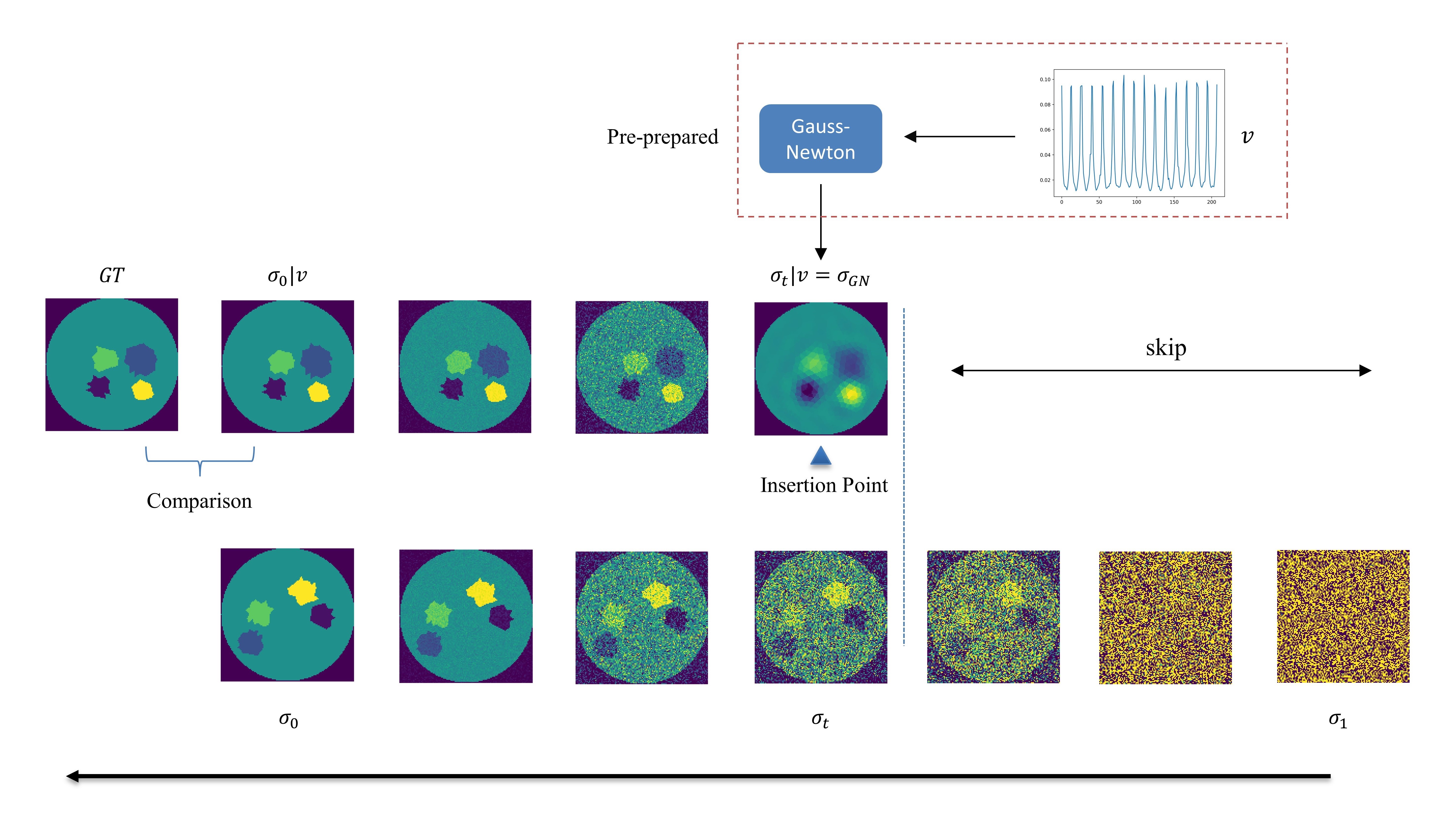}
    \caption{\textbf{Sampling in CSD vs sampling in $\mathrm{CSD}^*$.} Sampling in $\mathrm{CSD}^*$ is shown in the top line, which can be seen as the $\sigma_t \mid v$ substitutes $\sigma_t$ in the sampling of CSD and abandons states before $t$. GT is the ground truth.}\label{fig:Score}
\end{figure}

\noindent \textbf{Score-based diffusion as post-processing operator.}
We intend to hijack the reverse diffusion process by inserting a state which is a relatively clear outline of the target sample\textemdash the result of the Gauss-Newton method. The reason is that although it is still far from the target, the location of the anomaly is very clearly shown, and much physics information is included. Here a local optimal point is enough, which can therefore avoid exhaustive searching. On the other hand, it also avoids the long sampling process in the vanilla score-based diffusion model, which has been criticized since it
was proposed. The condition $v$ is therefore incorporated naturally.

It is worthwhile to note that the Gauss-Newton method often needs a relatively long time to generate not-bad results.
If we incorporate the Gauss-Newton method directly into the score-based diffusion model, the resulting end-to-end model will be highly time-consuming, which contradicts our initial goal. To address this issue, we decide to finish this phase ahead of schedule by performing the Gauss-Newton iteration first and then passing the results $\sigma_{GN}$ into the trained score-based diffusion model, shown in the red frame of Figure \ref{fig:Score}.

\begin{paracol}{2}
  \begin{algorithm}[htbp]
  \setcounter{algorithm}{3}
    \caption{Sampling in CSD}
    \label{alg:uncon}
    \begin{algorithmic}[1]
        \REQUIRE{$K$}
        \ENSURE{$\sigma_0 \mid v$}
        \vspace{1 mm}
        \STATE $\sigma_1 \sim p_1(\sigma_1), \Delta t \leftarrow \frac{1}{K}$
        \FOR {$i = K-1$ to $0$}
        \STATE $ t \leftarrow \frac{i+1}{K} $ 
        \vspace{28 mm}
       \STATE $ \sigma_{t-\Delta t} \leftarrow \sigma_t+g(t)^2 \log p\left(v \mid \sigma_t\right) \Delta t $
        \STATE $z \sim \mathcal{N}(0, I) $
        \STATE $ \sigma_{t-\Delta t} \leftarrow \sigma_{t-\Delta t}+g(t) \sqrt{\Delta t} z$
        \vspace{5 mm}
	\ENDFOR    	
    \end{algorithmic}
  \end{algorithm}
\switchcolumn
  \begin{algorithm}[htbp]
  \setcounter{algorithm}{4}
    \caption{Sampling in $\mathrm{CSD}^*$} 
    \label{alg:contest}
    \begin{algorithmic}[1]
        \REQUIRE{$K$,$K^{\prime},\sigma_{GN}$}
        \ENSURE{$\sigma_0 \mid v$}
        \vspace{1 mm}
        \STATE $\Delta t \leftarrow \frac{1}{K}$
        \FOR {$i = K^{\prime}-1$ to $0$}
        \STATE $ t \leftarrow \frac{i+1}{K} $ 
        \IF{$i=K^{\prime}-1$}
        \STATE { $\sigma_{t-\Delta t} \leftarrow \sigma_{GN}+g(t)^2 s_{\theta^*}\left(\sigma_{GN}, t\right) \Delta t$}
        \STATE $z \sim \mathcal{N}(0, I) $
        \STATE $ \sigma_{t-\Delta t} \leftarrow \sigma_{t-\Delta t}+g(t) \sqrt{\Delta t} z$
        \ELSE
        \STATE {$ \sigma_{t-\Delta t} \leftarrow \sigma_t+g(t)^2 s_{\theta^*}\left(\sigma_t, t\right) \Delta t $}
        \STATE $z \sim \mathcal{N}(0, I) $
        \STATE $ \sigma_{t-\Delta t} \leftarrow \sigma_{t-\Delta t}+g(t) \sqrt{\Delta t} z$
        \ENDIF
	\ENDFOR    	
    \end{algorithmic}
  \end{algorithm}
\end{paracol}

The mathematical model can be depicted as given presumed initial values of $\sigma_{0}$, $v$, $\lambda > 0$, reaching $\sigma_{GN}$ using Levenberg-Marquardt algorithm at first through Algorithm~\ref{alg:gn} shown in the red frame of Figure~\ref{fig:Score}. Then it will be inserted into the reverse diffusion process at $t$ point (or $K^{\prime}$ point in the original $K$ iterations)
shown in Figure~\ref{fig:Score} and experience a new round of updates

\begin{subequations}
\begin{align}
    &\sigma_{t-\Delta t} \leftarrow \sigma_{GN}+g(t)^2 s_{\theta^*}\left(\sigma_{GN}, t\right) \Delta t, \label{eq:sample1}
    \\
    &\sigma_{t-\Delta t} \leftarrow \sigma_t+g(t)^2 s_{\theta^*}\left(\sigma_t, t\right) \Delta t, \label{eq:sample2}
\end{align}
\end{subequations}
 updating $\sigma_t$ according to \eqref{eq:sample1} at time $t \to t-\Delta t$ and all subsequent updates are based on \eqref{eq:sample2}. The specific sampling process can be seen in Algorithm \ref{alg:contest}.

\noindent \textbf{Insertion point.} Another question is when to hijack, in other words, which state in the reverse diffusion process is the most suitable point to replace the original state? Actually, we don't have a universal method that can be directly used to locate the cut-off point, instead, we have to do some experiments to determine replacement in which point can generate the best sample quality. Although we can not put forward a fixed point, we notice that the parameter of starting point ranging from $K^{\prime}=500$ to $K^{\prime}=650$  usually can generate good results when it reaches $K=1000$ step, which conventionally needs a total of 1000 steps
to walk from the Gaussian noise to a clear image, and we usually check the final result every 10 steps. So it is not too burdensome for us to reach the best sample quality.

Subsequently, we will showcase the impact of this model in Section \ref{sec:experiments}.
 For the model architecture, we use the noise conditional score network (NCSN++cont) model straight proposed in~\cite{song2021scorebased}, which is continuous and has the double depth of the original NCSN network. In the sampling procedure, we adopt the predictor-corrector sampler with reverse diffusion as the predictor and annealed Langevin dynamics as the corrector.

\section{Datasets and Experiments} \label{sec:experiments}

\subsection{Datasets}
By using pyEIT~\cite{liu2018pyeit}, a tool based on FEM, we can calculate the corresponding voltage value for the electrical impedance. Specifically, we use a simplified version of the complete electrode model, referenced in Eq. \eqref{eq:EIT}. In this model, we assume the electrical current flows in and out of the imaging area, $\Omega$, via boundary nodes. The complete electrode model is preferred. Then to avoid inverse crime, which occurs when the same theoretical components—or ones that are very close to them—are used to synthesize and invert data in an inverse problem, we convert the original finite element data into 128 $\times$128 pixels data by inverse distance weighted interpolation~\cite{liu2018pyeit} and do various experiments based on this.

We mainly focus on two kinds of samples: 2 anomalies and 4 anomalies. The anomalies consist of two random circles with radii generated from the uniform distribution $U(-0.55,0.55)$ and the conductivity values are 0.5 and 1.5 respectively in each circle. The anomalies with four random circles with the same radii distribution but the conductivity values are 0.01, 0.1, 0.5, and 1.5 respectively. We train these models on them separately. In each dataset, we have a total of 23810 pairs of data, 21600 for training, 2160 for validation, and the remaining for testing. Different noise levels are also in consideration, which is measured by the signal-to-noise ratio (SNR) below
\begin{equation}
SNR = 10 \log_{10}\frac{\sum \sigma_{i}^2}{\sum (\sigma_{i}-\sigma_{GT})^2},
\end{equation}
where $\sigma_i$ is the noisy image and $\sigma_{GT}$ represents the ground truth.
\begin{itemize}
    \item \textbf{Test 1}  40dB Gaussian noise is added to measurements to make a comprehensive reconstruction comparison among these three generative models at the beginning in 2 anomalies, and 4 anomalies settings respectively.
    \item \textbf{Test 2} We then test the robustness of these models in a higher level noise, 25dB, with the same setting as the above dataset.
    \item \textbf{Test 3} To verify the generalization performance, we also apply these methods trained on low-level noise with four anomalies to the same and high-level noise with only two anomalies.
\end{itemize}

\subsection{Metrics}
The following metrics are included to compare the performances among these models: mean squared error (MSE), peak signal-to-noise ratio (PSNR), structural similarity (SSIM)~\cite{wang2004image}, and relative error (RE)

\begin{equation}
RE = \frac{\Vert \sigma^{\prime} - \sigma_{GT} \Vert_1}{\Vert \sigma_{GT} \Vert_1},
\end{equation}
absolute error(AE)
\begin{equation}
AE = \Vert \sigma^{\prime} - \sigma_{GT} \Vert_1,
\end{equation}
as well as dynamic range (DR)~\cite{herzberg2021graph} 
\begin{equation}
DR = \frac{max(\sigma^{\prime})-min(\sigma^{\prime})}{max(\sigma_{GT})-min(\sigma_{GT})} \times 100\%,
\end{equation}

Of the six metrics presented, MSE, AE, and RE indicate better performance when they have lower values. Conversely, SSIM and PSNR indicate superior with higher values. DR, on the other hand, suggests better results when it approaches 1.

\subsection{Experiments}
We mainly run our experiments on these three different conditional models (CVAE, CNF, $\mathrm{CSD}^*$) introduced in Section \ref{sec:method} and compare their performances according to the indexes mentioned above with the ground truth (GT) and the Gauss-Newton method (GN) visually and quantitatively. One key point needs to be noticed, the CVAE model and CNF are trained with condition together directly, in other words, $\{\sigma_i,v_i\},i=1,2,3 \cdots, 21600$ both will be used in the training stage. However, considering the condition $v$ is incorporated in $\mathrm{CSD}^*$ only in the test stage, we train it using $\{\sigma_i\},i=1,2,3 \cdots, 21600$ only and see how it performs when incorporating conditional information $v$ in the test stage. The implementation of all the experiments is accessible at \url{https://github.com/adahfbch/DGM-EIT}.

\subsubsection{Performance comparison}
\begin{figure}[htbp]
\centering
\subfigure[GT]{
    \begin{minipage}[b]{0.19\linewidth} 
    \centering
    \includegraphics[width=1in]{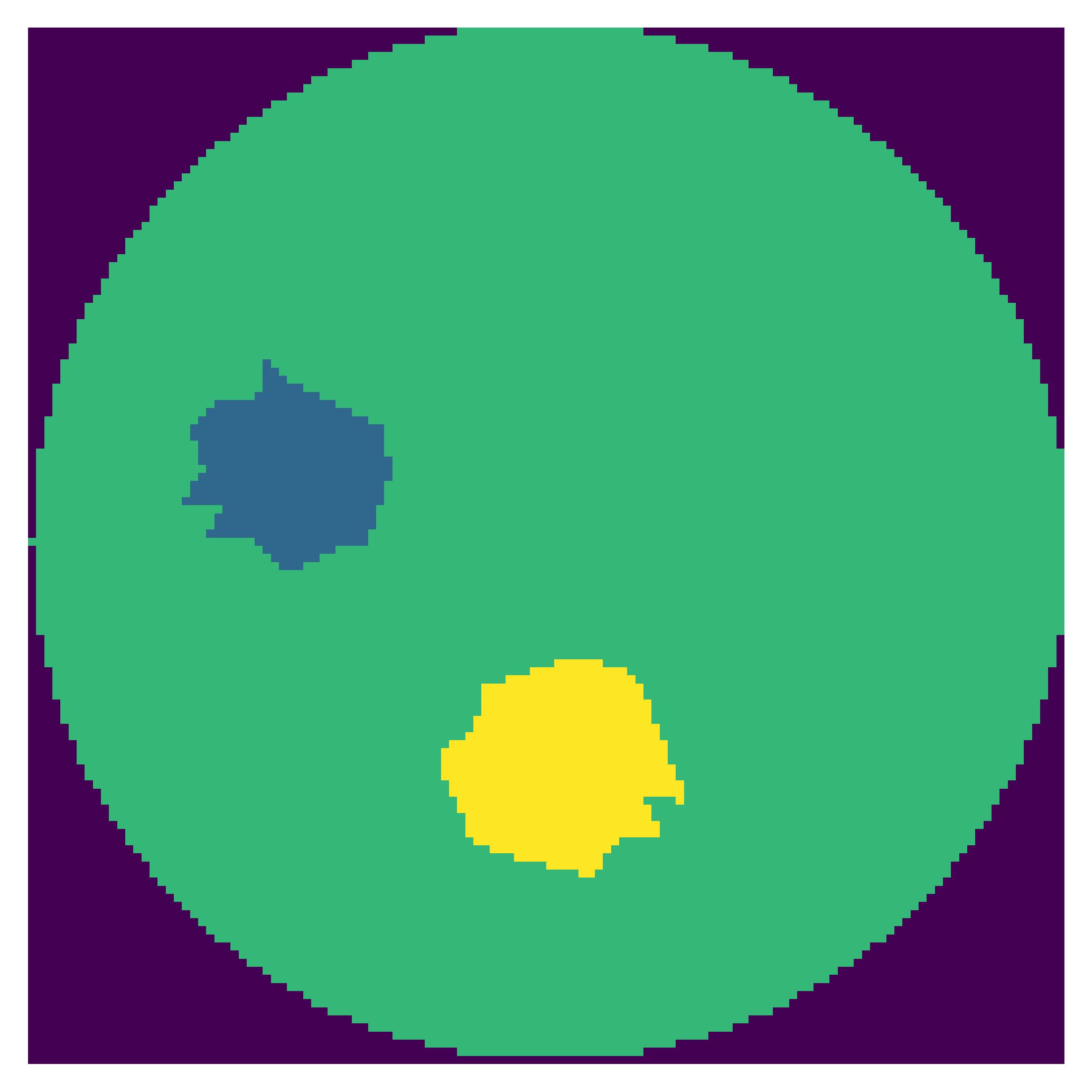}\vspace{5pt} 
    \includegraphics[width=1in]{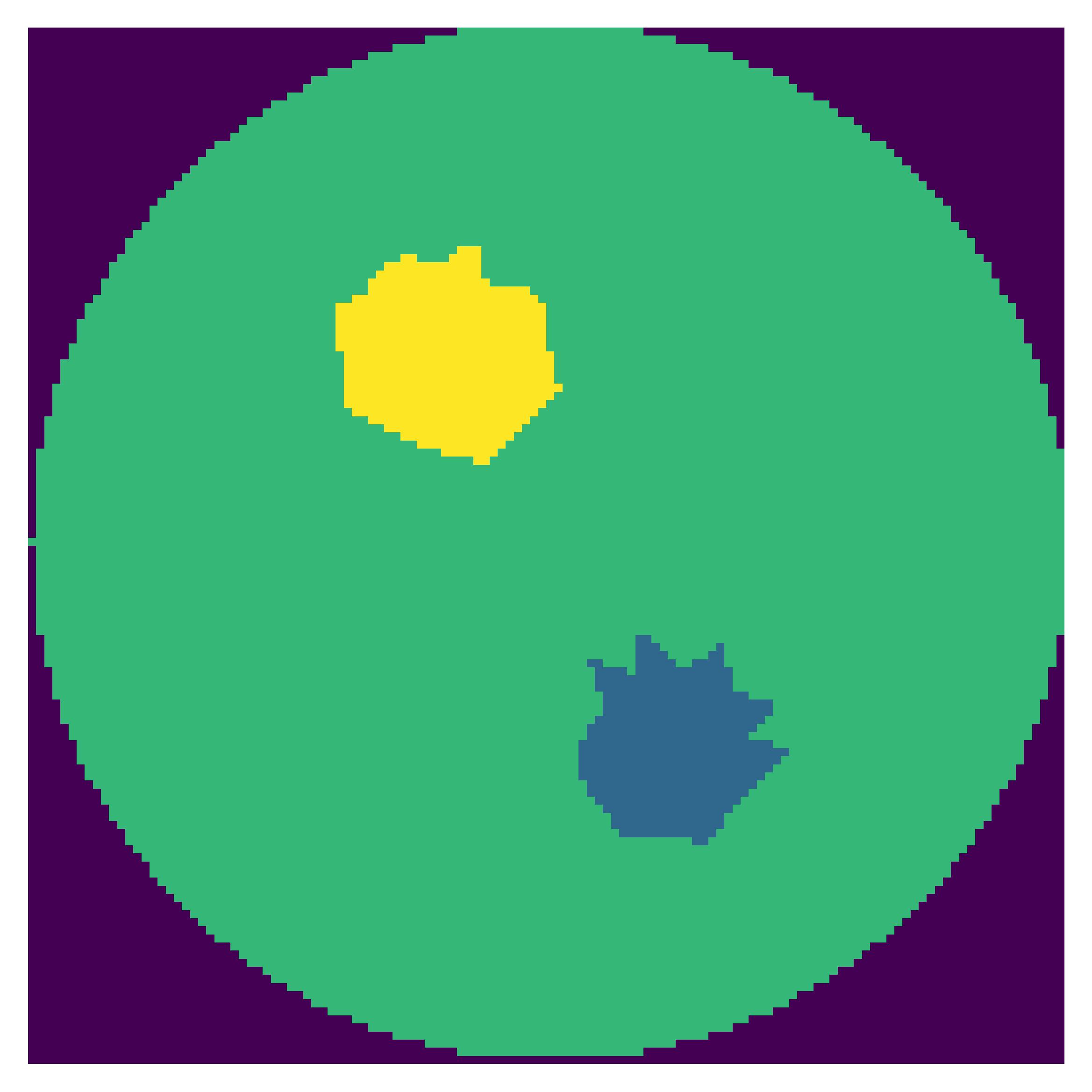}\vspace{5pt}
    \end{minipage}
}
\hspace{-7mm} 
\subfigure[GN]{
    \begin{minipage}[b]{0.19\linewidth}
    \centering
    \includegraphics[width=1in]{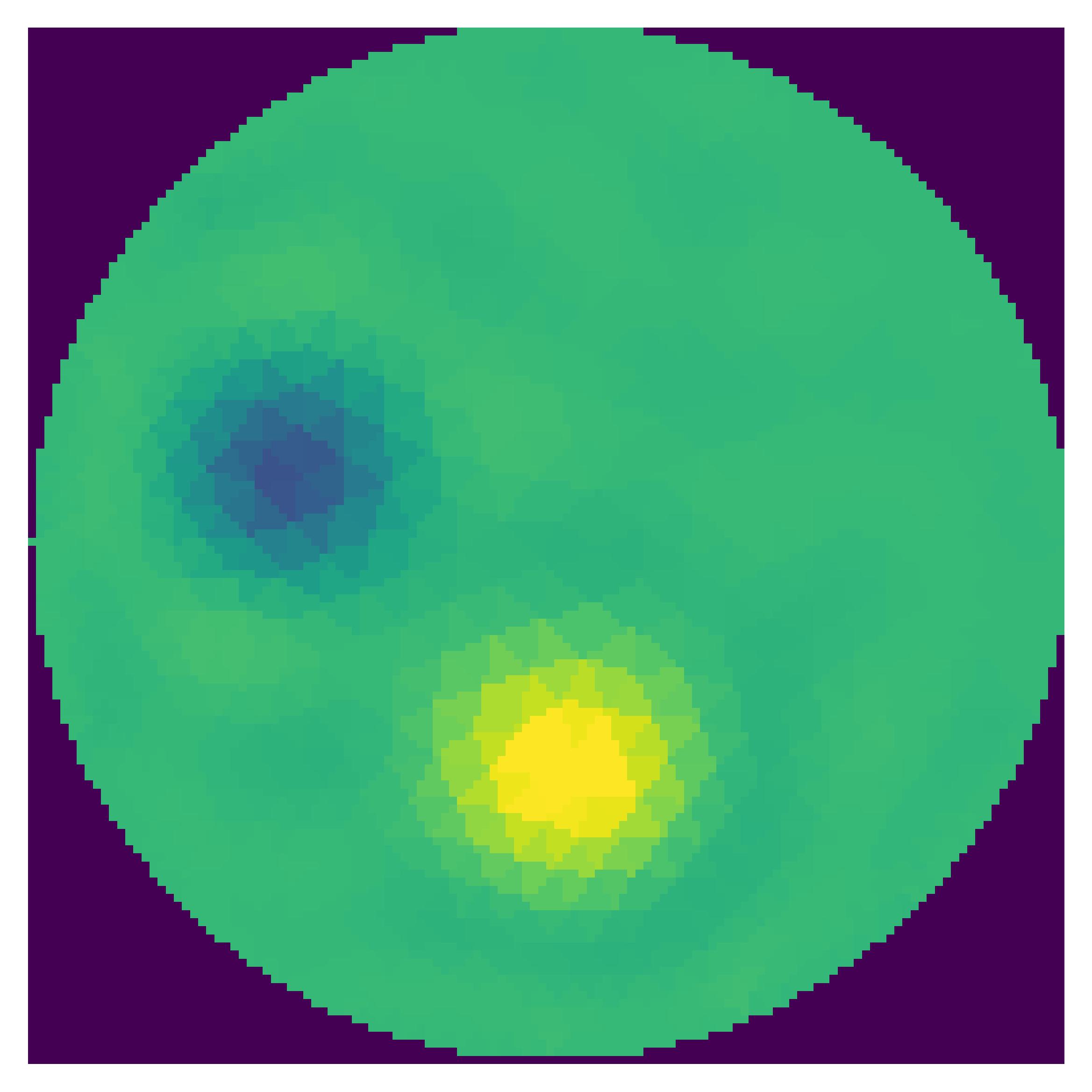}\vspace{5pt} 
    \includegraphics[width=1in]{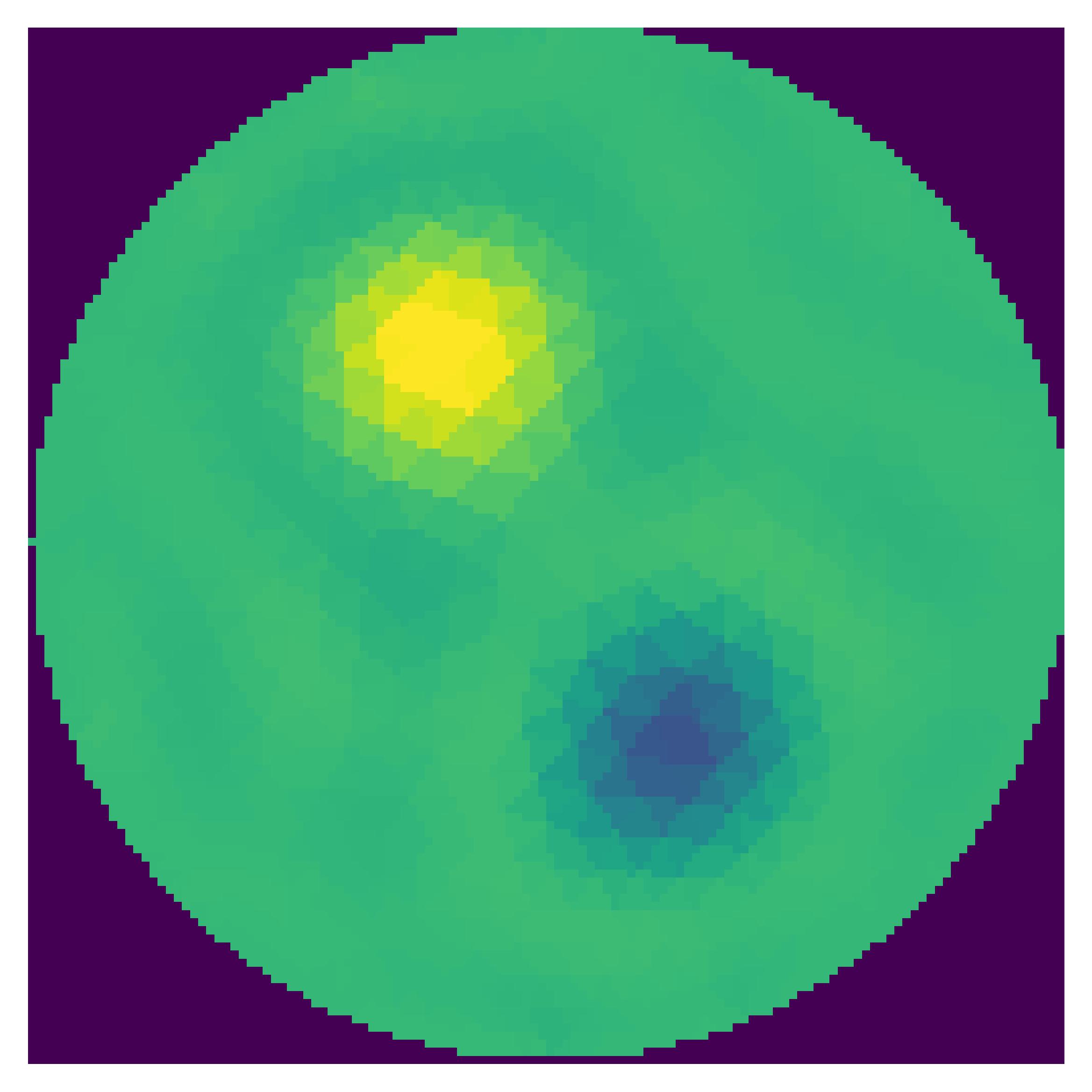}\vspace{5pt}
    \end{minipage}
}
\hspace{-7mm} 
\subfigure[CVAE]{
    \begin{minipage}[b]{0.19\linewidth}
    \centering
    \includegraphics[width=1in]{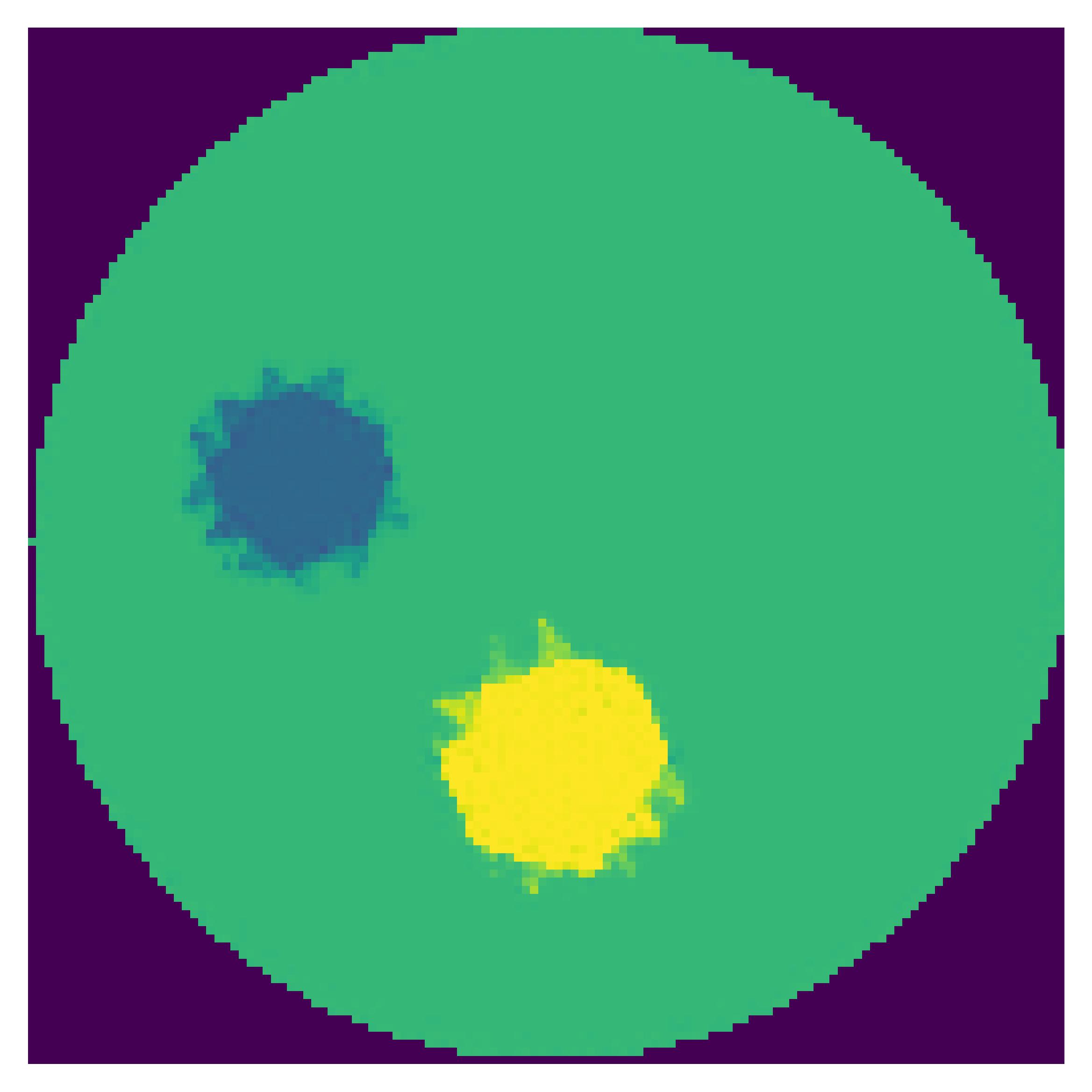}\vspace{5pt} 
    \includegraphics[width=1in]{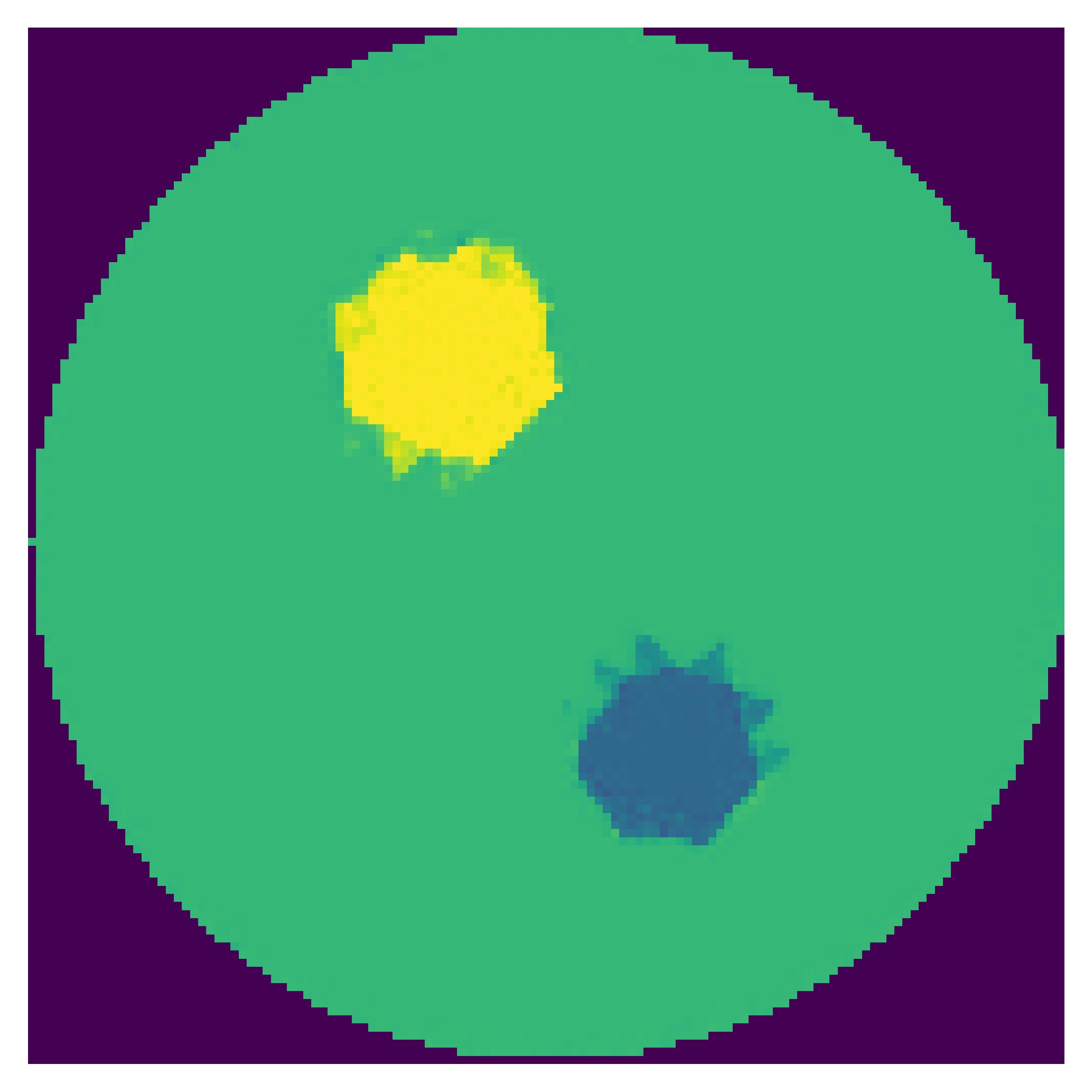}\vspace{5pt}
    \end{minipage}
}
\hspace{-7mm} 
\subfigure[CNF]{
    \begin{minipage}[b]{0.19\linewidth}
    \centering
    \includegraphics[width=1in]{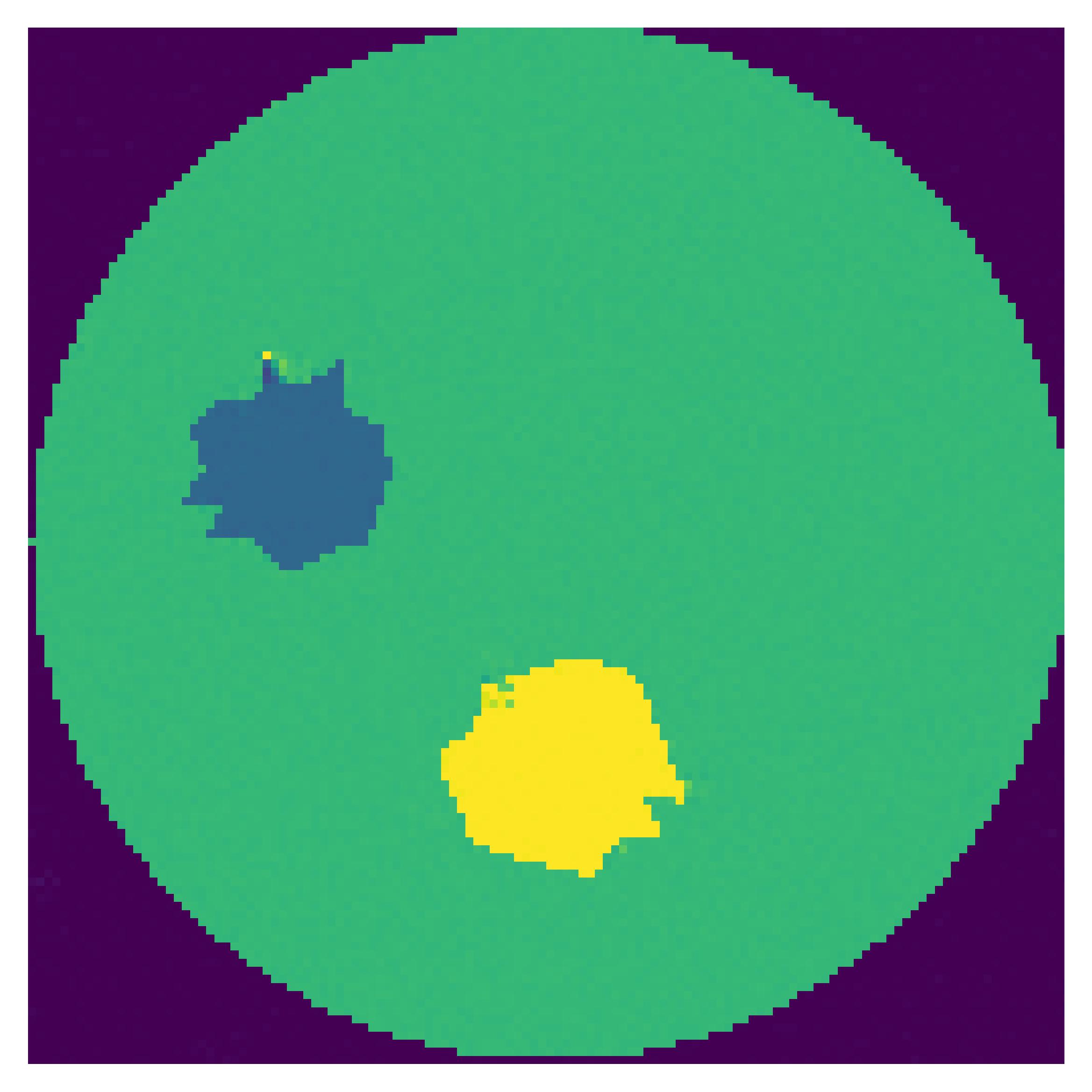}\vspace{5pt} 
    \includegraphics[width=1in]{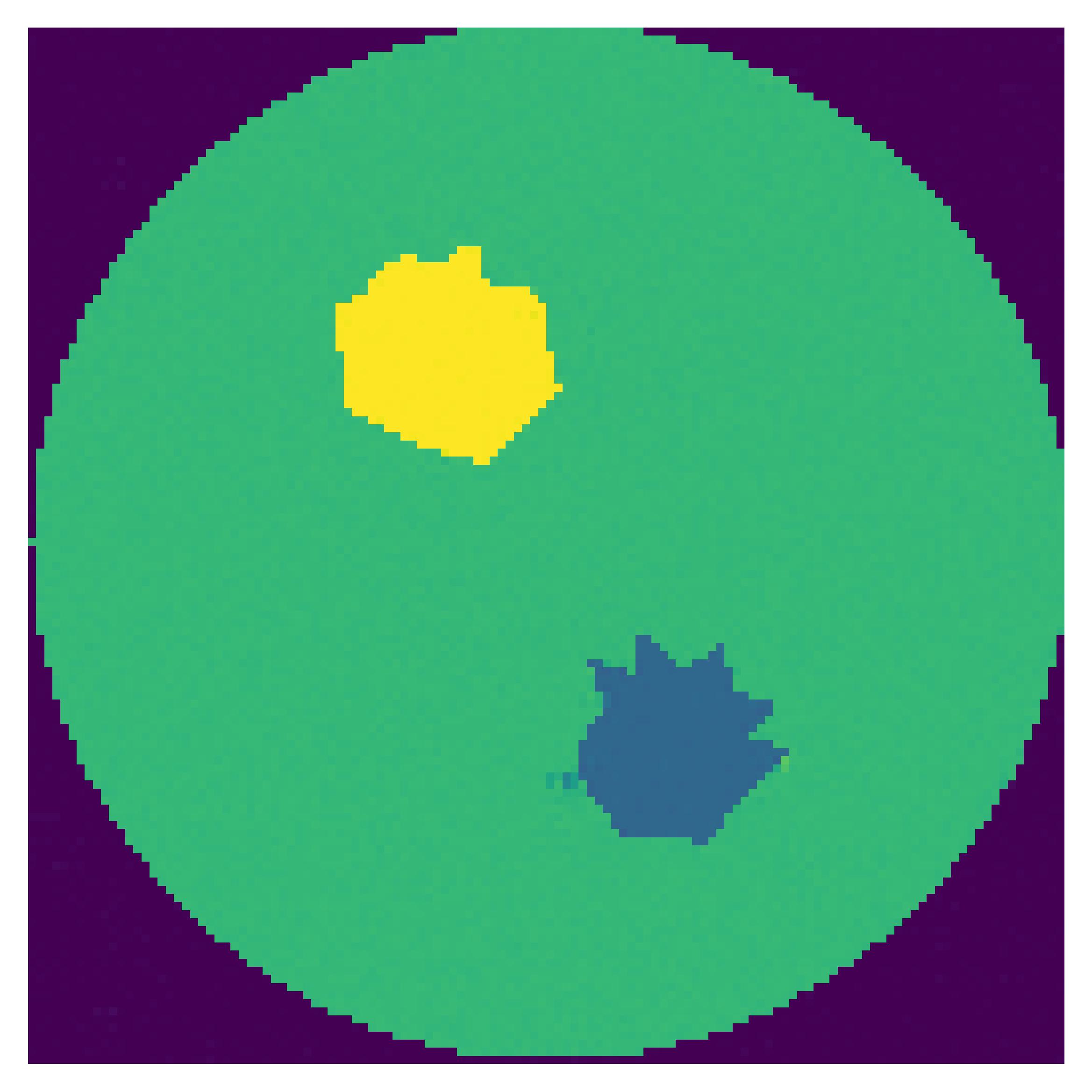}\vspace{5pt}
    \end{minipage}
}
\hspace{-6mm} 
\subfigure[$\mathrm{CSD}^*$]{
    \begin{minipage}[b]{0.18\linewidth}
    \centering
    \includegraphics[width=1in]{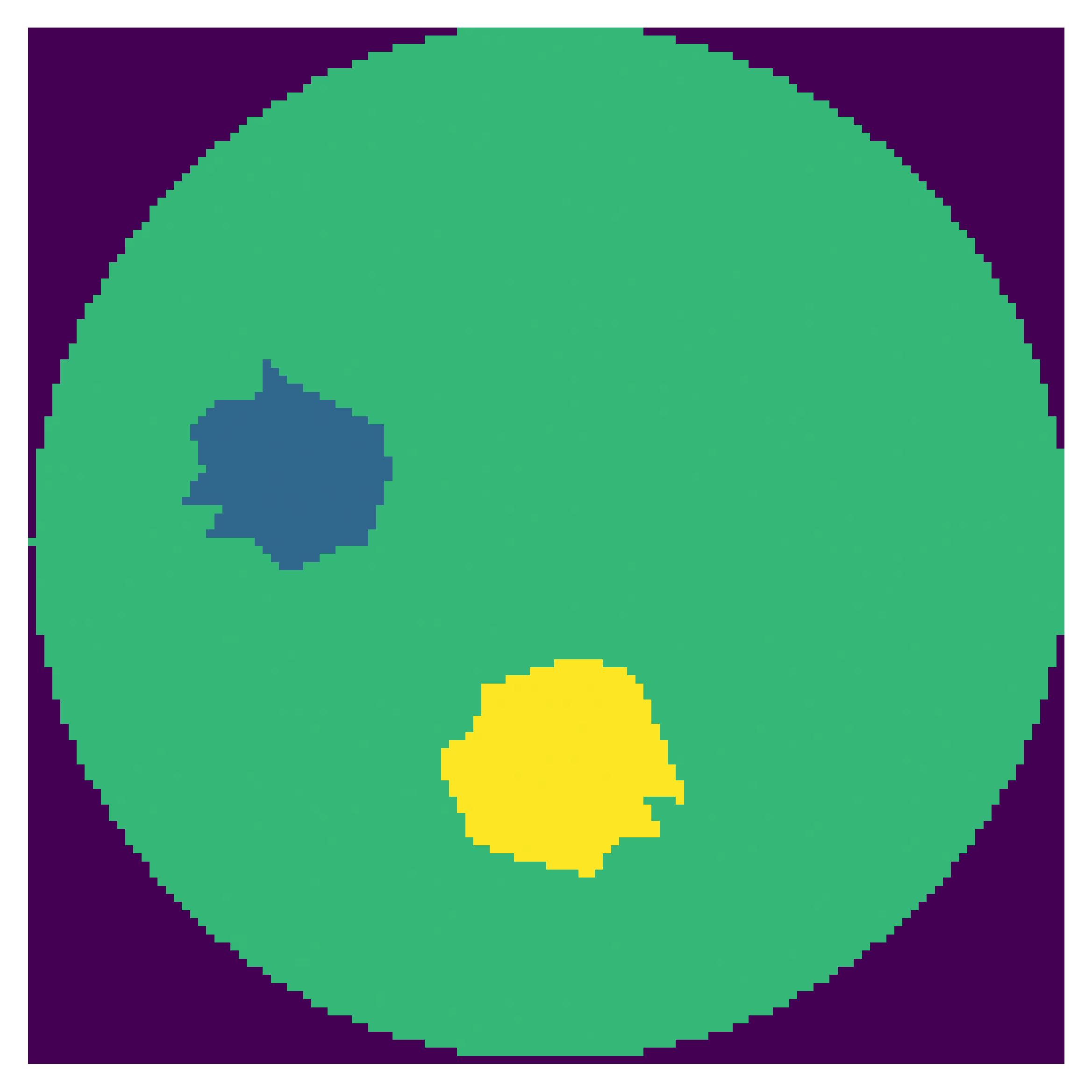}\vspace{5pt} 
    \includegraphics[width=1in]{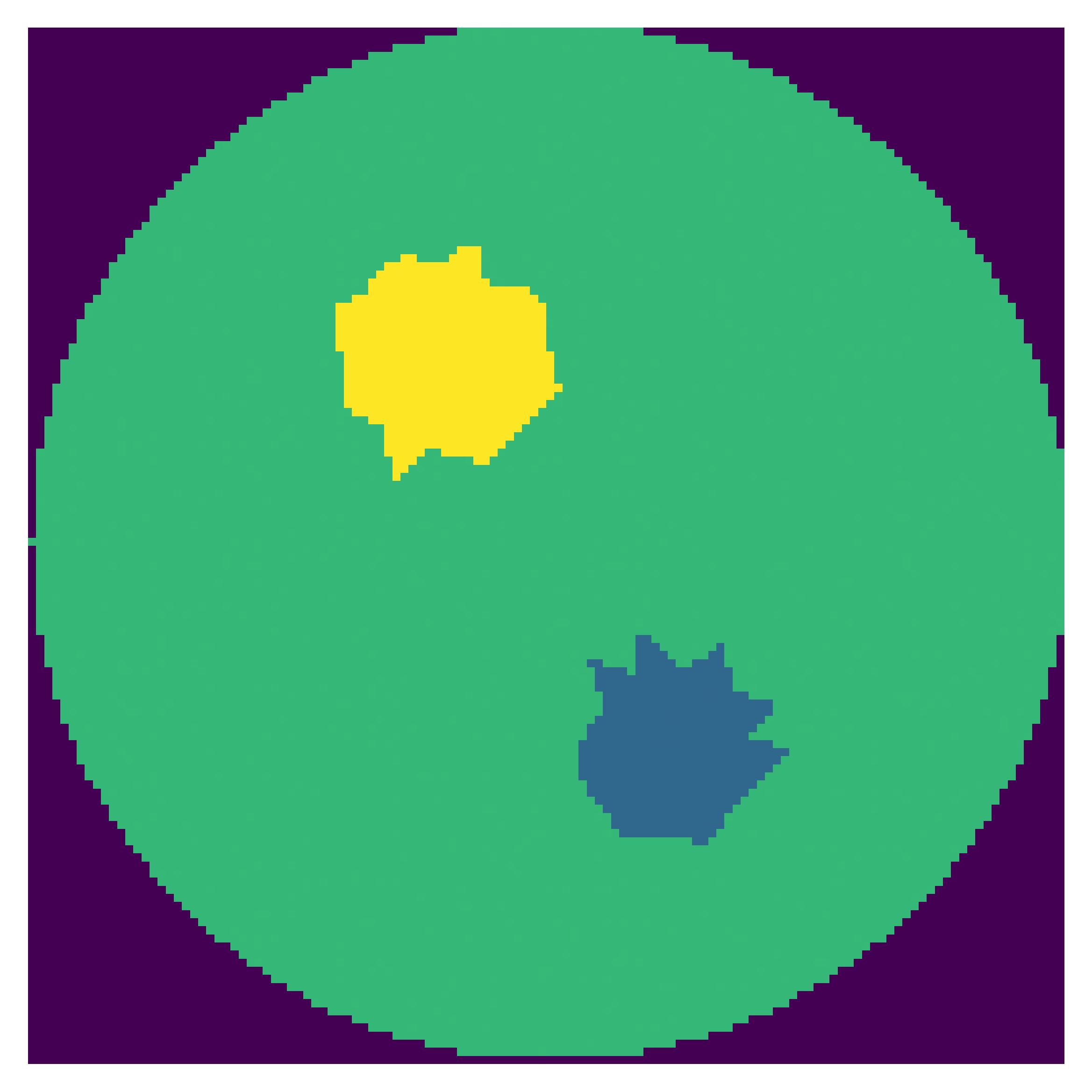}\vspace{5pt}
    \end{minipage}
}
\caption{\textbf{2 anomalies and noise in 40dB}, (a) original image of $128\times 128$ pixels, (b) result of Gauss-Newton, reconstructed image from (c) CVAE, and (d) CNF, as well as (e) $\mathrm{CSD}^*$. }
\label{fig:40db_2}
\end{figure}

Firstly, we compare these models in the 40dB noise dataset with 2 anomalies in each image, the quantitative results are shown in Table \ref{tab:40db2}. It is clear that CNF has the best value in MSE, which is 65.6\% better than GN compared to 56.3\% for CVAE and 53.1\% for $\mathrm{CSD}^*$, and PSNR. $\mathrm{CSD}^*$ achieves the highest performance among the remaining four metrics, with an increase of 11.7\% in SSIM, and a fall of 82.6\% and 80.8\% in RE and AE respectively compared with GN, DR is nearly the best. Although CVAE is slightly better than $\mathrm{CSD}^*$ in MSE and PSNR, it is worse in the other. Details in images can be seen in Figure \ref{fig:40db_2}, GN and CVAE have poorer performances in terms of the shape and clarity of the anomalies. CNF and $\mathrm{CSD}^*$ behave similarly, with some sharp bumps CNF can generate but not $\mathrm{CSD}^*$, and vice versa.

\begin{table} [!ht]
\resizebox{\linewidth}{!}{
\begin{threeparttable}
    \centering
    \caption{2 inclusions with noise in 40dB.}
    \label{tab:40db2}
    \begin{tabular}{lcccccc}
    \toprule
        \textbf{Method} & \textbf{MSE} & \textbf{PSNR} & \textbf{SSIM} & \textbf{RE} & \textbf{AE} & \textbf{DR} \\ \midrule
        GN &0.0032$\pm$0.0004 &28.850$\pm$0.675 &0.874$\pm$0.011 &0.034$\pm$0.003 &0.026$\pm$0.002 &1.045$\pm$0.066 \\ [0.8ex]
        CVAE &0.0014$\pm$0.0005 &32.804$\pm$1.631 &0.969$\pm$0.008 &0.009$\pm$0.002 &0.007$\pm$0.001 &1.058$\pm$0.022 \\[0.8ex]
        CNF &\textbf{0.0011$\pm$0.0005} &\textbf{33.789$\pm$2.715} &0.969$\pm$0.007 &0.008$\pm$0.001 &0.006$\pm$0.001 &0.920$\pm$0.020  \\[0.8ex]
        $\mathrm{CSD}^*$ &0.0015$\pm$0.0006 &32.673$\pm$4.670 &\textbf{0.976$\pm$0.009} &\textbf{0.006$\pm$0.002} &\textbf{0.005$\pm$0.001} &\textbf{1.001$\pm$0.002} \\
    \bottomrule
    \end{tabular}    
\end{threeparttable}}
\end{table}

\begin{figure}[htbp]
\centering
\subfigure[GT]{
    \begin{minipage}[b]{0.19\linewidth} 
    \centering
    \includegraphics[width=1in]{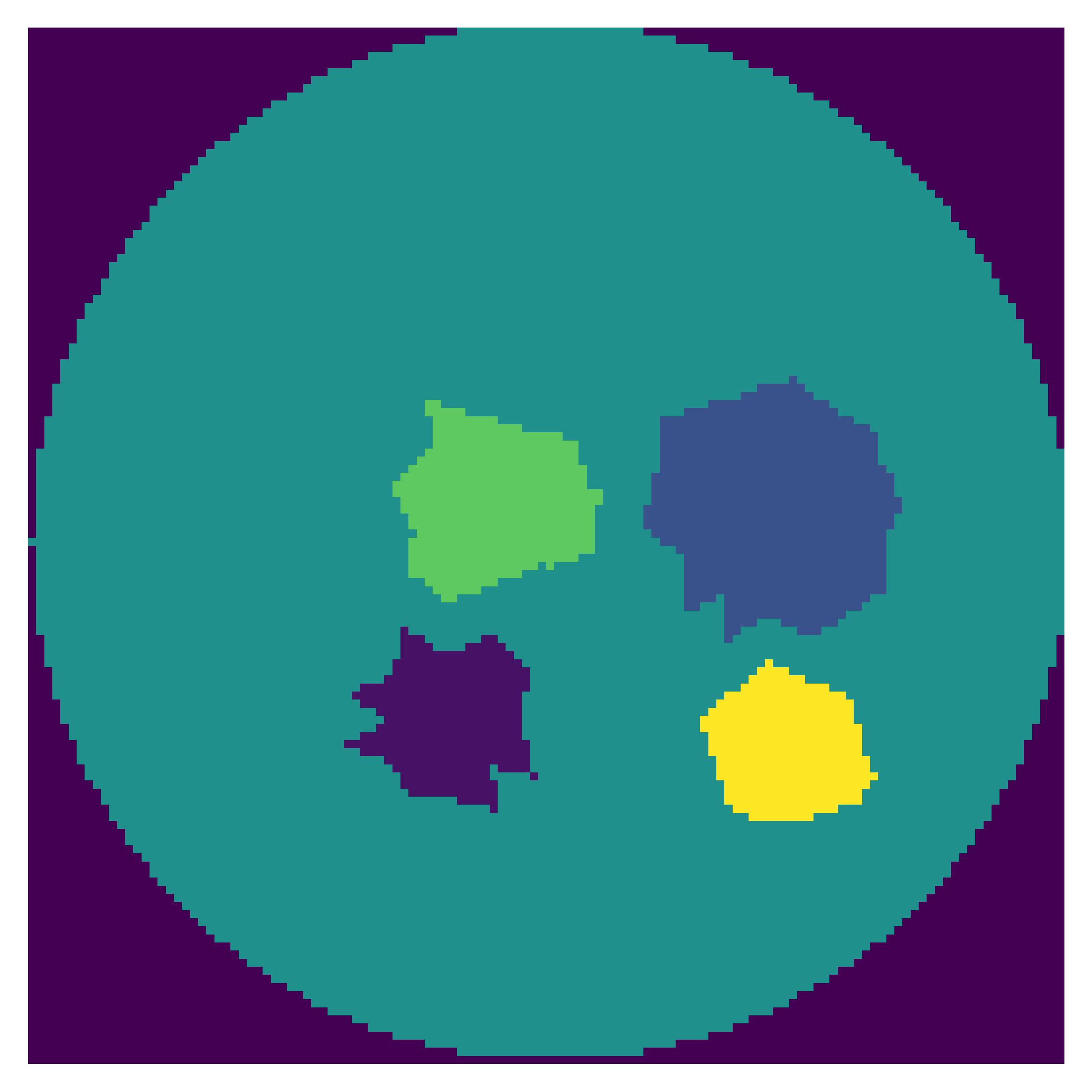}\vspace{5pt} 
    \includegraphics[width=1in]{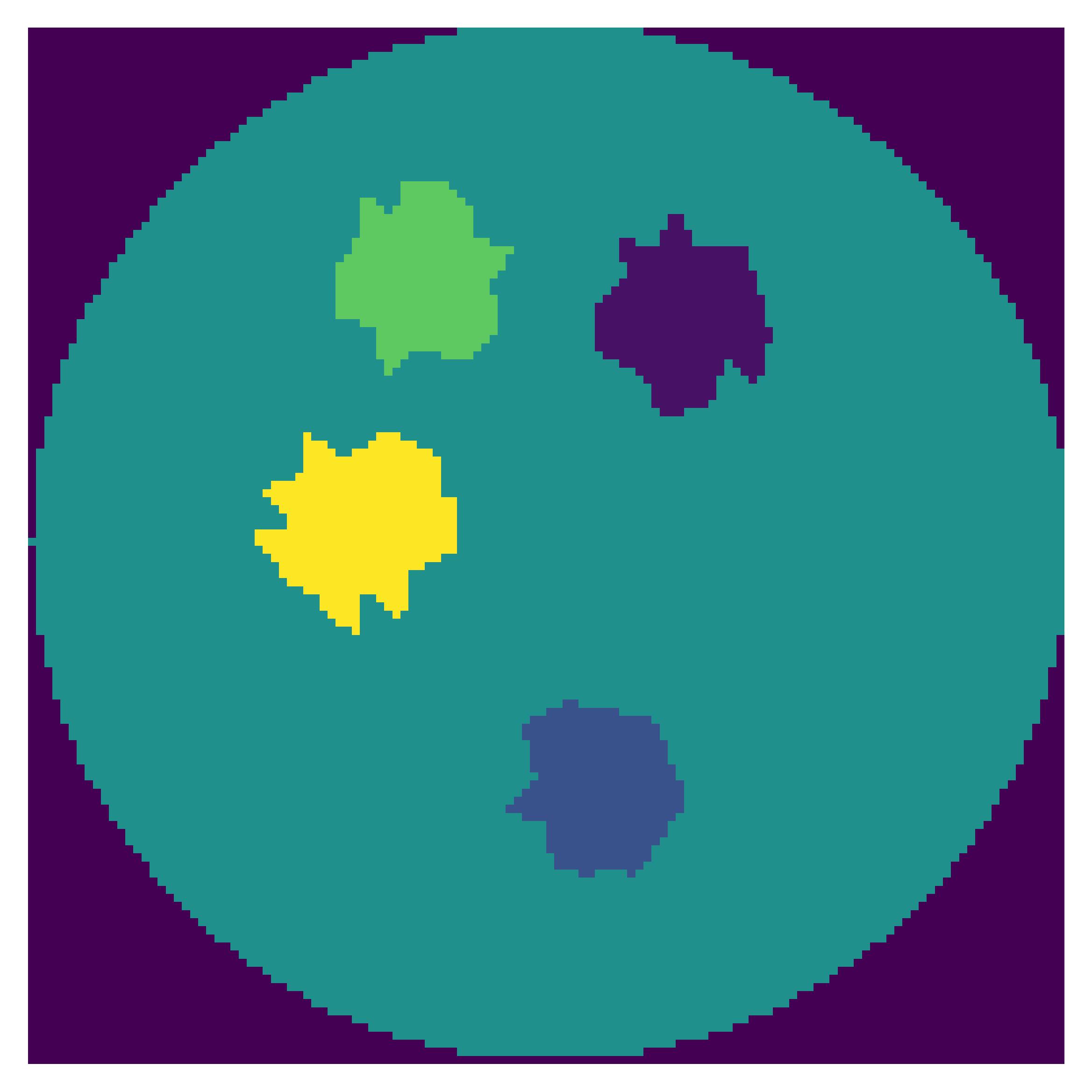}\vspace{5pt}
    \end{minipage}
}
\hspace{-7mm} 
\subfigure[GN]{
    \begin{minipage}[b]{0.19\linewidth}
    \centering
    \includegraphics[width=1in]{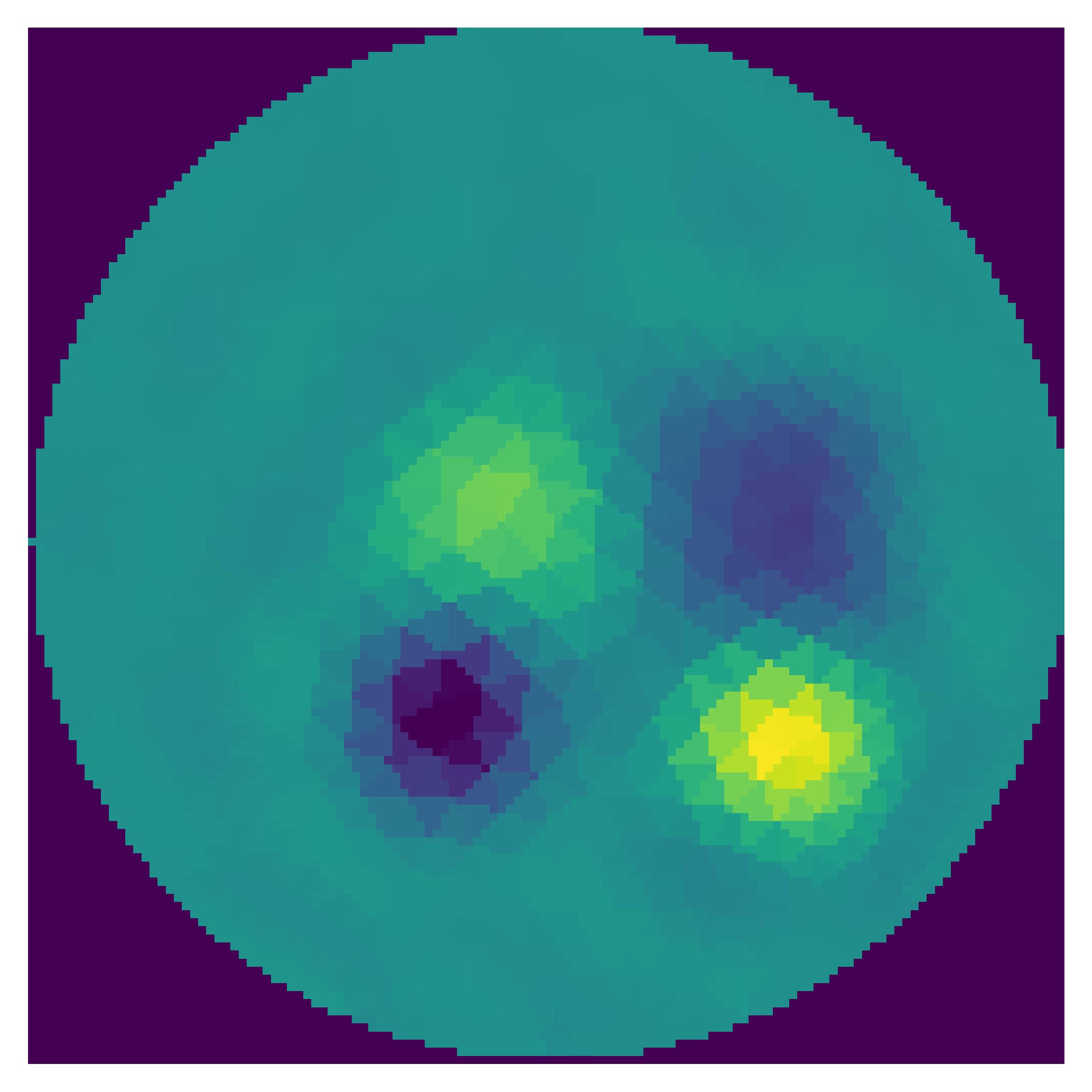}\vspace{5pt} 
    \includegraphics[width=1in]{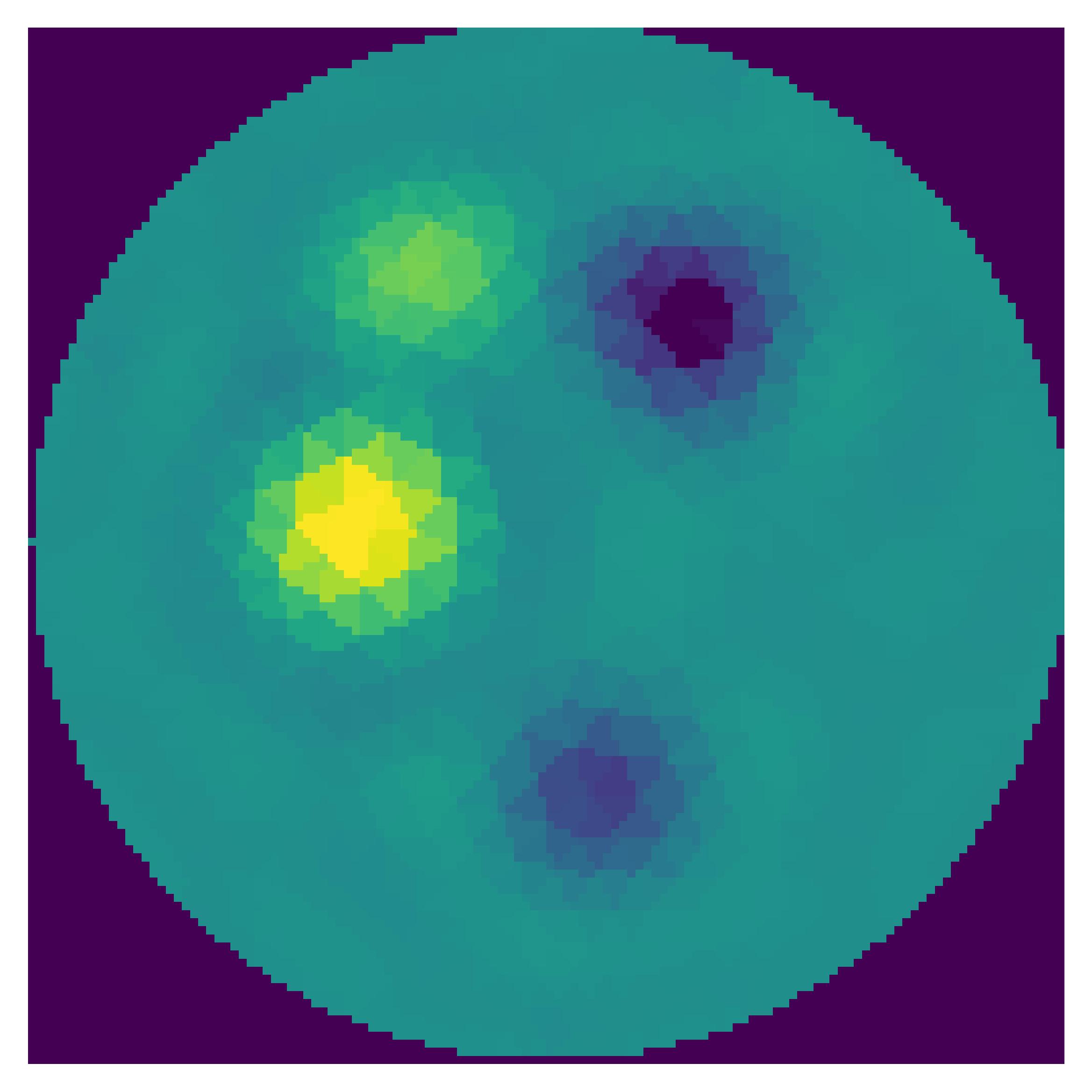}\vspace{5pt}
    \end{minipage}
}
\hspace{-7mm} 
\subfigure[CVAE]{
    \begin{minipage}[b]{0.19\linewidth}
    \centering
    \includegraphics[width=1in]{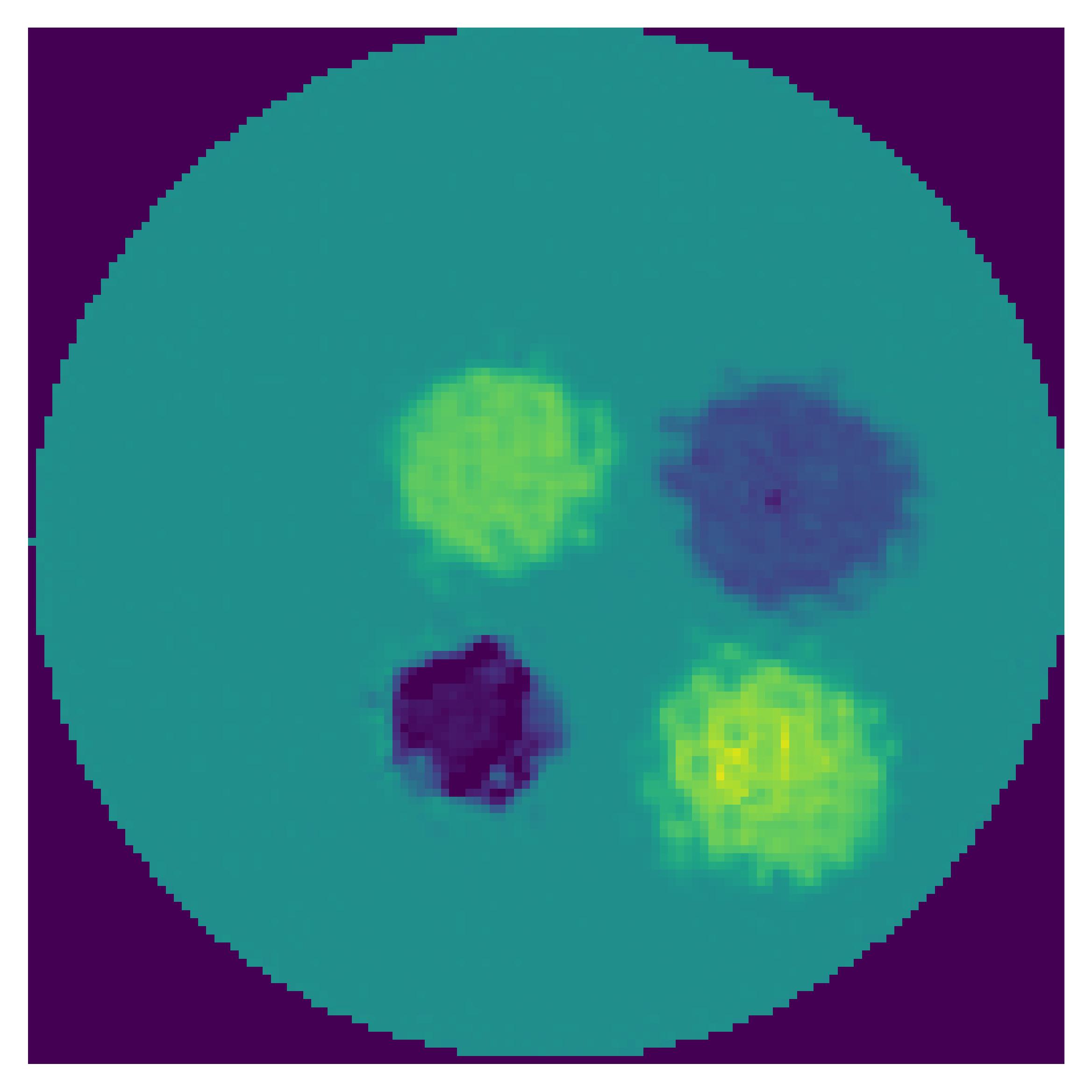}\vspace{5pt} 
    \includegraphics[width=1in]{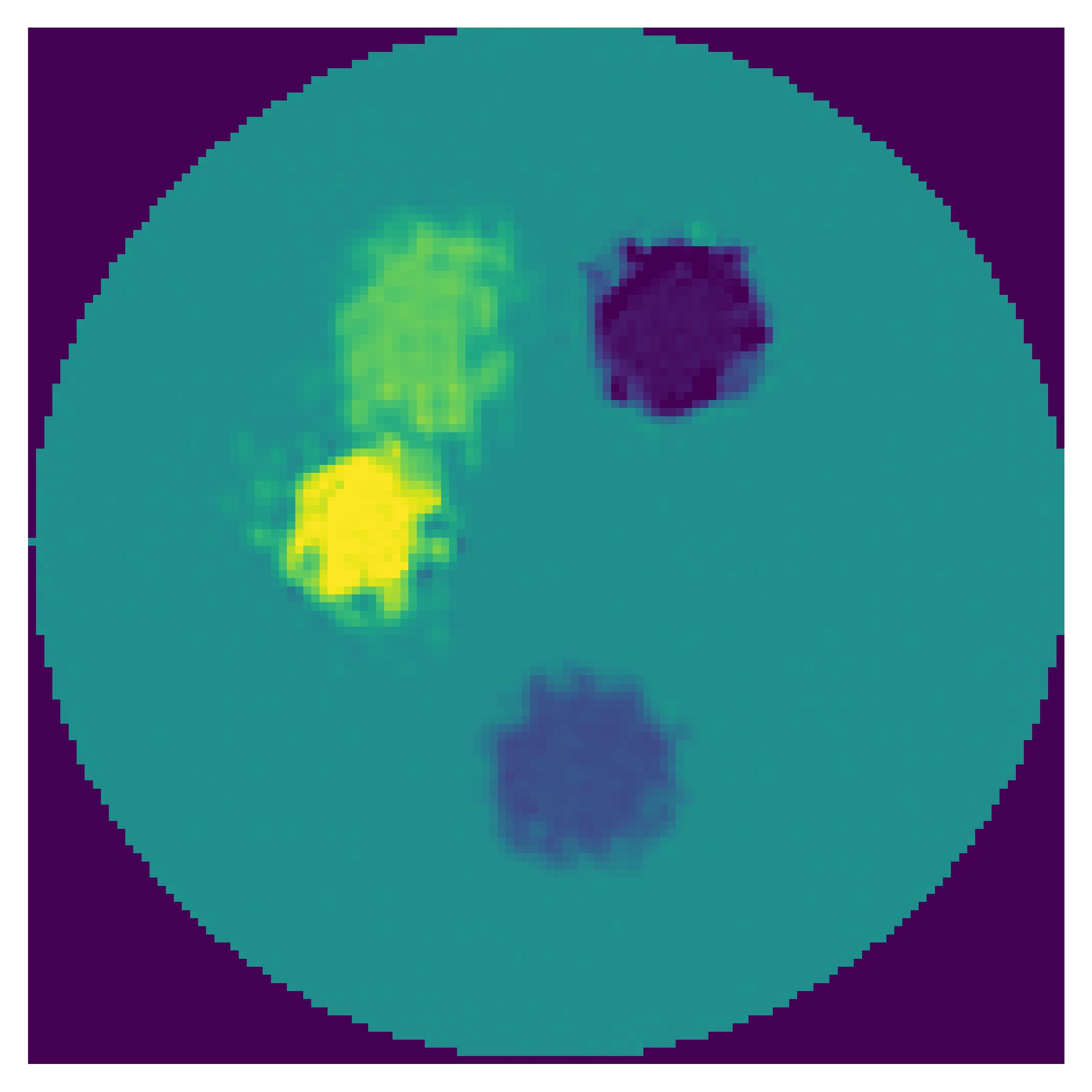}\vspace{5pt}
    \end{minipage}
}
\hspace{-7mm} 
\subfigure[CNF]{
    \begin{minipage}[b]{0.19\linewidth}
    \centering
    \includegraphics[width=1in]{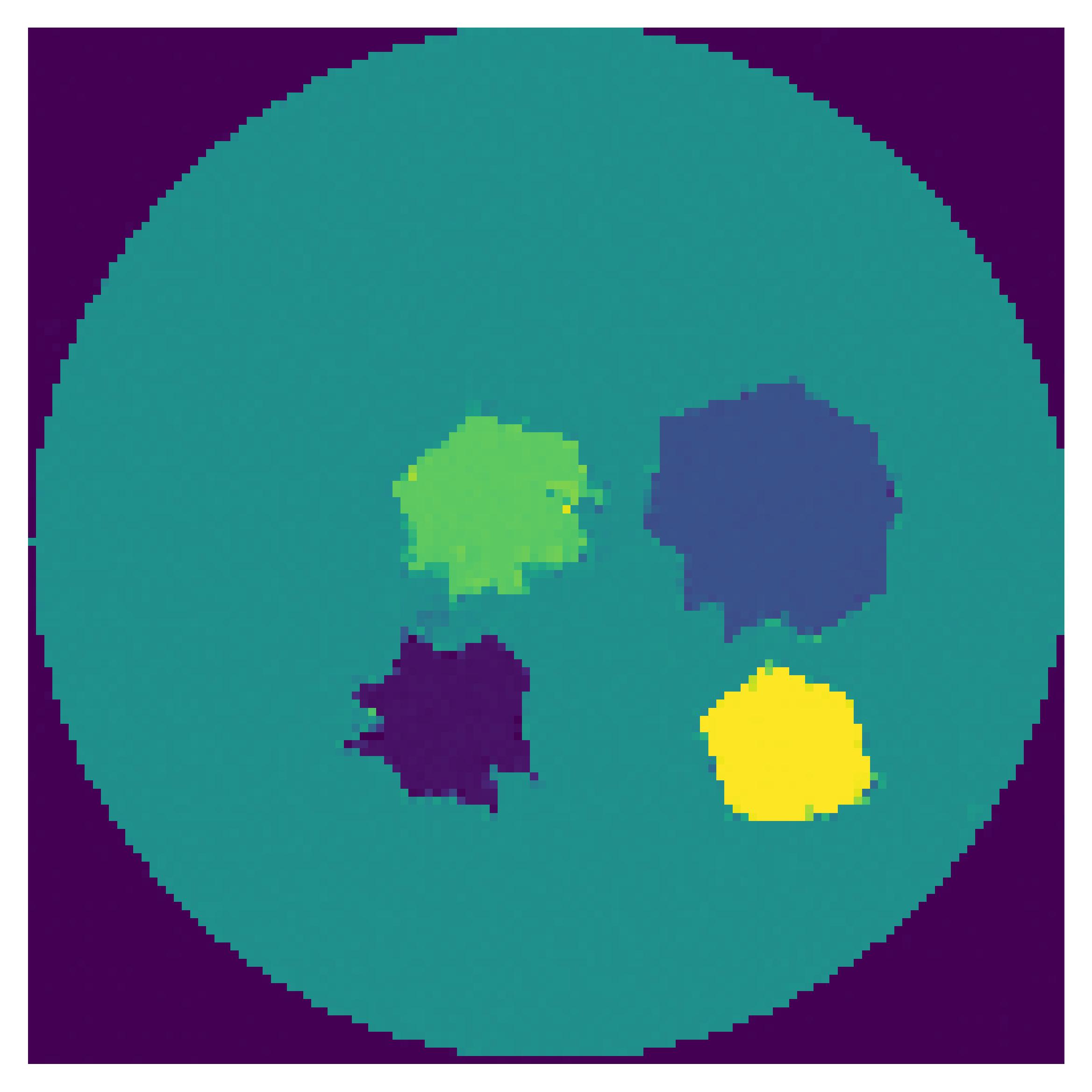}\vspace{5pt} 
    \includegraphics[width=1in]{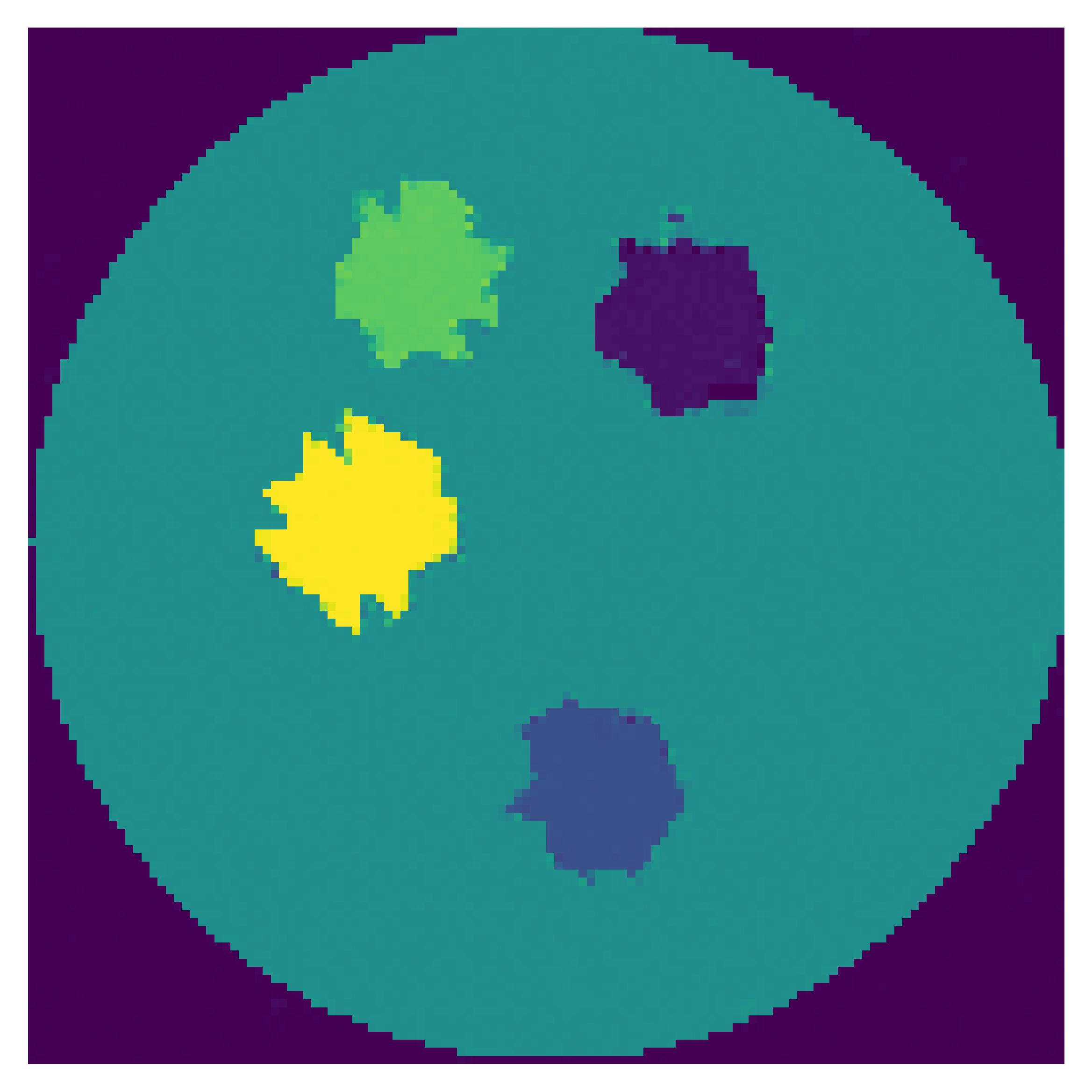}\vspace{5pt}
    \end{minipage}
}
\hspace{-6mm} 
\subfigure[$\mathrm{CSD}^*$]{
    \begin{minipage}[b]{0.18\linewidth}
    \centering
    \includegraphics[width=1in]{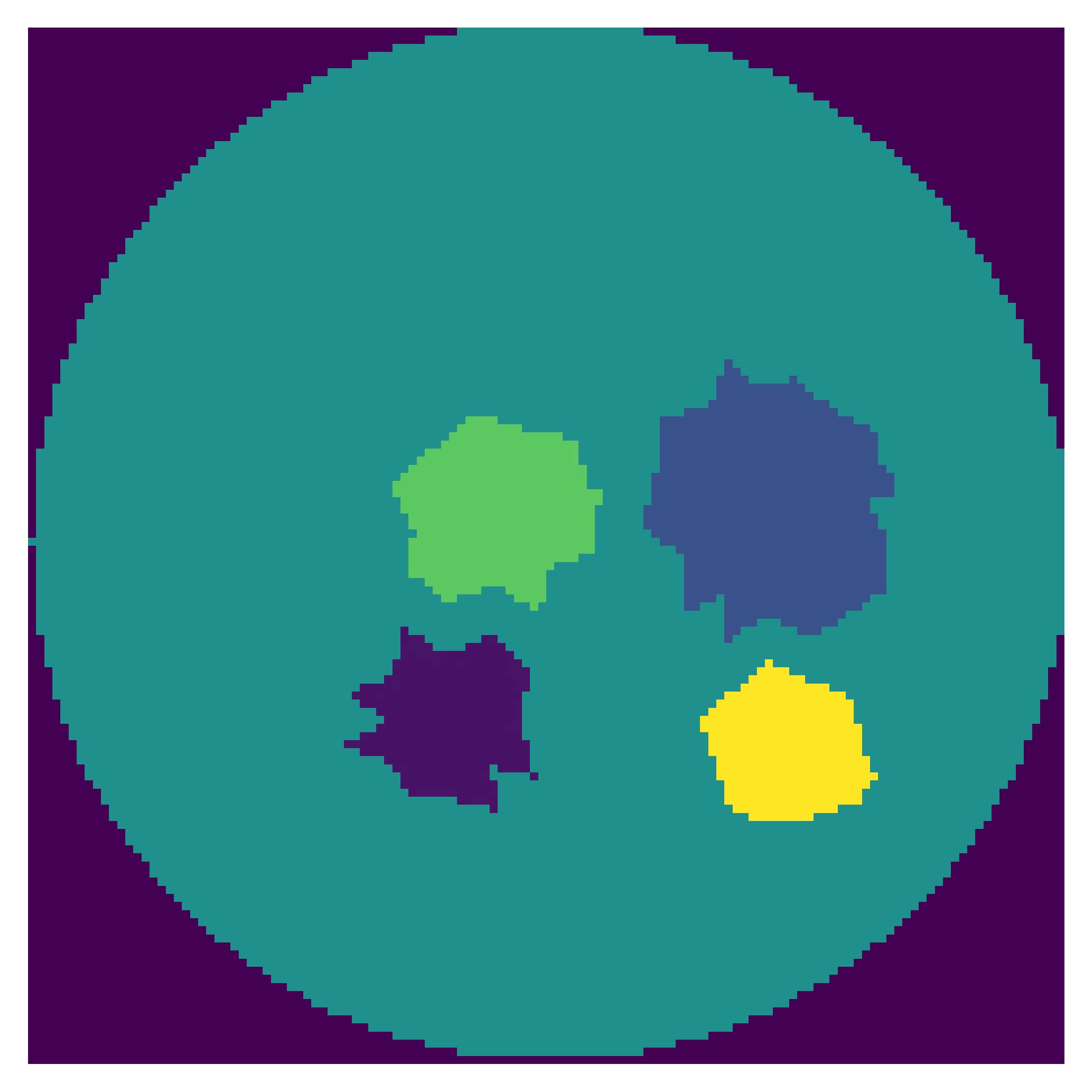}\vspace{5pt} 
    \includegraphics[width=1in]{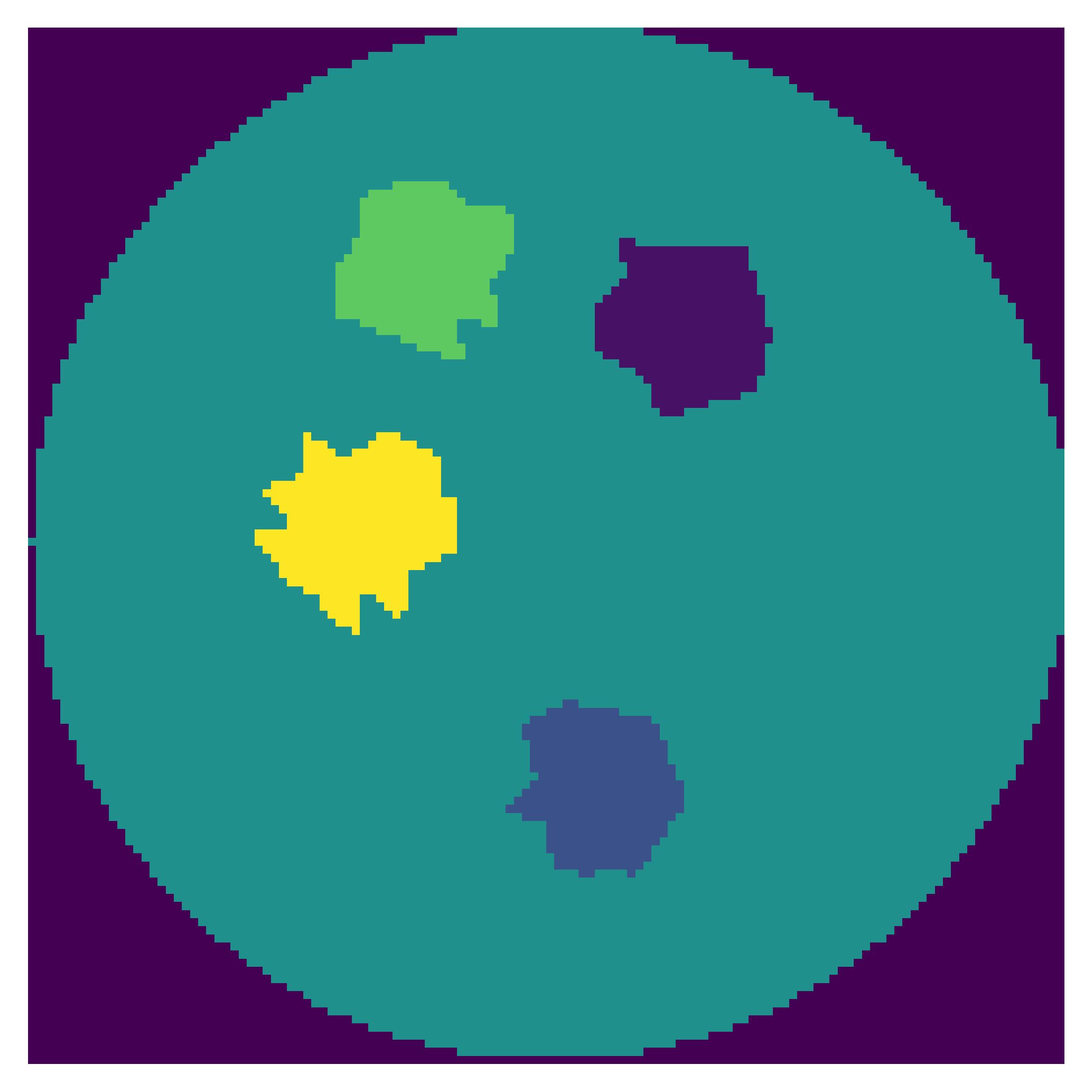}\vspace{5pt}
    \end{minipage}
}
\caption{\textbf{4 anomalies with noise in 40dB}, (a) original image of $128\times 128$ pixels, (b) result of Gauss-Newton,  reconstructed image from (c) CVAE, and (d) CNF, as well as (e) $\mathrm{CSD}^*$. }
\label{fig:40db_4}
\end{figure}

To evaluate the models' performance as the number of anomalies increases, we also run these models on 4 anomalies dataset with the same level of noise as the 2 one. According to Table \ref{tab:40db4}, CNF has overwhelmingly good results in almost all of the indexes, especially when measured in MSE, we can see CNF is 71\% lower than GN while CVAE and $\mathrm{CSD}^*$ only outperform GN 13.6\% and 50\% respectively. However, $\mathrm{CSD}^*$ can also be on par with CNF in terms of RE and AE metrics and is best in terms of DR indicator at 1. Although it can not be competent to CNF in other aspects, it can far surpass the results of CVAE and GN largely. We can also see some visual differences in Figure \ref{fig:40db_4}. From the shape of internal anomalies, discerning the shape of anomalies in GN and VAE is challenging, as they both exhibit unclear shapes and lack clear boundaries. But VAE shows slightly better performance than GN in this regard. In contrast, CNF and $\mathrm{CSD}^*$ can accurately generate the shape of internal anomalies, including small bumps. However, both models may fail to generate some individual bumps, and the boundaries of CNF are susceptible to noise, while $\mathrm{CSD}^*$ does not exhibit this issue.

\begin{table*}[ht]
\resizebox{\linewidth}{!}{
\begin{threeparttable}
    \centering
    \caption{4 inclusions with noise in 40dB.}
    \begin{tabular}{lcccccc}
    \toprule
        \textbf{Method} & \textbf{MSE} & \textbf{PSNR} & \textbf{SSIM} & \textbf{RE} & \textbf{AE} & \textbf{DR} \\ \midrule
        GN &0.0140$\pm$0.006 & 26.317$\pm$1.034 & 0.794$\pm$0.047 & 0.061$\pm$0.015 &0.078$\pm$0.019 & 1.199$\pm$0.107  \\ [0.8ex]
        CVAE &0.0121$\pm$0.004 &26.282$\pm$1.507 &0.882$\pm$0.019  &0.033$\pm$0.006  &0.042$\pm$0.008  &1.110$\pm$0.068  \\[0.8ex]
        CNF &\textbf{0.0040$\pm$0.003} &\textbf{30.981$\pm$2.764} &\textbf{0.962$\pm$0.021} &\textbf{0.014$\pm$0.006} &\textbf{0.011$\pm$0.005} &0.752$\pm$0.044 \\[0.8ex]
        $\mathrm{CSD}^*$ &0.0070$\pm$0.004 &28.217$\pm$2.499&0.952$\pm$0.025&\textbf{0.014$\pm$0.007} &\textbf{0.011$\pm$0.006} &\textbf{1.000$\pm$0.004} \\
    \bottomrule
    \end{tabular}
    \label{tab:40db4} 
\end{threeparttable}}
\end{table*}

In short, CNF performs well in MSE and PSNR, but $\mathrm{CSD}^*$ tends to have better results in SSIM, AE, RE, and DR. In addition, when it comes to data with more anomalies, GN and VAE are not good choices for EIT reconstruction.

\subsubsection{Robustness to the noise}
How do these models perform when the noise in datasets increases? We go on to give the results of different models on 25dB noise datasets to provide a detailed comparison. 
\begin{table*}[ht]
\resizebox{\linewidth}{!}{
\begin{threeparttable}
    \centering
    \caption{2 inclusions with noise in 25dB.}
    \label{tab:25db2}
    \begin{tabular}{lcccccc}
    \toprule
        \textbf{Method} & \textbf{MSE} & \textbf{PSNR} & \textbf{SSIM} & \textbf{RE} & \textbf{AE} & \textbf{DR} \\ \midrule
        GN &0.0067$\pm$0.0009 &24.271$\pm$0.992 &0.804$\pm$0.018 &0.059$\pm$0.004 &0.046$\pm$0.003 &0.891$\pm$0.068  \\ [0.8ex]
        CVAE &\textbf{0.0035$\pm$0.0015} &\textbf{28.996$\pm$1.812} &\textbf{0.945$\pm$0.014} &\textbf{0.014$\pm$0.004} &\textbf{0.011$\pm$0.003}
        &\textbf{1.062$\pm$0.025}  \\[0.8ex]
        CNF &0.0063$\pm$0.0014 &25.652$\pm$2.807 &0.932$\pm$0.019 &0.018$\pm$0.006 &0.014$\pm$0.005 &0.933$\pm$0.131  \\[0.8ex]
        ${\mathrm{CSD}^{*}}_{40dB}^2$ &0.0060$\pm$0.0025 &25.453$\pm$2.807 &0.932$\pm$0.019 &0.018$\pm$0.006 &0.014$\pm$0.005 &0.933$\pm$0.132 \\
    \bottomrule
    \end{tabular}
    \begin{tablenotes}    
        \footnotesize              
        \item[1] ${\mathrm{CSD}^{*}}_{40dB}^2$ in the chart is the trained model in 40dB noise with 2 anomalies.        
       \end{tablenotes}      
\end{threeparttable}}
\end{table*}

\begin{figure}[htbp]
\centering
\subfigure[GT]{
    \begin{minipage}[b]{0.19\linewidth} 
    \centering
    \includegraphics[width=1in]{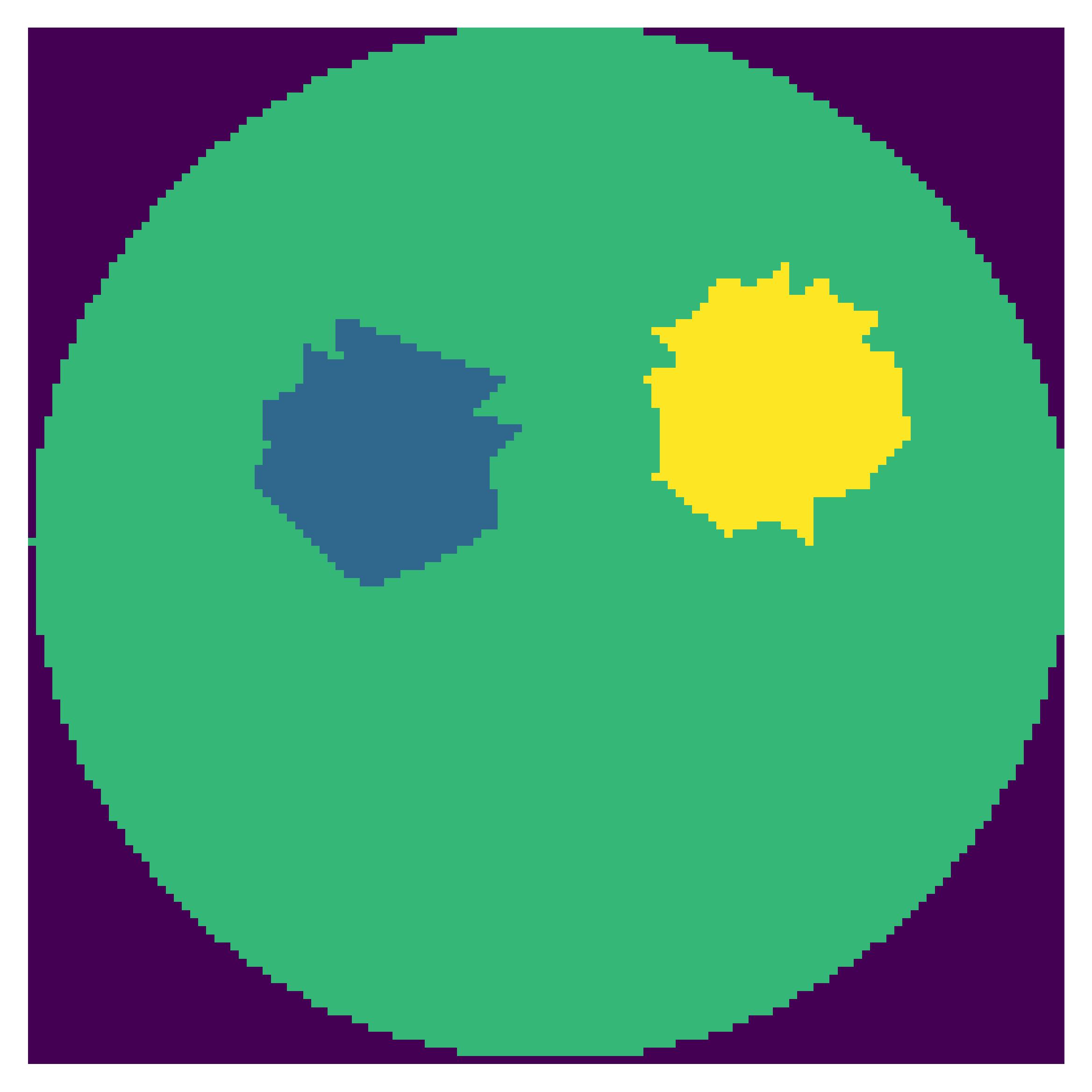}\vspace{5pt} 
    \includegraphics[width=1in]{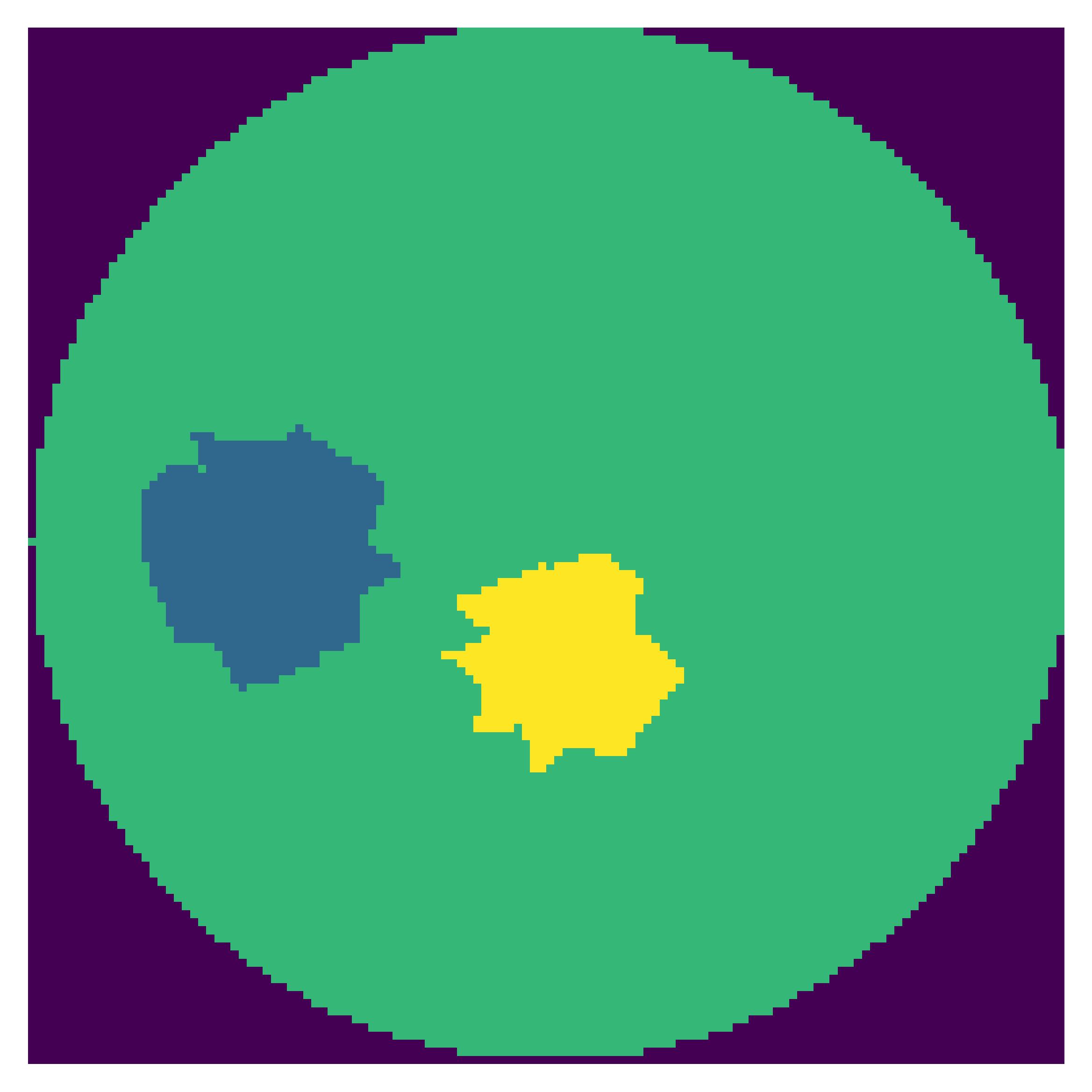}\vspace{5pt}
    \end{minipage}
}
\hspace{-7mm} 
\subfigure[GN]{
    \begin{minipage}[b]{0.19\linewidth}
    \centering
    \includegraphics[width=1in]{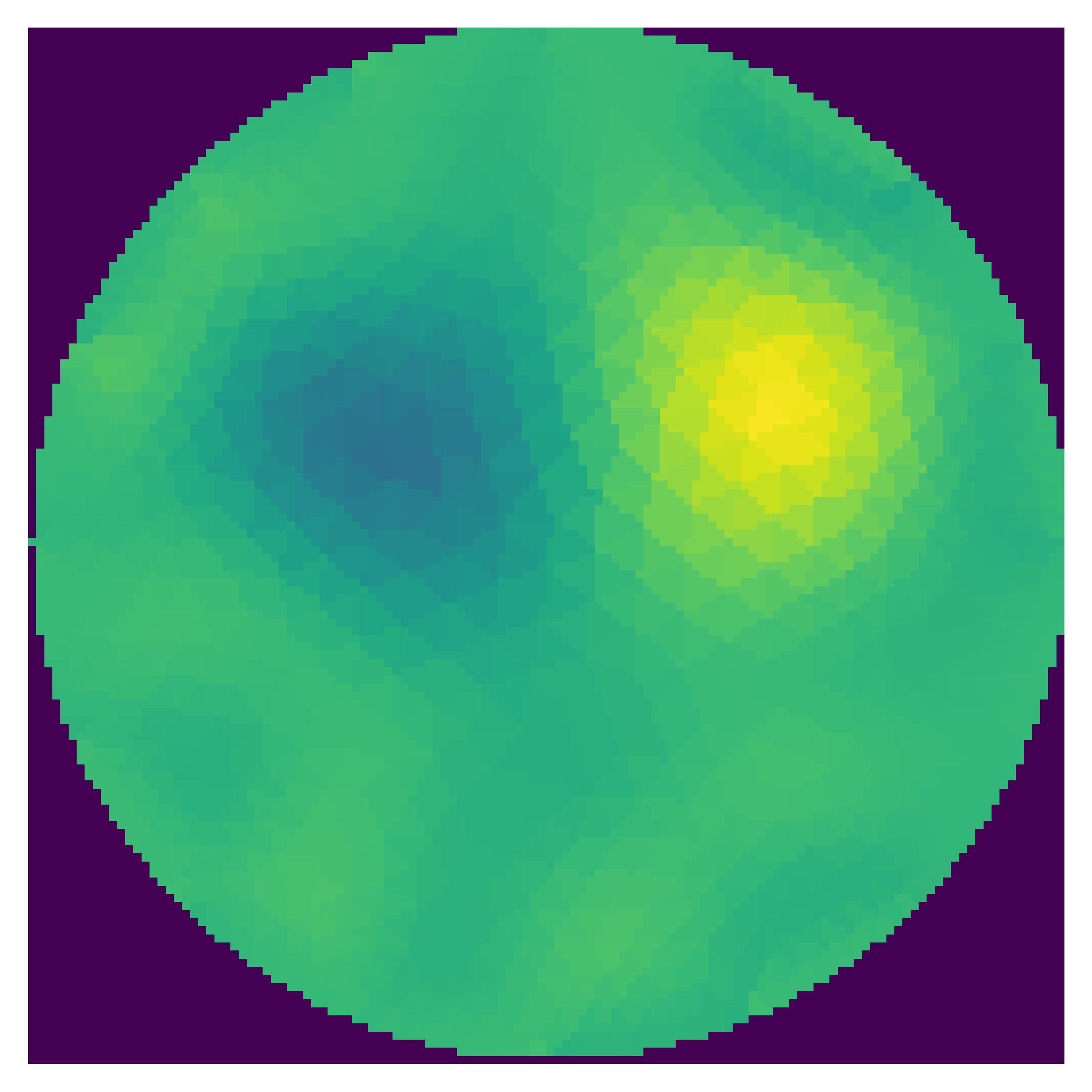}\vspace{5pt} 
    \includegraphics[width=1in]{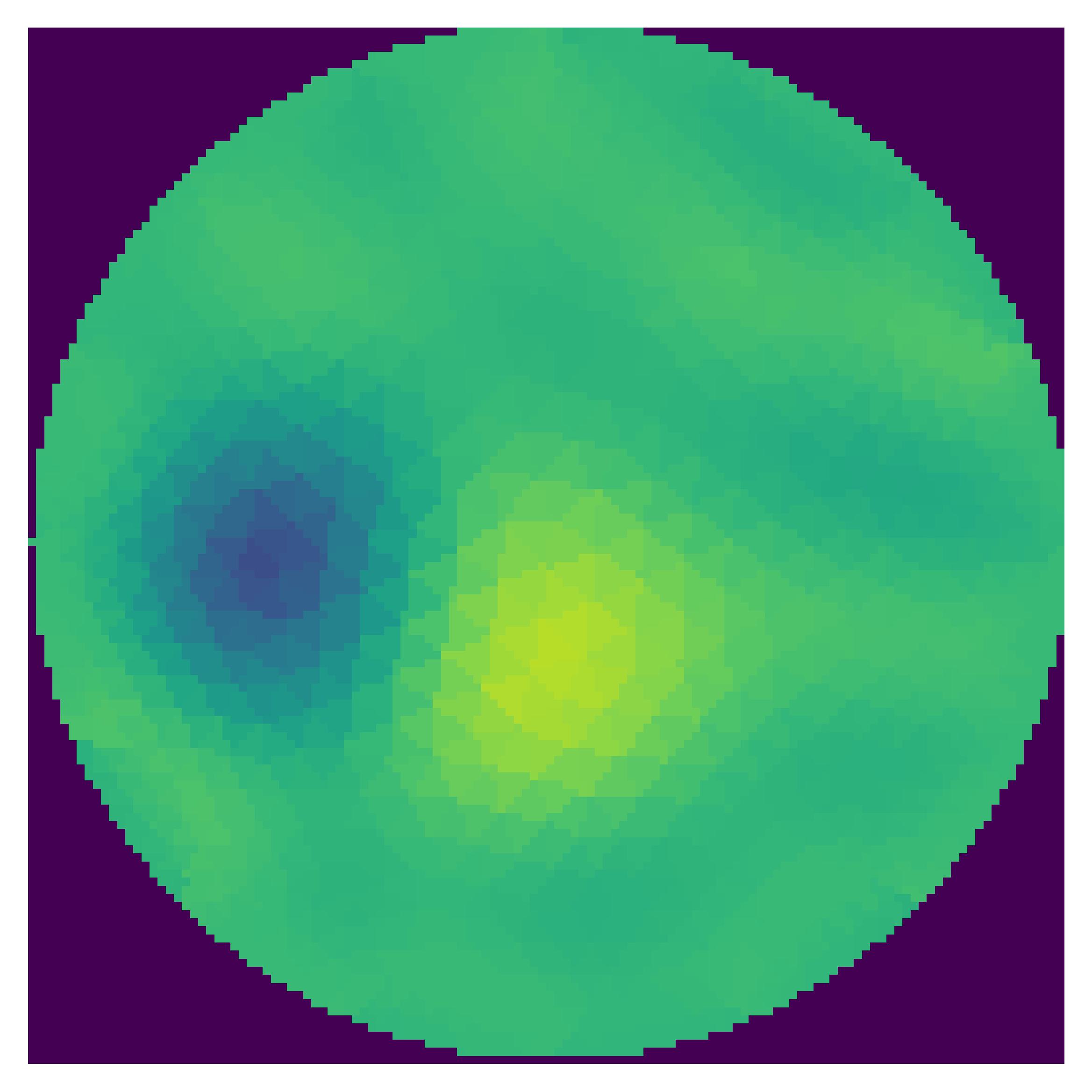}\vspace{5pt}
    \end{minipage}
}
\hspace{-7mm} 
\subfigure[CVAE]{
    \begin{minipage}[b]{0.19\linewidth}
    \centering
    \includegraphics[width=1in]{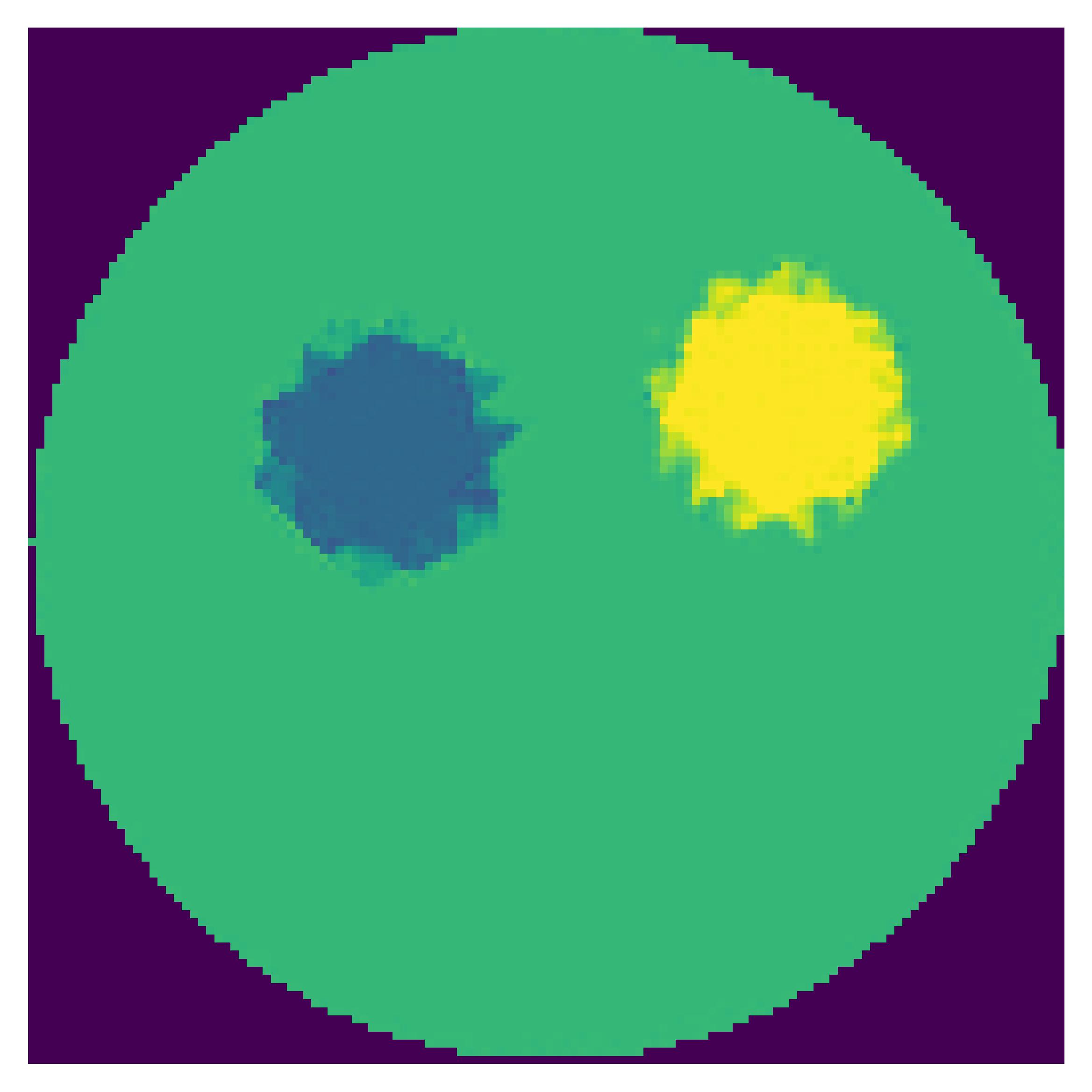}\vspace{5pt} 
    \includegraphics[width=1in]{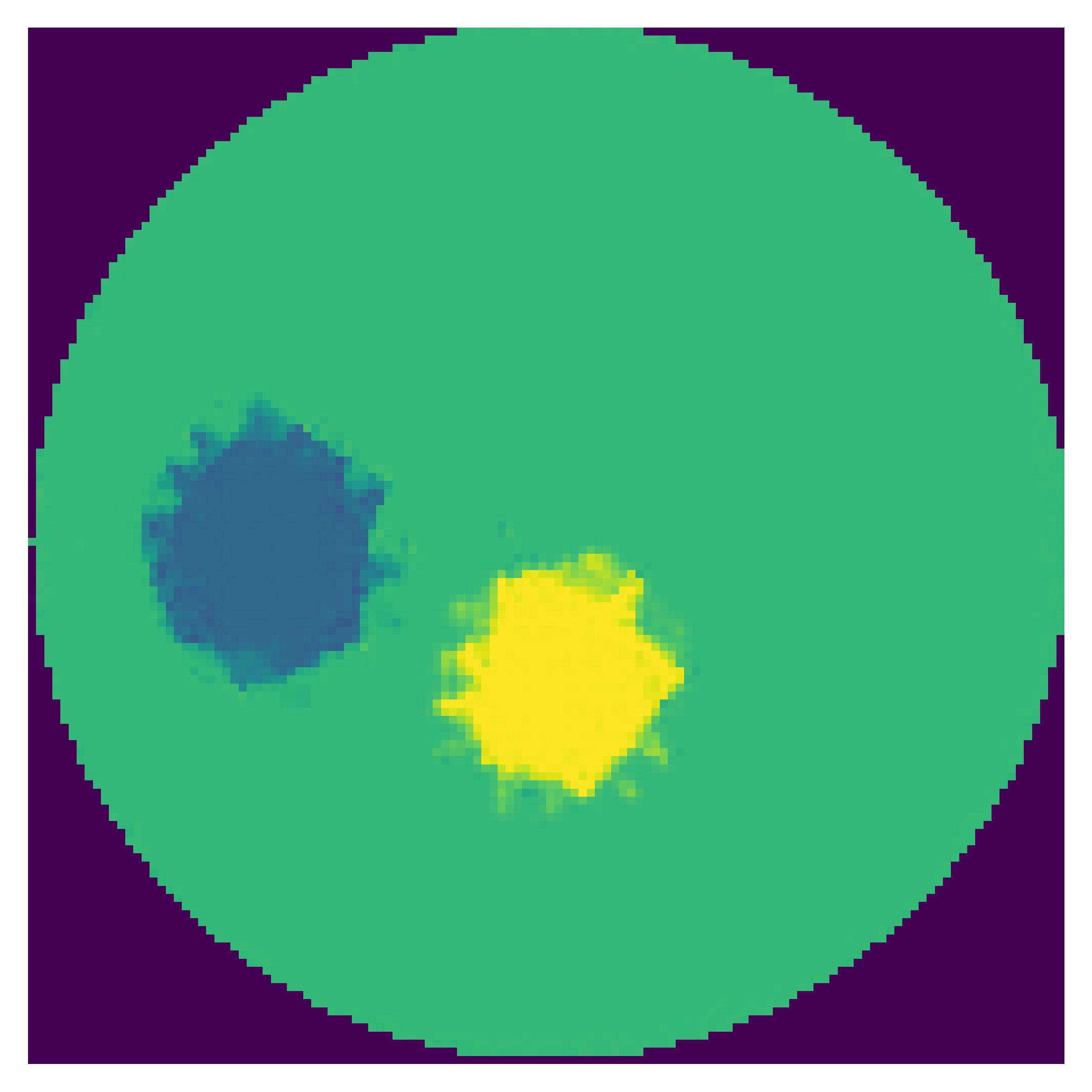}\vspace{5pt}
    \end{minipage}
}
\hspace{-7mm} 
\subfigure[CNF]{
    \begin{minipage}[b]{0.19\linewidth}
    \centering
    \includegraphics[width=1in]{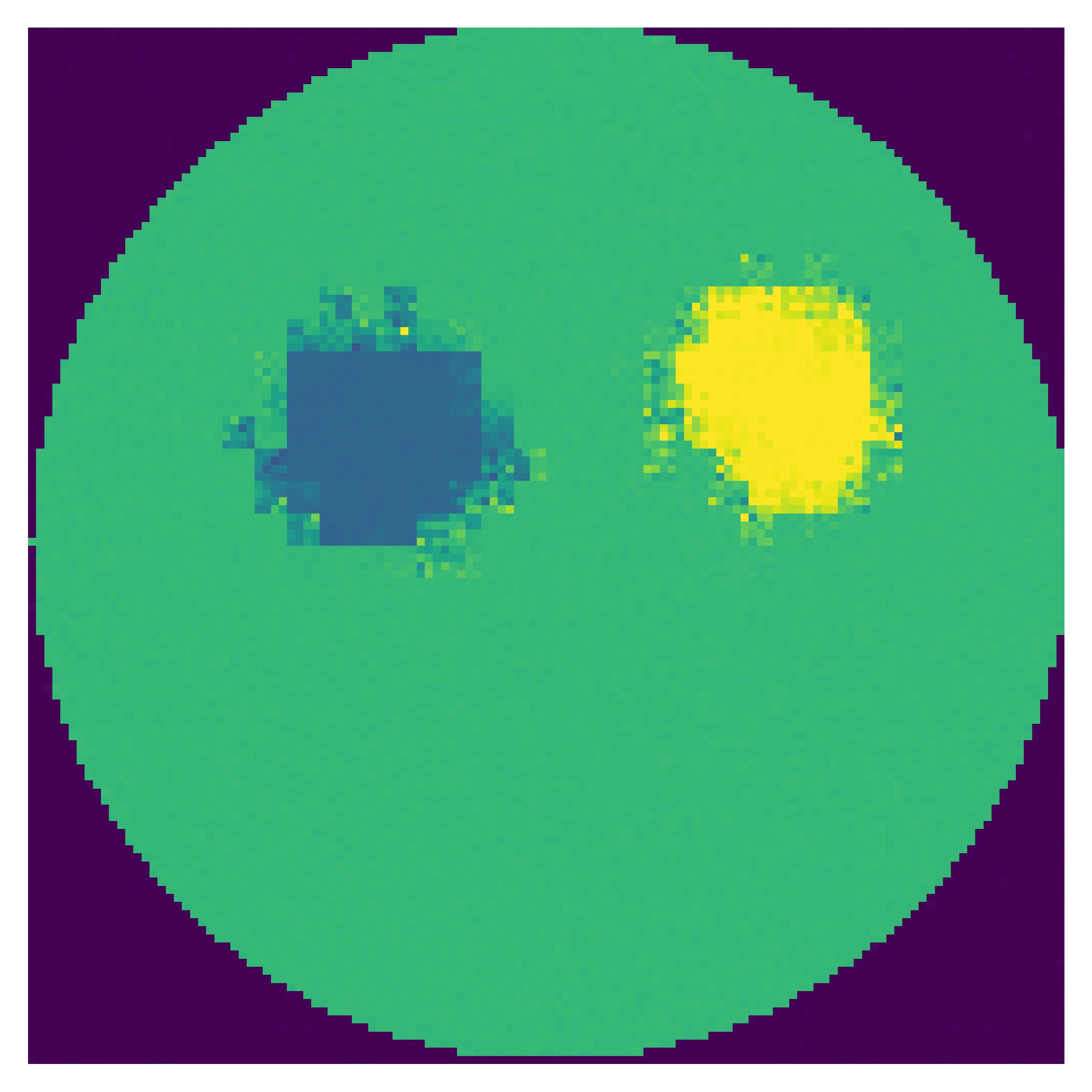}\vspace{5pt} 
    \includegraphics[width=1in]{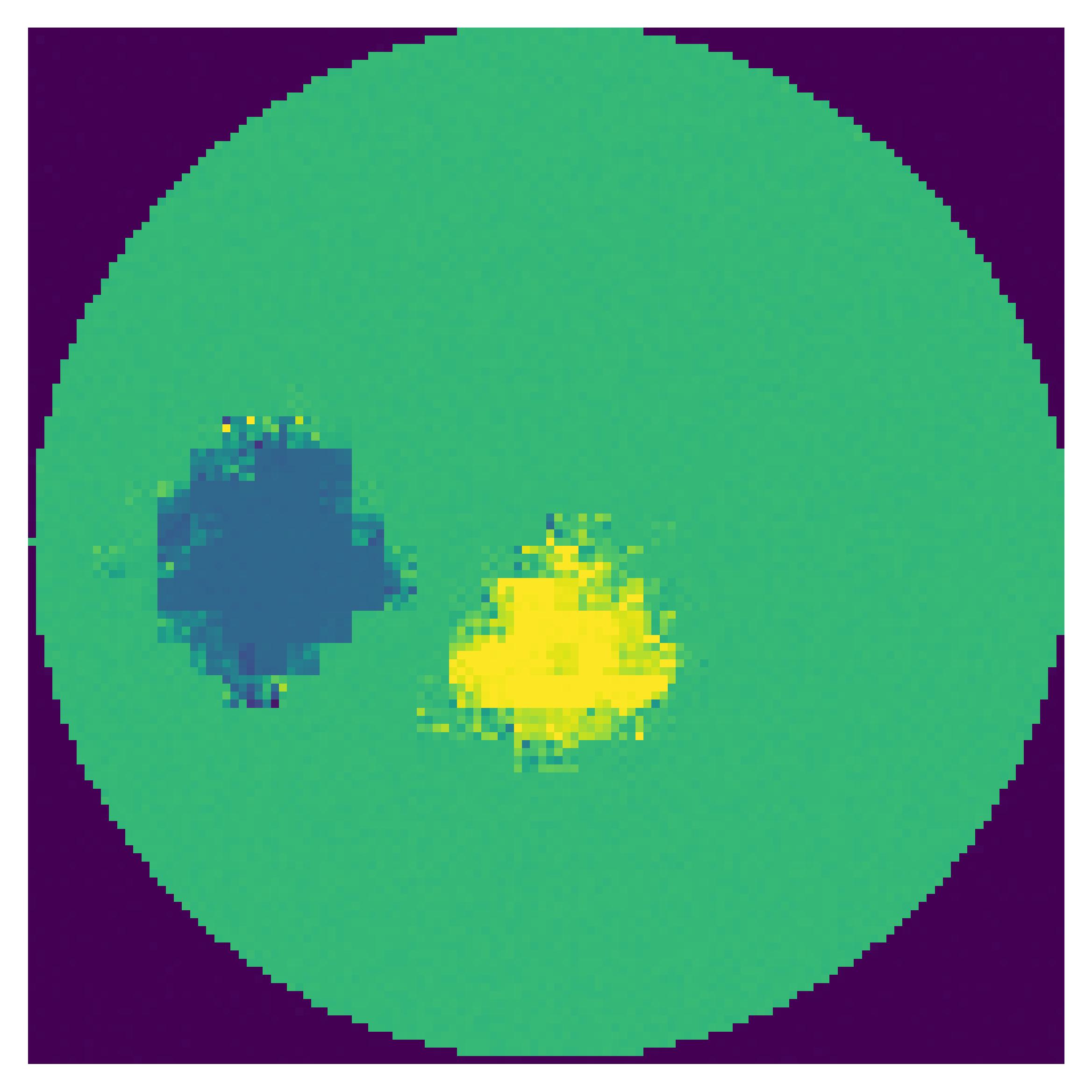}\vspace{5pt}
    \end{minipage}
}
\hspace{-6mm} 
\subfigure[${\mathrm{CSD}^{*}}_{40dB}^2$]{
    \begin{minipage}[b]{0.18\linewidth}
    \centering
    \includegraphics[width=1in]{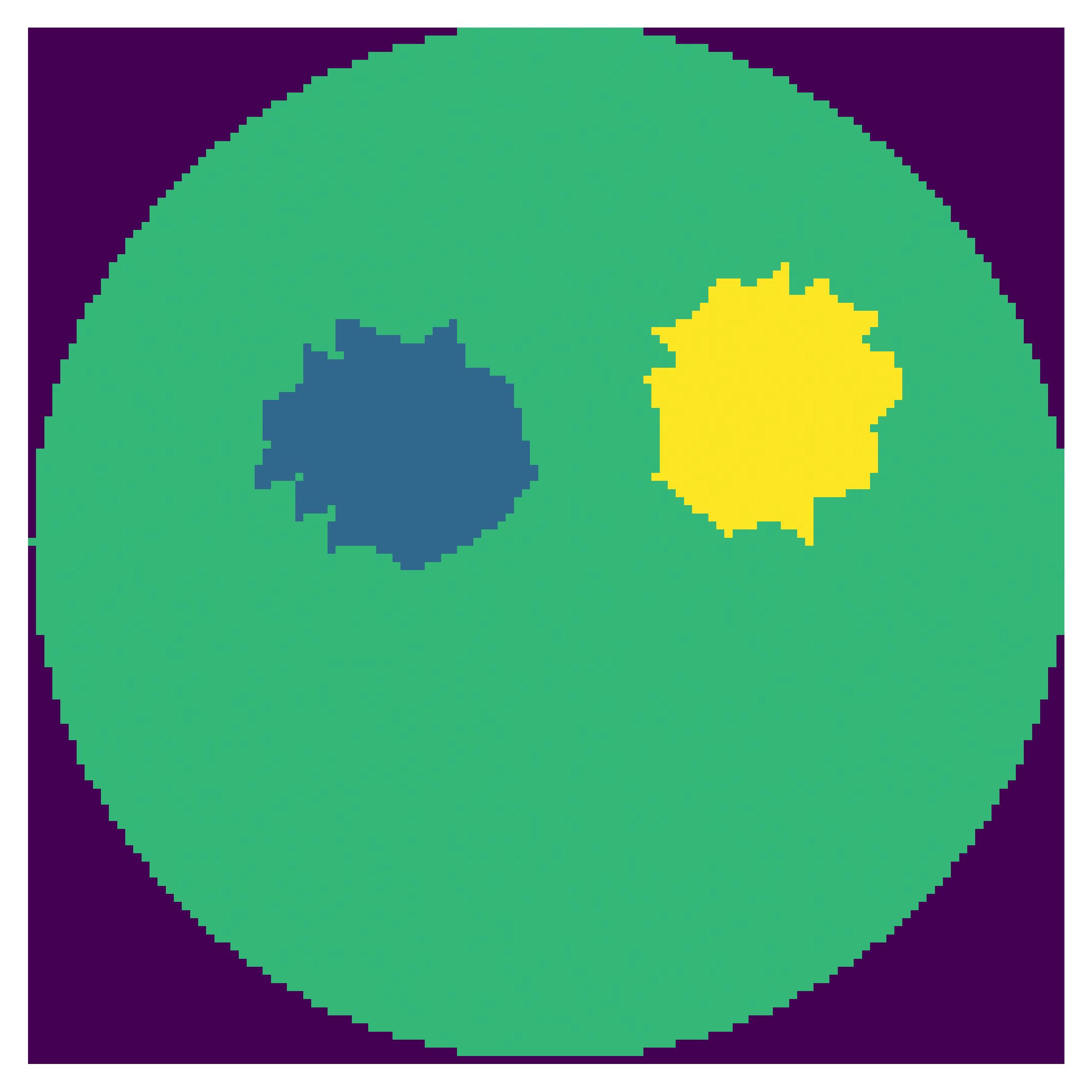}\vspace{5pt} 
    \includegraphics[width=1in]{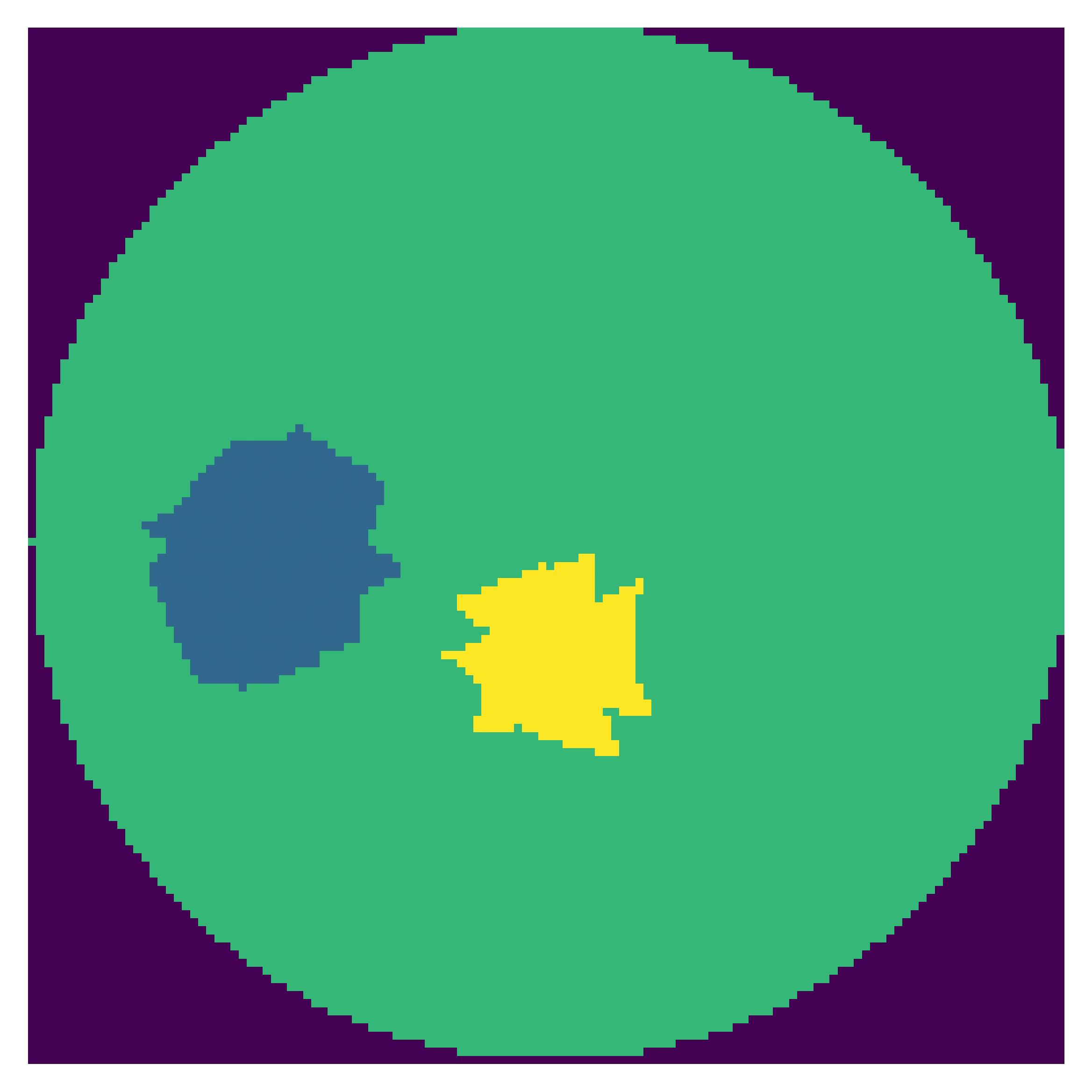}\vspace{5pt}
    \end{minipage}
}
\caption{\textbf{2 anomalies with noise in 25dB}, (a) original image of $128\times 128$ pixels, (b) result of Gauss-Newton,  reconstructed image from (c) CVAE, and (d) CNF, as well as (e) ${\mathrm{CSD}^{*}}_{40dB}^2$.}
\label{fig:25db_2}
\end{figure}

Based on the results shown in Table \ref{tab:25db2}, CVAE exhibits the best performance in all measures. However, as seen in Figure \ref{fig:25db_2}, the boundaries are still unclear, although the general outlines are close to the ground truth. CNF and $\mathrm{CSD}^*$ perform similarly, with CNF slightly outperforming $\mathrm{CSD}^*$ in terms of MSE and PSNR. However, the boundaries of CNF are not clear and they are only possible to determine approximate locations of anomalies. In contrast, $\mathrm{CSD}^*$ produces clear boundaries for anomalies. Nevertheless, in some cases, the shape of anomalies generated by $\mathrm{CSD}^*$ deviates significantly from the ground truth.

\begin{table*}[ht]
\resizebox{\linewidth}{!}{
\begin{threeparttable}
    \centering
    \caption{4 inclusions with noise in 25dB.}
    \label{tab:25db4}
    \begin{tabular}{lcccccc}
    \toprule
        \textbf{Method} & \textbf{Mse} & \textbf{PSNR} & \textbf{SSIM} & \textbf{RE} & \textbf{AE} & \textbf{DR} \\\midrule
        GN &0.0164$\pm$0.003 &24.752$\pm$1.018 &0.747$\pm$0.031 &0.092$\pm$0.012 &0.072$\pm$0.009 &1.108$\pm$0.129  \\ [0.8ex]
        CVAE &0.0134$\pm$0.005 &25.838$\pm$1.416 &0.868$\pm$0.023 &0.044$\pm$0.010 &0.035$\pm$0.007 &1.104$\pm$0.031  \\[0.8ex]
        CNF &\textbf{0.0093$\pm$0.004} &\textbf{26.839$\pm$1.760} &0.922$\pm$0.019  &0.026$\pm$0.007 &0.020$\pm$0.006 &0.631$\pm$0.065  \\[0.8ex]
        ${\mathrm{CSD}^*}_{40dB}^4$ &0.0100$\pm$0.004 &26.347$\pm$1.768 &\textbf{0.928$\pm$0.022} &\textbf{0.020$\pm$0.009} &\textbf{0.016$\pm$0.007} &\textbf{0.999$\pm$0.002} \\
    \bottomrule
    \end{tabular}
    \begin{tablenotes}    
        \footnotesize              
        \item[1] ${\mathrm{CSD}^*}_{40dB}^4$ in the chart is the trained model in 40dB noise with 4 anomalies.                
       \end{tablenotes}      
\end{threeparttable}}
\end{table*}

\begin{figure}[htbp]
\centering
\subfigure[GT]{
    \begin{minipage}[b]{0.19\linewidth} 
    \centering
    \includegraphics[width=1in]{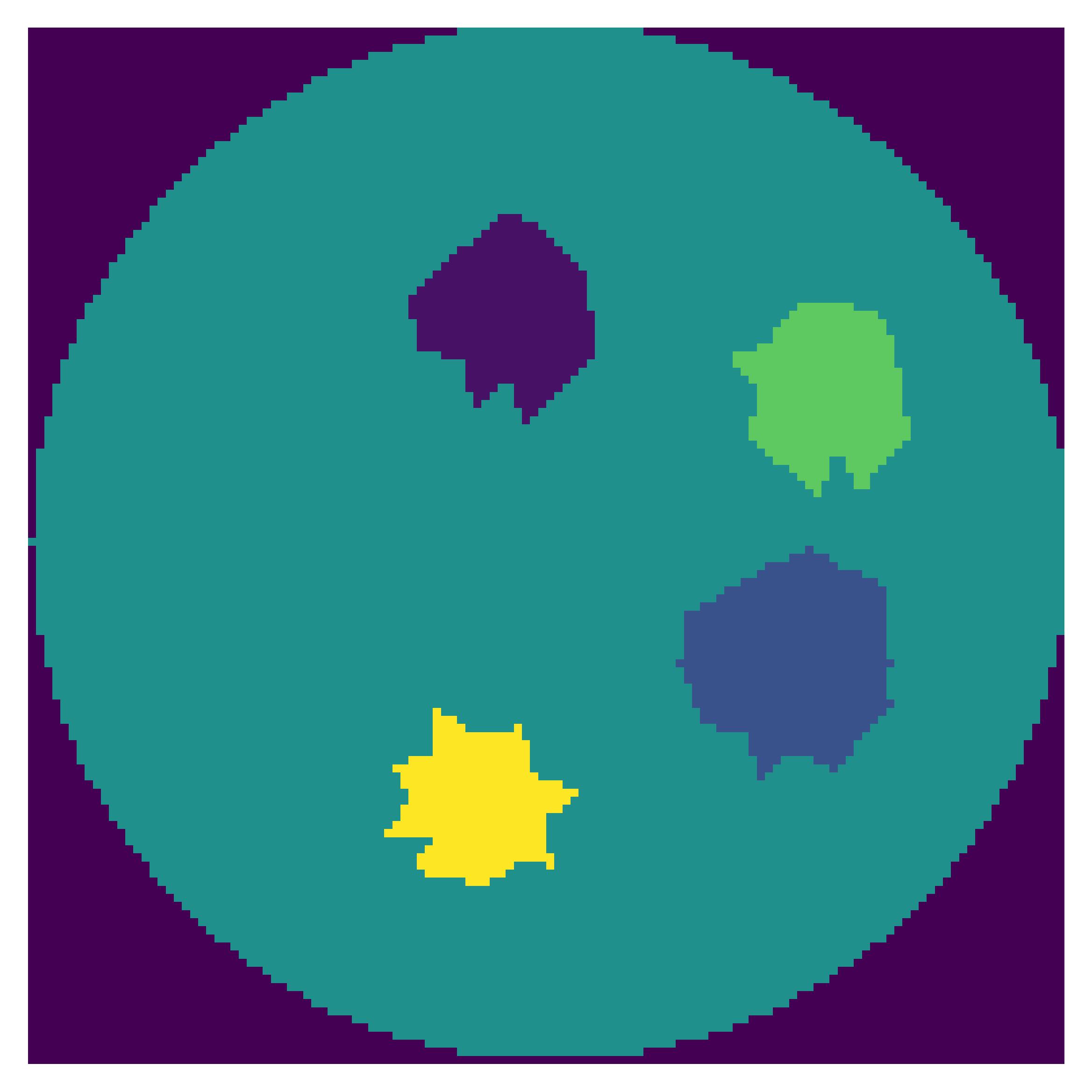}\vspace{5pt} 
    \includegraphics[width=1in]{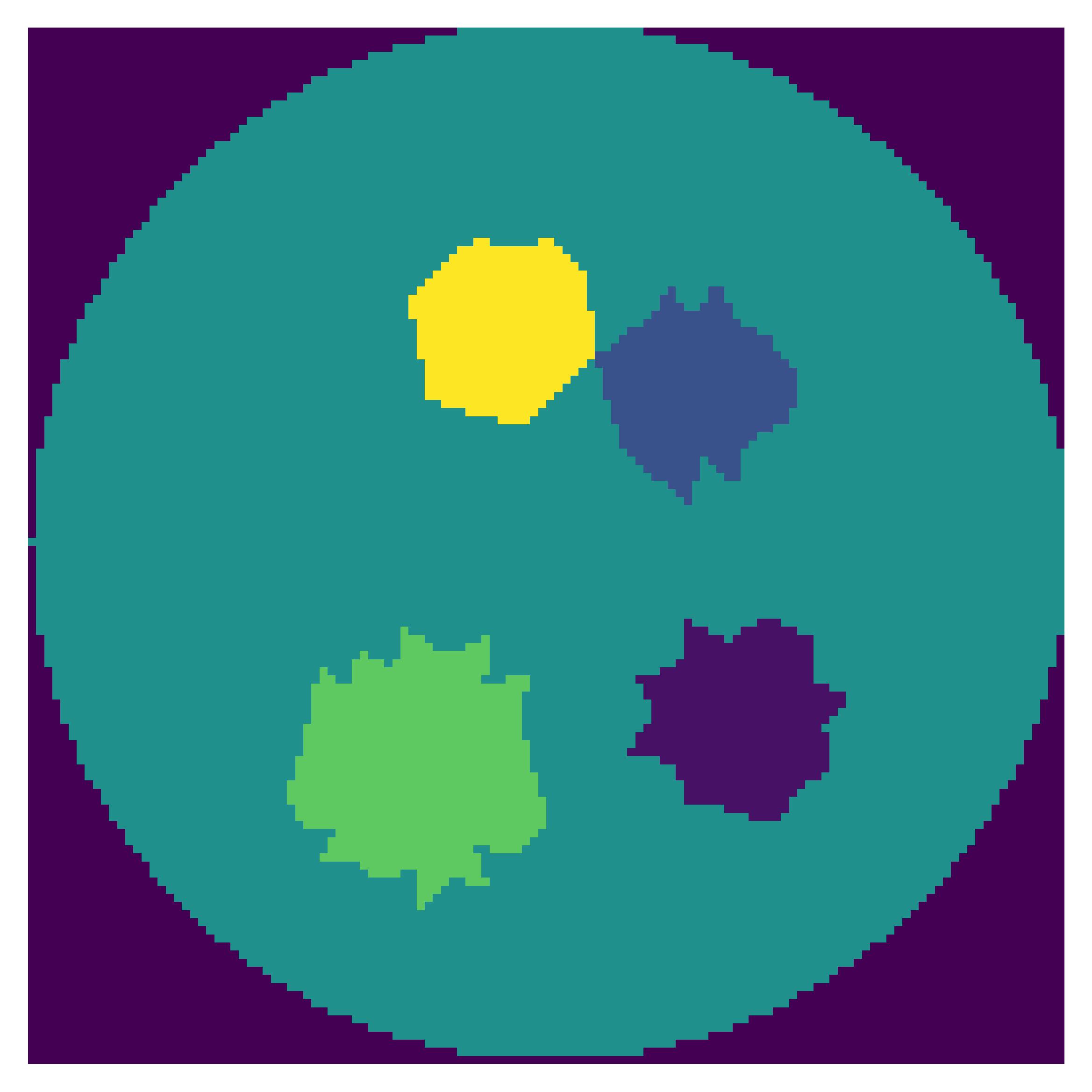}\vspace{5pt}
    \end{minipage}
}
\hspace{-7mm} 
\subfigure[GN]{
    \begin{minipage}[b]{0.19\linewidth}
    \centering
    \includegraphics[width=1in]{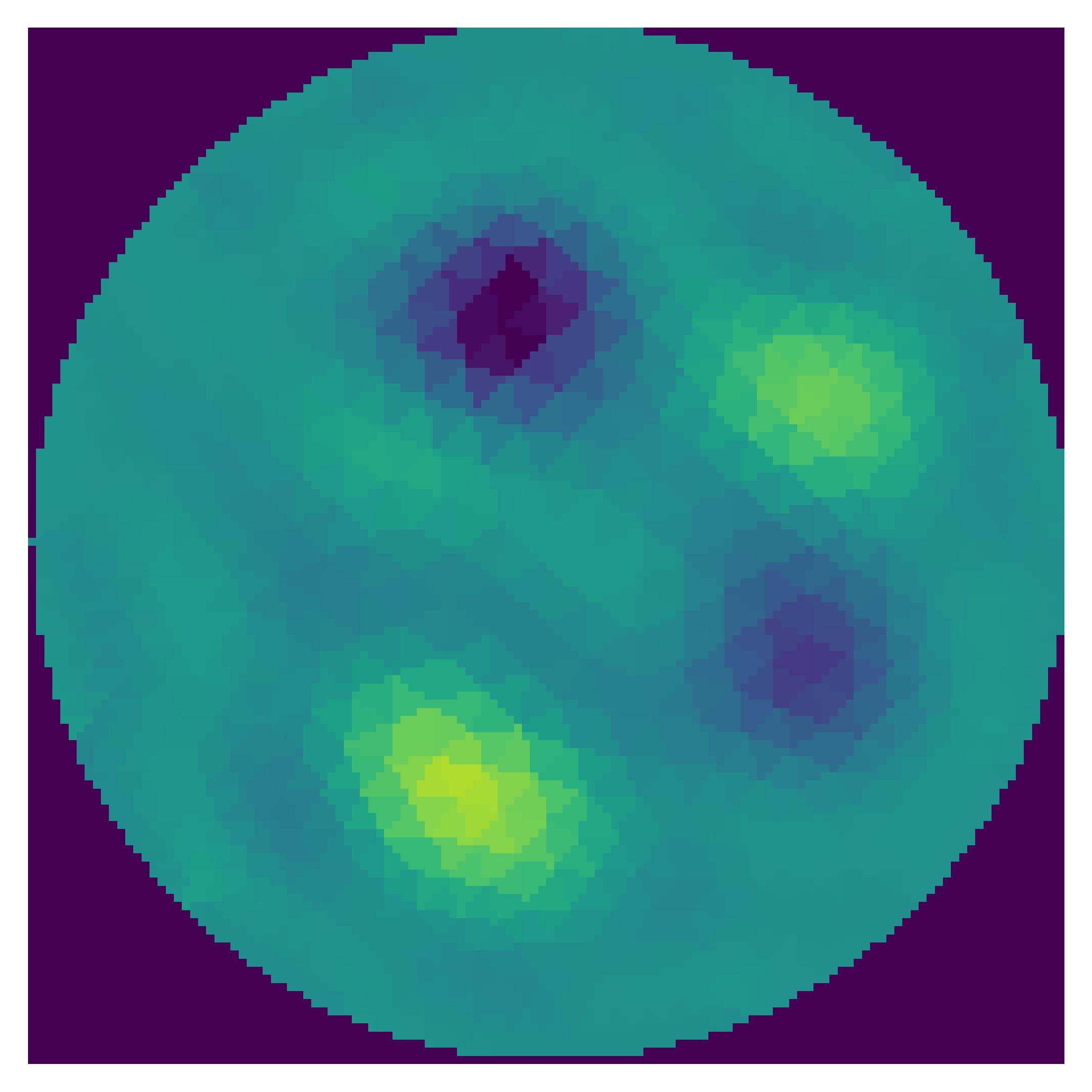}\vspace{5pt} 
    \includegraphics[width=1in]{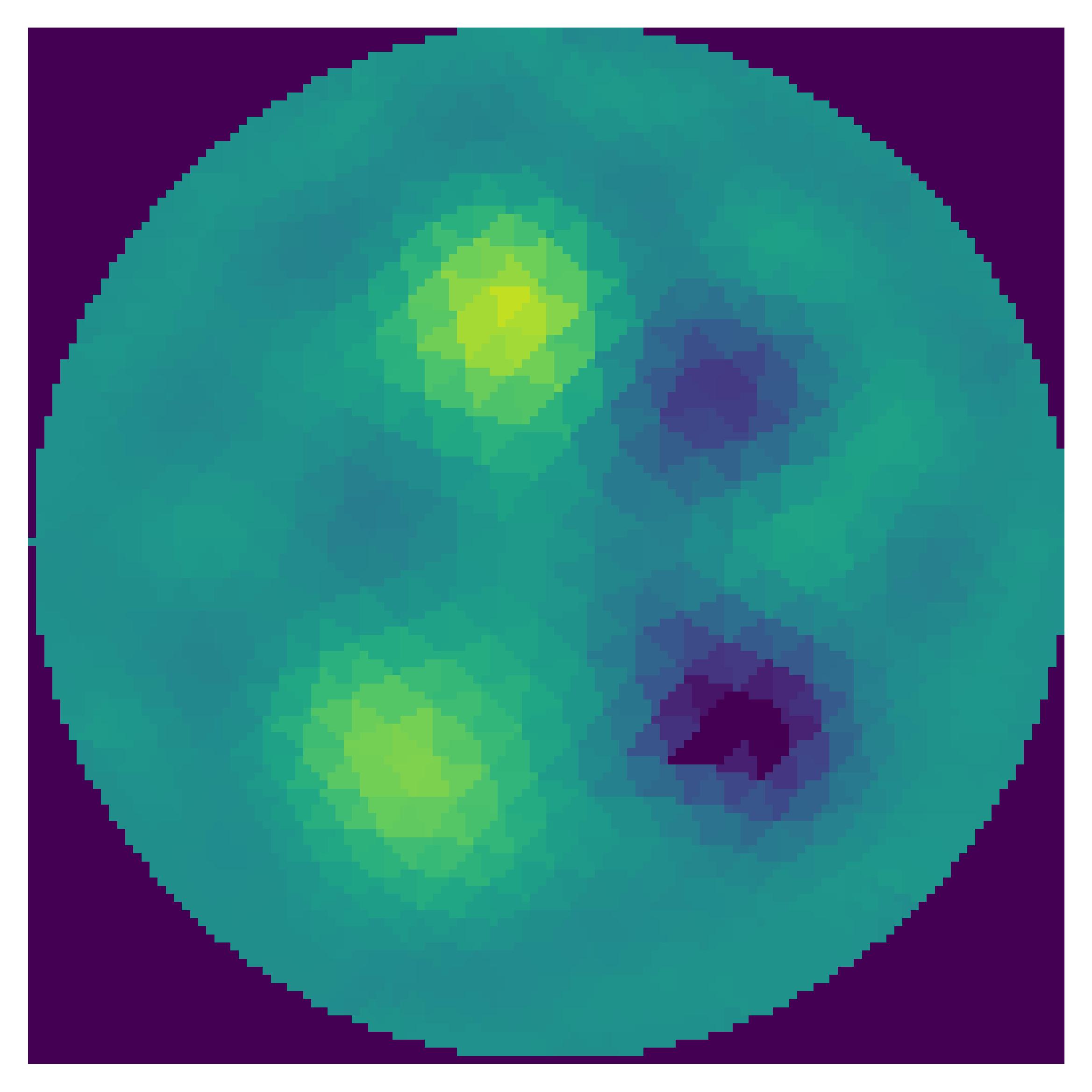}\vspace{5pt}
    \end{minipage}
}
\hspace{-7mm} 
\subfigure[CVAE]{
    \begin{minipage}[b]{0.19\linewidth}
    \centering
    \includegraphics[width=1in]{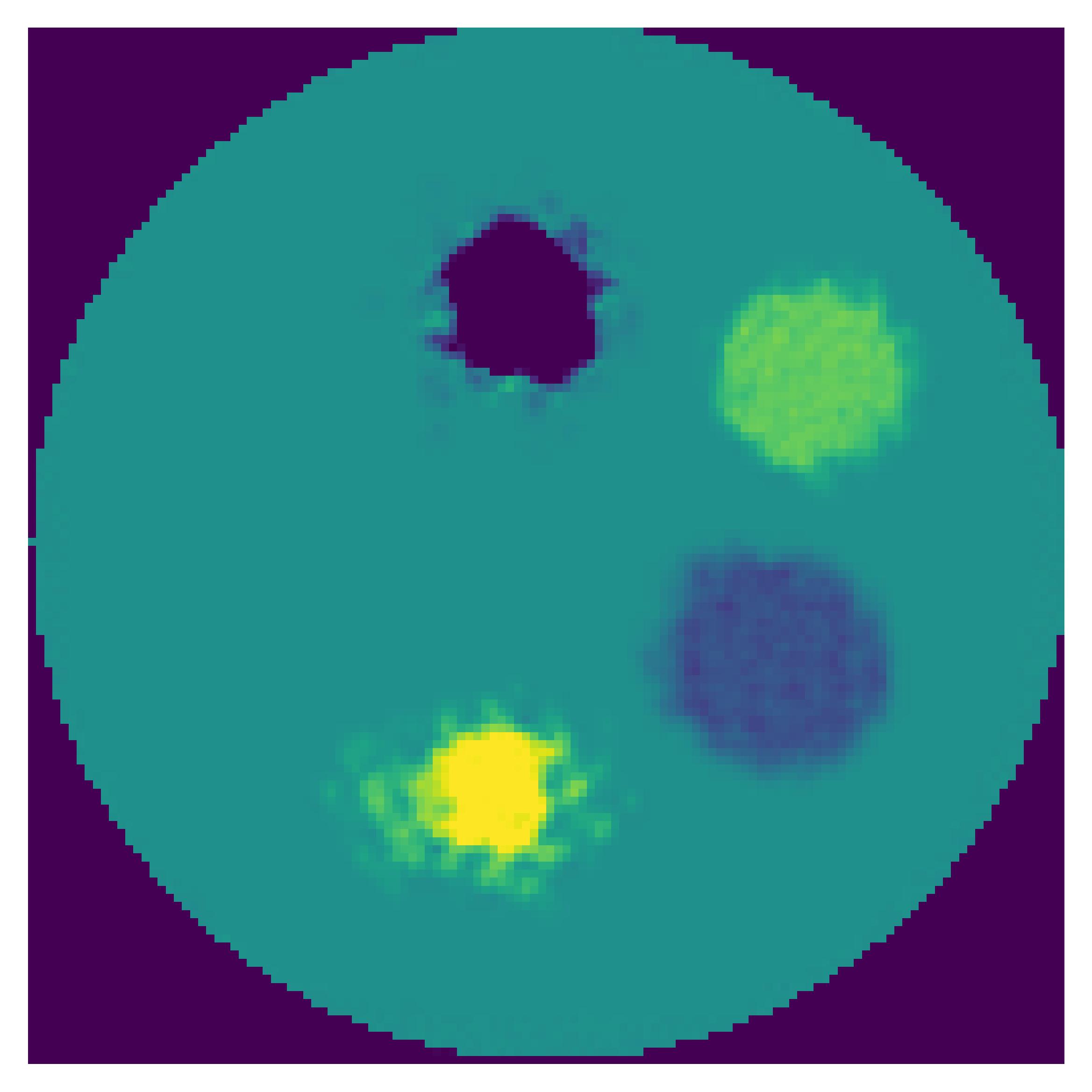}\vspace{5pt} 
    \includegraphics[width=1in]{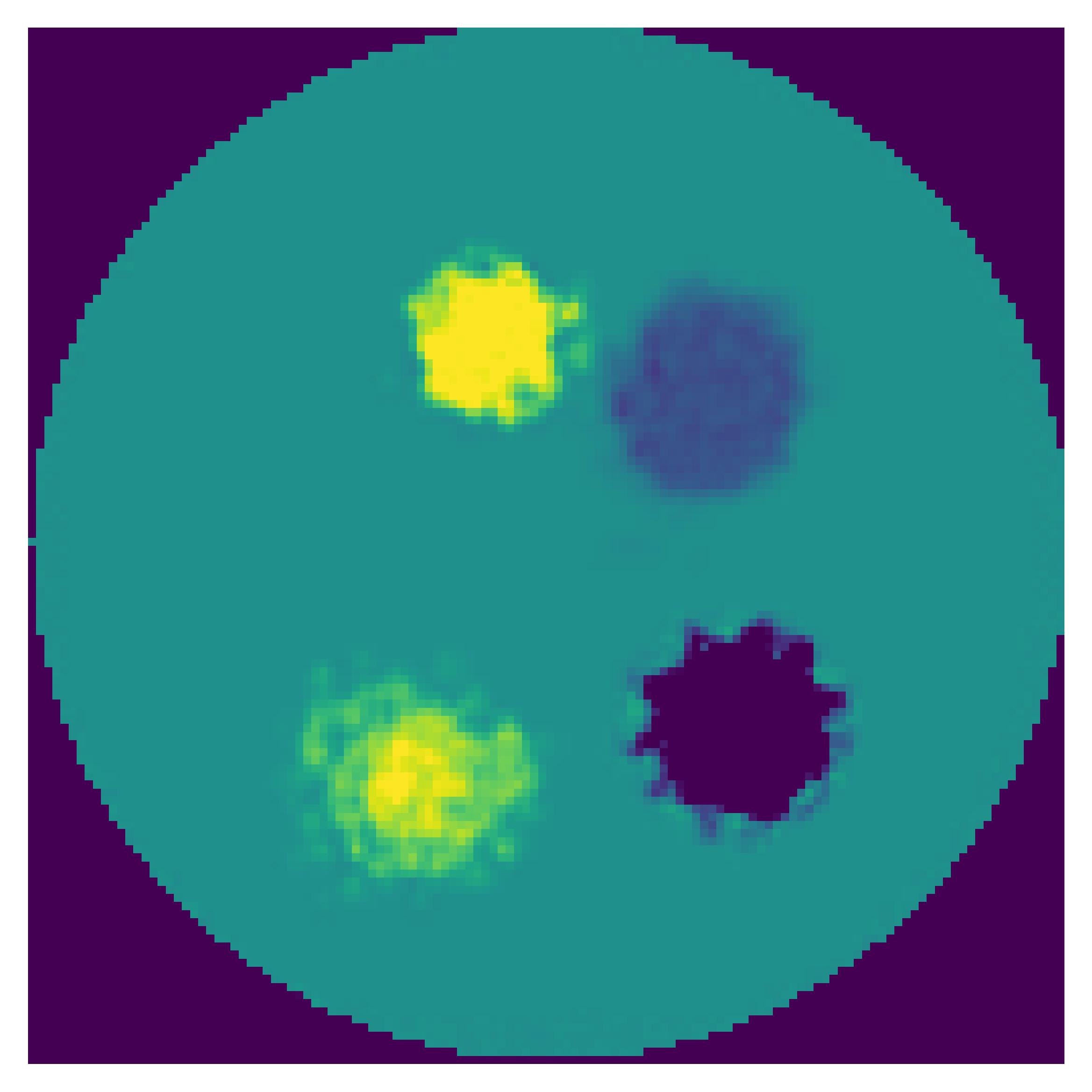}\vspace{5pt}
    \end{minipage}
}
\hspace{-7mm} 
\subfigure[CNF]{
    \begin{minipage}[b]{0.19\linewidth}
    \centering
    \includegraphics[width=1in]{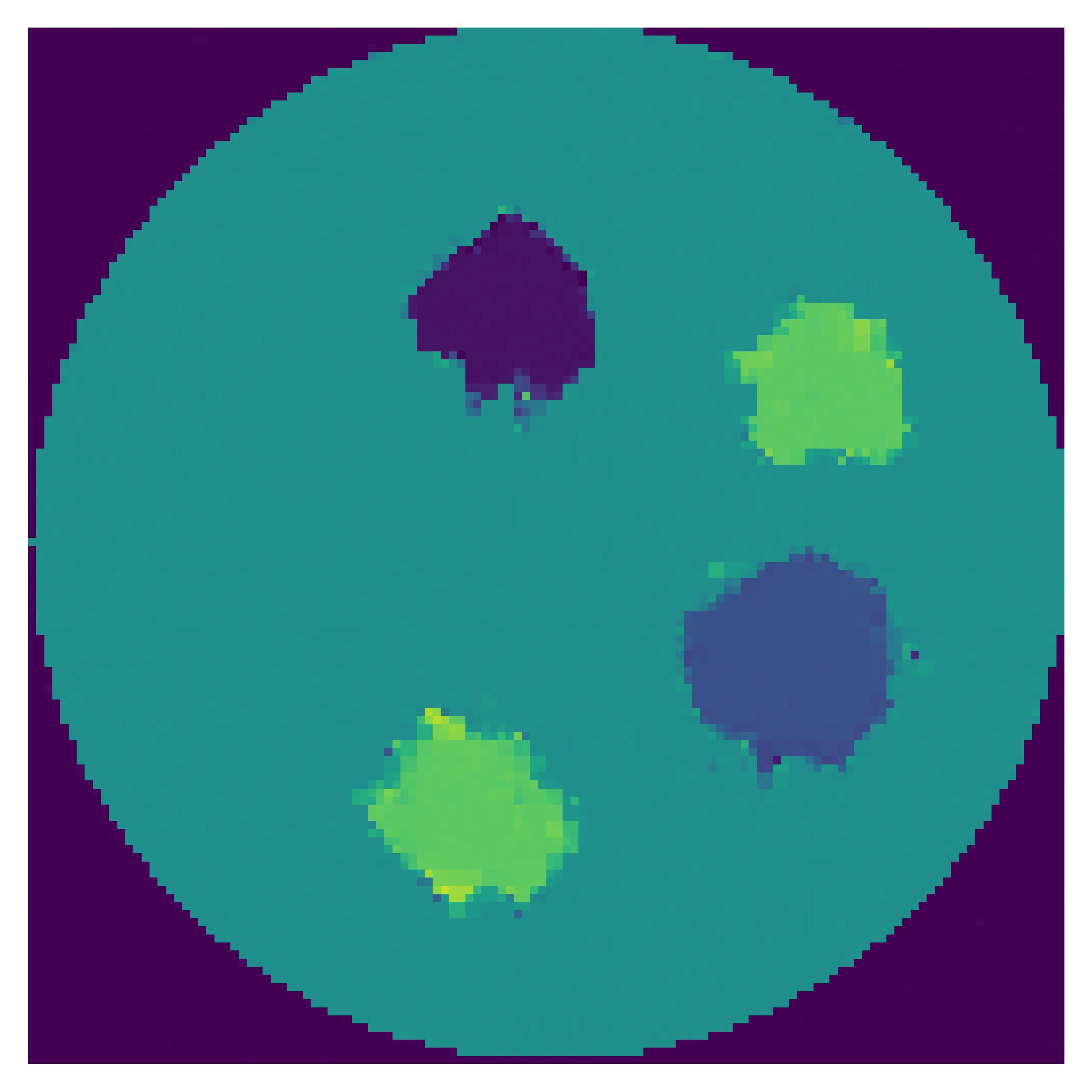}\vspace{5pt} 
    \includegraphics[width=1in]{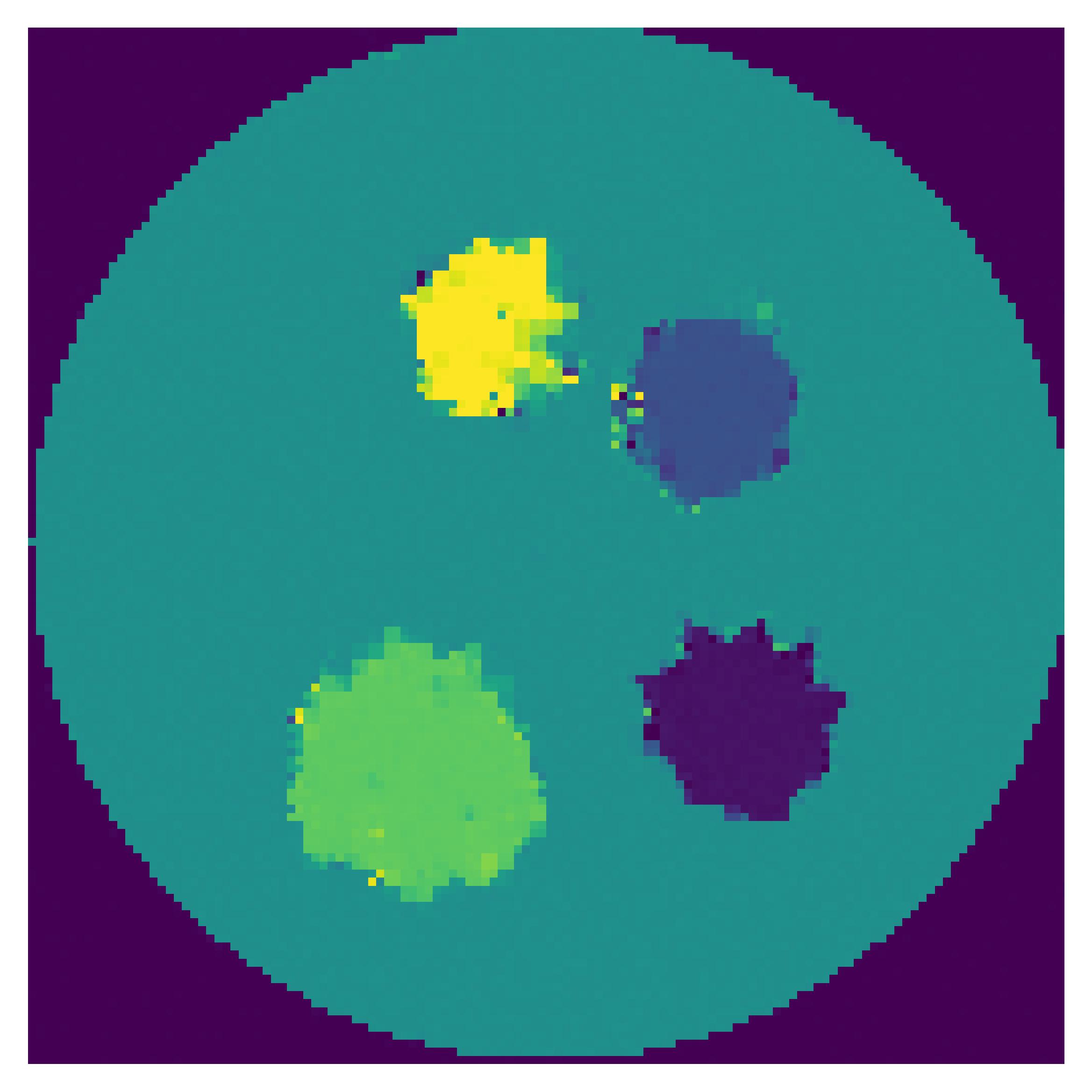}\vspace{5pt}
    \end{minipage}
}
\hspace{-6mm} 
\subfigure[${\mathrm{CSD}^*}_{40dB}^4$]{
    \begin{minipage}[b]{0.18\linewidth}
    \centering
    \includegraphics[width=1in]{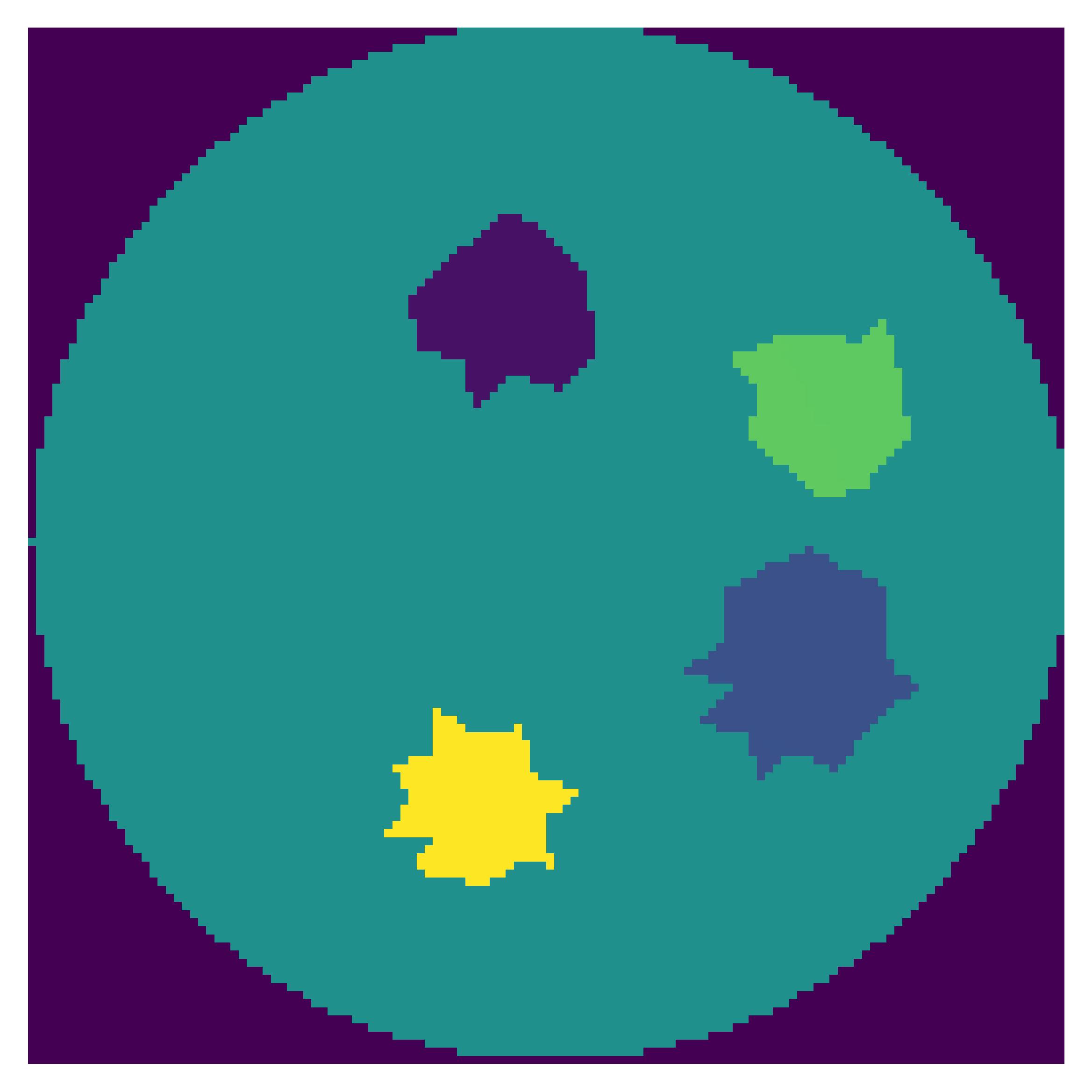}\vspace{5pt} 
    \includegraphics[width=1in]{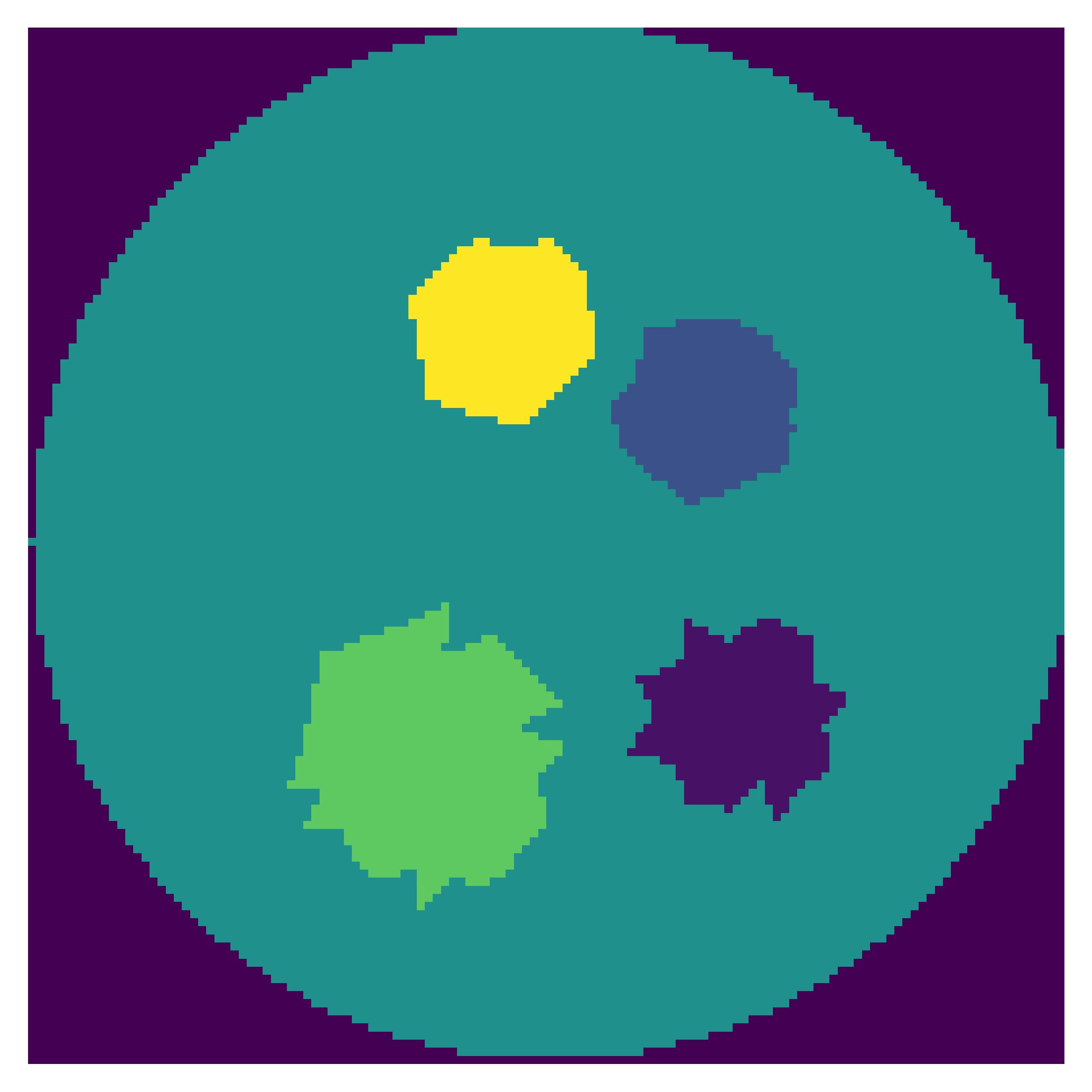}\vspace{5pt}
    \end{minipage}
}
\caption{\textbf{4 anomalies and noise in 25dB}, (a) original image of $128\times 128$ pixels, (b) result of Gauss-Newton,  reconstructed image from (c) CVAE, and (d) CNF, as well as (e) ${\mathrm{CSD}^*}_{40dB}^4$. }
\label{fig:25db_4}
\end{figure}

From Table \ref{tab:25db4} shown above, while CNF achieves best in MSE and PSNR compared to the other three models, with MSE 43.2\% lower and PSNR 8.4\% higher than GN, $\mathrm{CSD}^*$ model trained on 40dB again has better results in terms of SSIM, AE, RE, and DR, with SSIM 24\% higher, RE and AE around 75\% lower, as well as DR 9.8\% better than GN. Overall, the differences between the two models' indicator values are not very large. Two samples are shown in Figure \ref{fig:25db_4}. Performances of both GN and VAE do not improve significantly in the presence of four anomalies with increased noise. CNF also starts to exhibit inconsistent shapes, unclear boundaries, and inaccurate color representation of anomalies. Additionally, there is an increase in boundary noise. $\mathrm{CSD}^*$, on the other hand, demonstrates significantly better performance, although there is still a notable deviation from the ground truth in terms of shape.

\subsubsection{Generalization}
This section is to show whether the object with 2 anomalies can be generated by a model trained on 4 anomalies dataset in low-level noise, 40dB. We conduct tests on datasets at the same noise level of 40 dB and a higher noise level of 25 dB, respectively.
\begin{table*}[ht]
\resizebox{\linewidth}{!}{
\begin{threeparttable}
    \centering
    \caption{Generalization performance at the same noise level.}
    \label{tab:gener_40db2}
    \begin{tabular}{lcccccc}
    \toprule
        \textbf{Method} & \textbf{MSE} & \textbf{PSNR} & \textbf{SSIM} & \textbf{RE} & \textbf{AE} & \textbf{DR} \\ \midrule
        CVAE &0.0061$\pm$0.0028 &27.965$\pm$1.209 &0.919$\pm$0.013 &0.027 $\pm$0.007 &0.021$\pm$0.005 &1.262$\pm$0.192 \\[0.8ex]
        CNF &\textbf{0.0008$\pm$0.0004} &\textbf{35.031$\pm$2.691} &\textbf{0.973$\pm$0.006} &\textbf{0.008$\pm$0.001} &\textbf{0.006$\pm$0.001} &0.805$\pm$0.041  \\[0.8ex]
       $\mathrm{CSD}^*$ &0.0015$\pm$0.0009 &34.082$\pm$7.267 &0.966$\pm$0.023 &0.008$\pm$0.004 &0.006$\pm$0.003 &\textbf{1.003$\pm$0.005} \\
    \bottomrule
    \end{tabular}    
\end{threeparttable}}
\end{table*}

\begin{figure}[htbp]
\centering
\subfigure[GT]{
    \begin{minipage}[b]{0.22\linewidth} 
    \centering
    \includegraphics[width=1in]{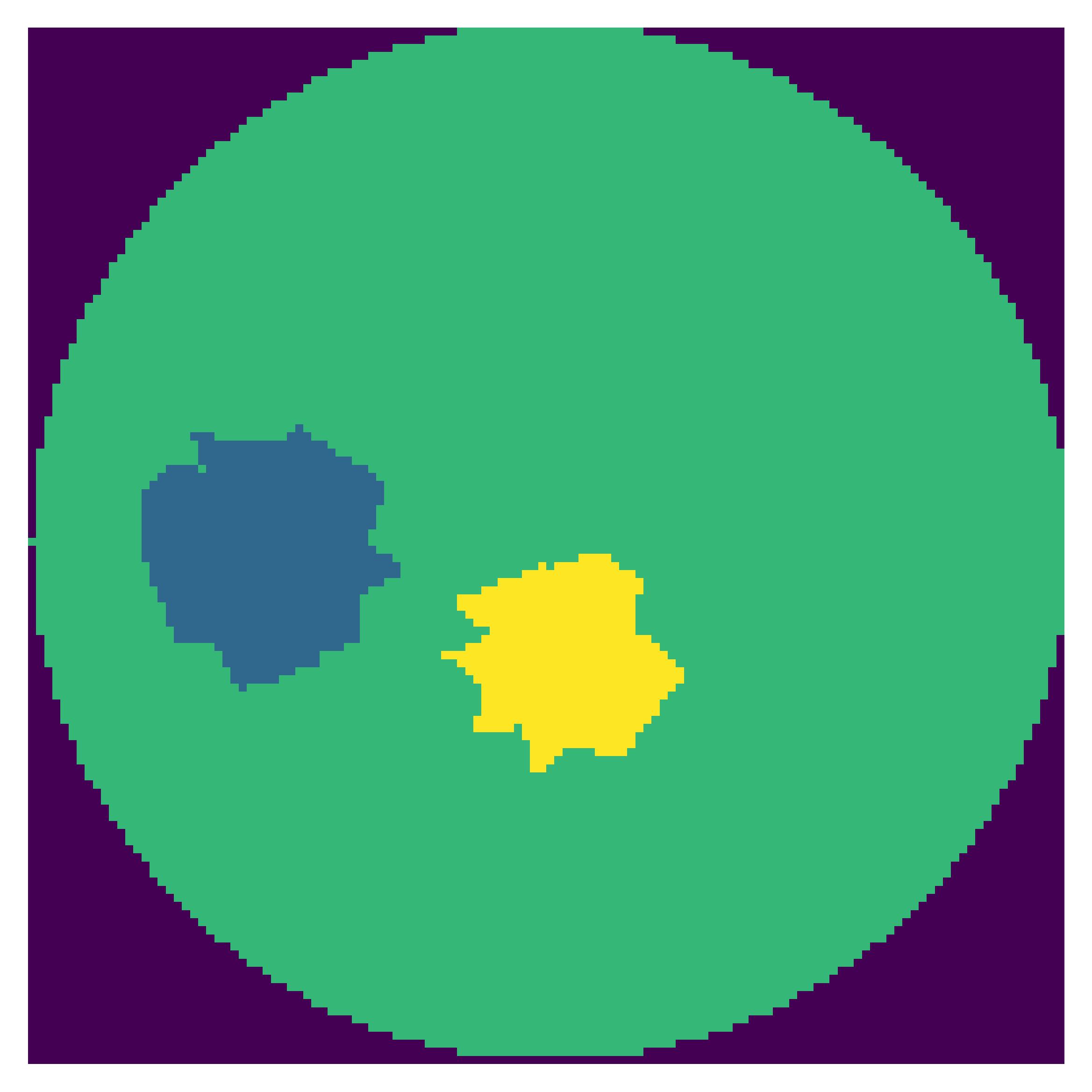}\vspace{5pt} 
    \includegraphics[width=1in]{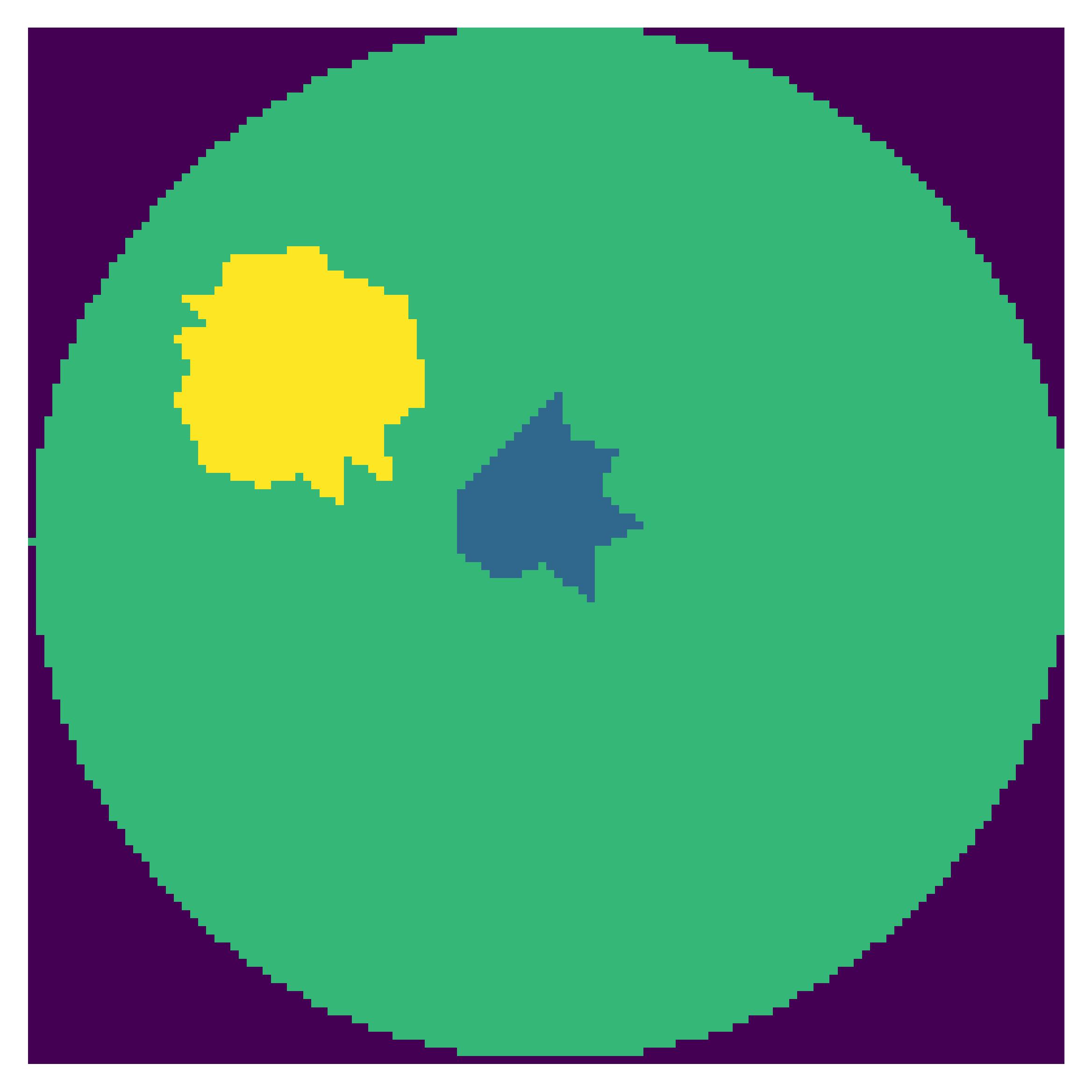}\vspace{5pt}
    \end{minipage}
}
\hspace{-12mm} 
\subfigure[CVAE]{
    \begin{minipage}[b]{0.22\linewidth}
    \centering
    \includegraphics[width=1in]{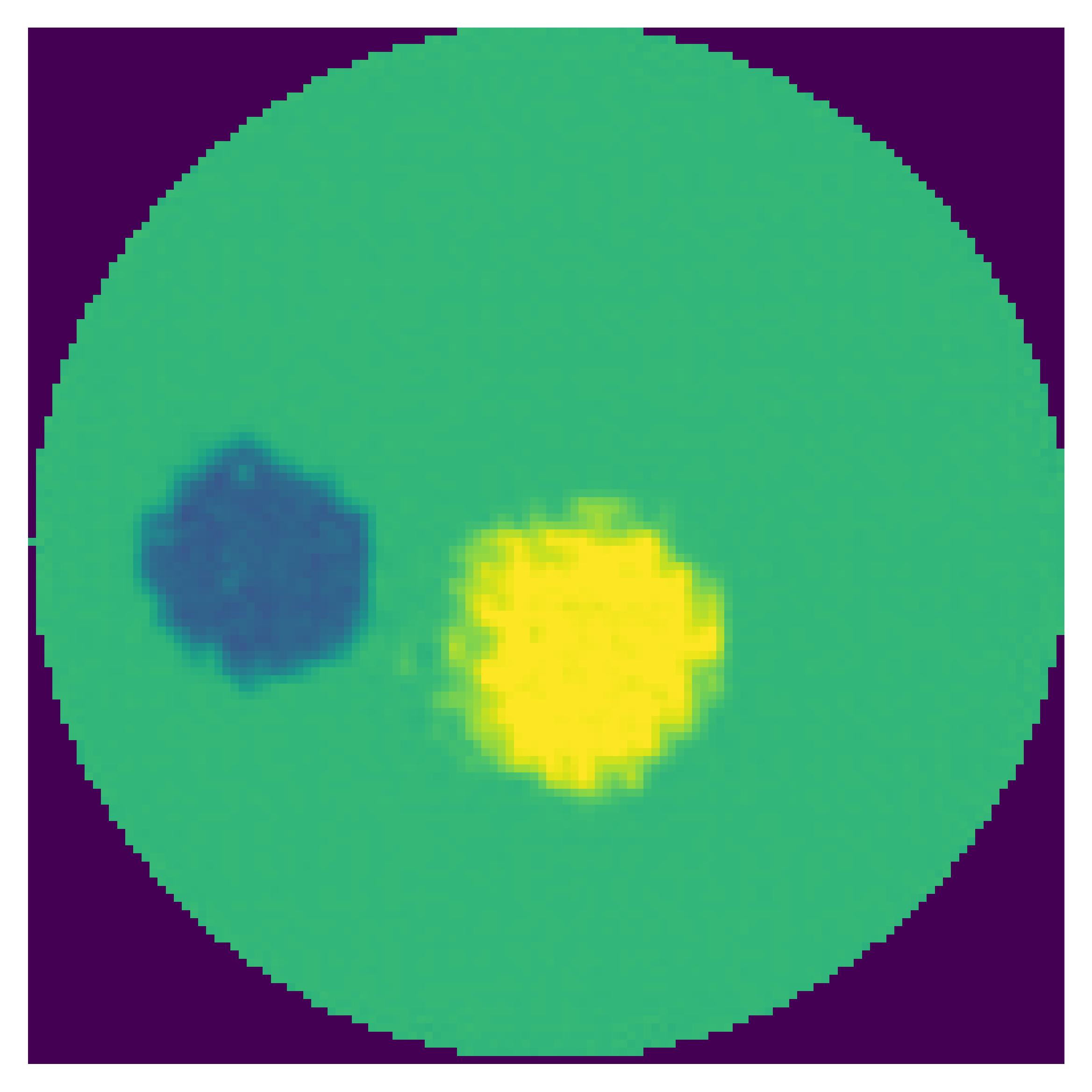}\vspace{5pt} 
    \includegraphics[width=1in]{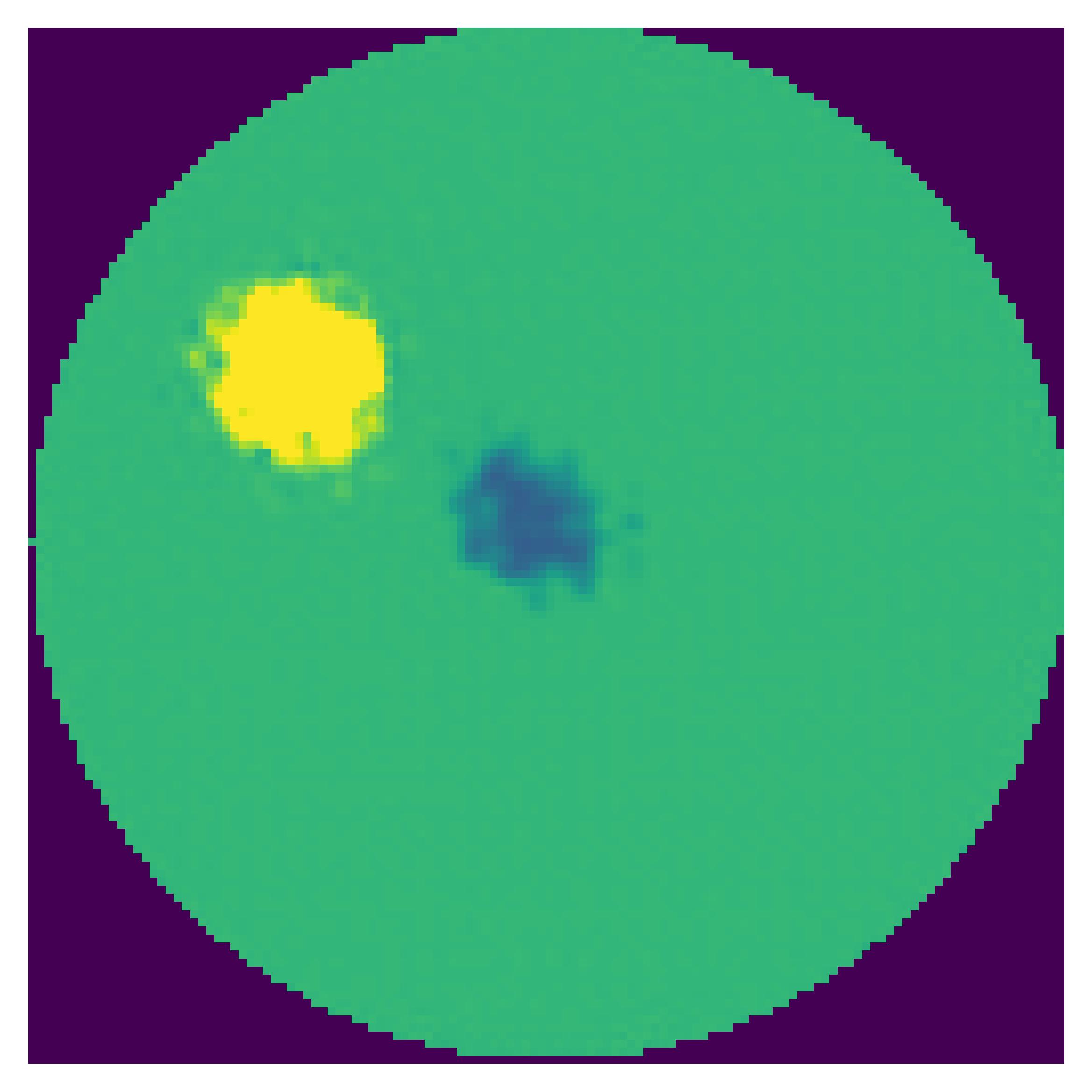}\vspace{5pt}
    \end{minipage}
}
\hspace{-12mm} 
\subfigure[CNF]{
    \begin{minipage}[b]{0.22\linewidth}
    \centering
    \includegraphics[width=1in]{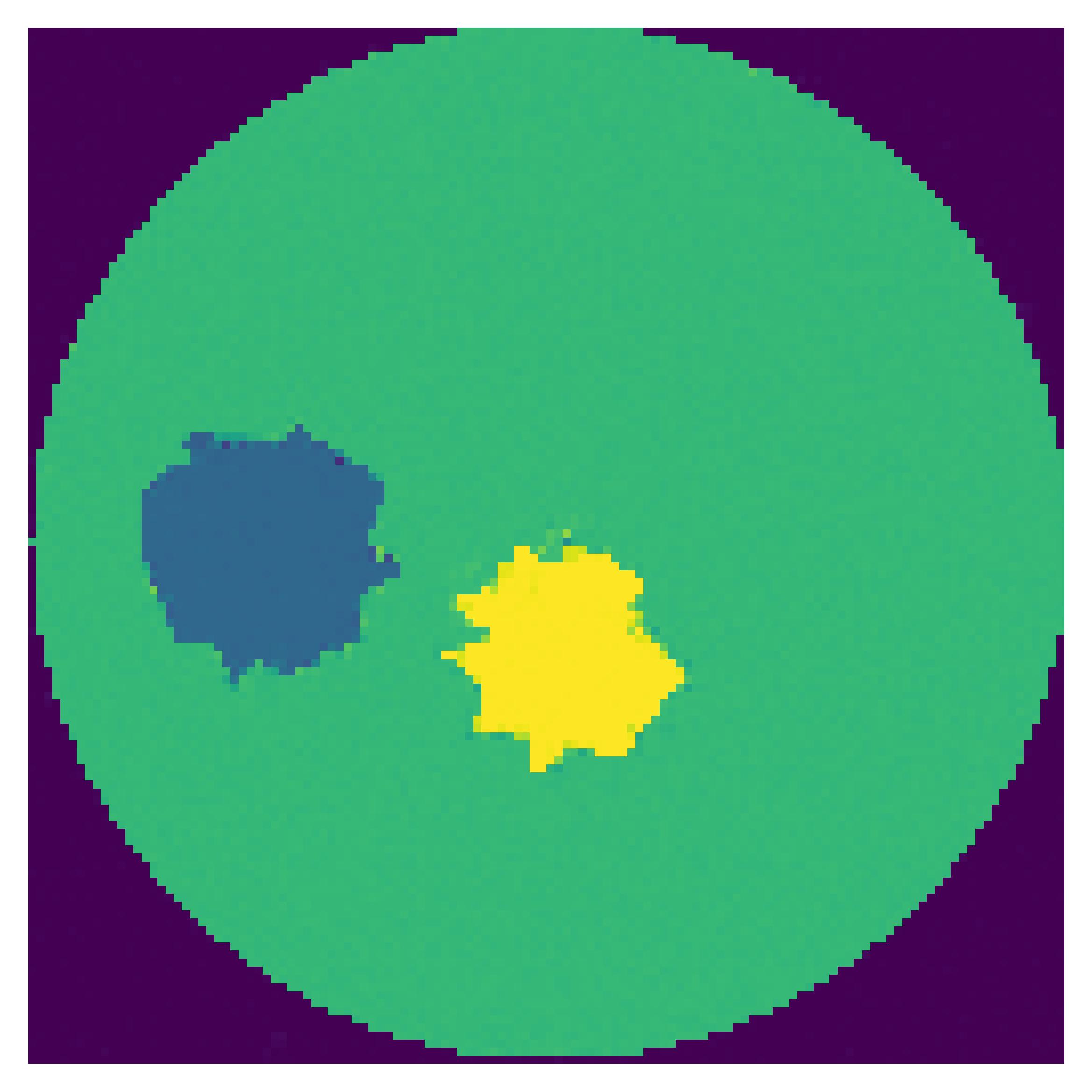}\vspace{5pt} 
    \includegraphics[width=1in]{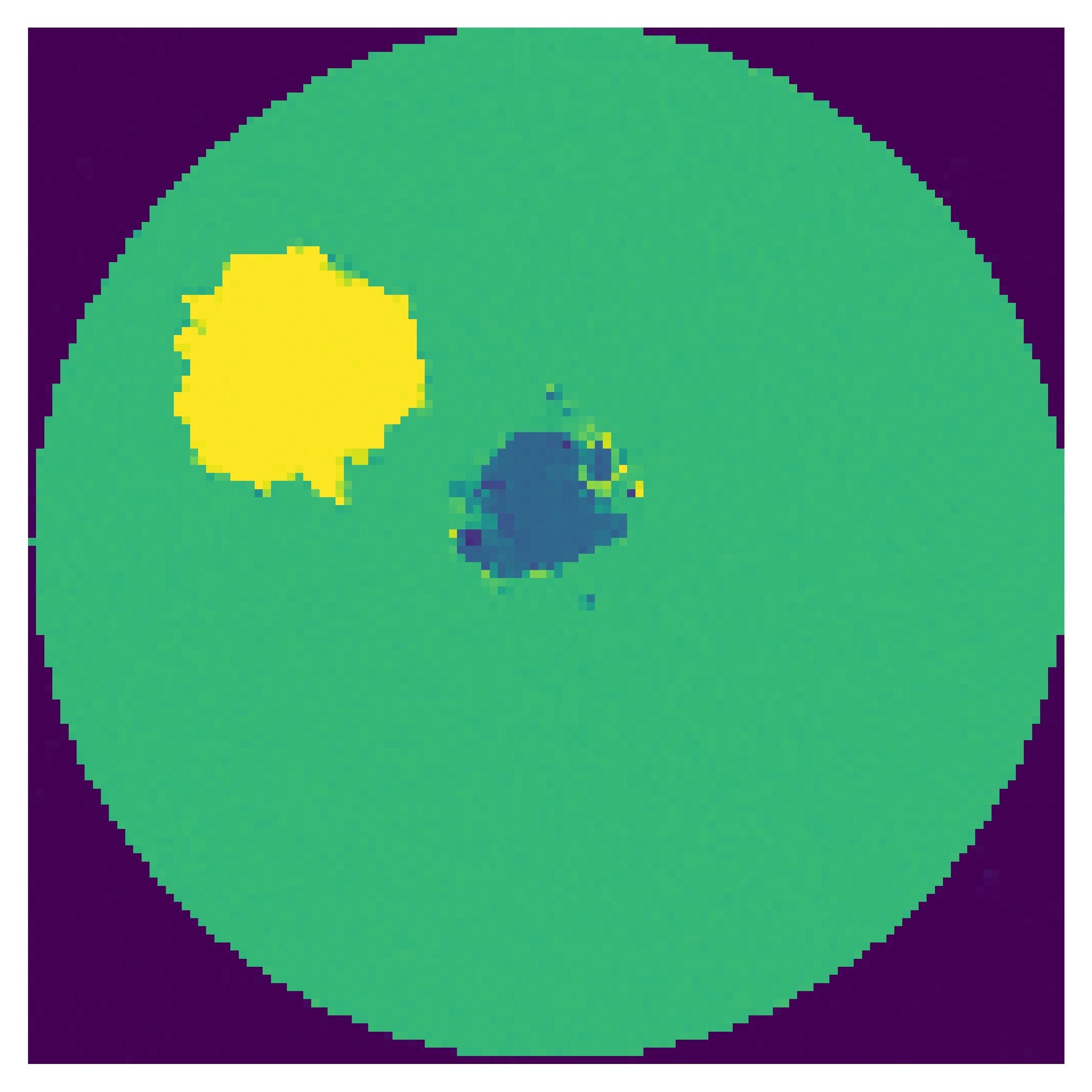}\vspace{5pt}
    \end{minipage}
}
\hspace{-12mm} 
\subfigure[$\mathrm{CSD}^*$]{
    \begin{minipage}[b]{0.22\linewidth}
    \centering
    \includegraphics[width=1in]{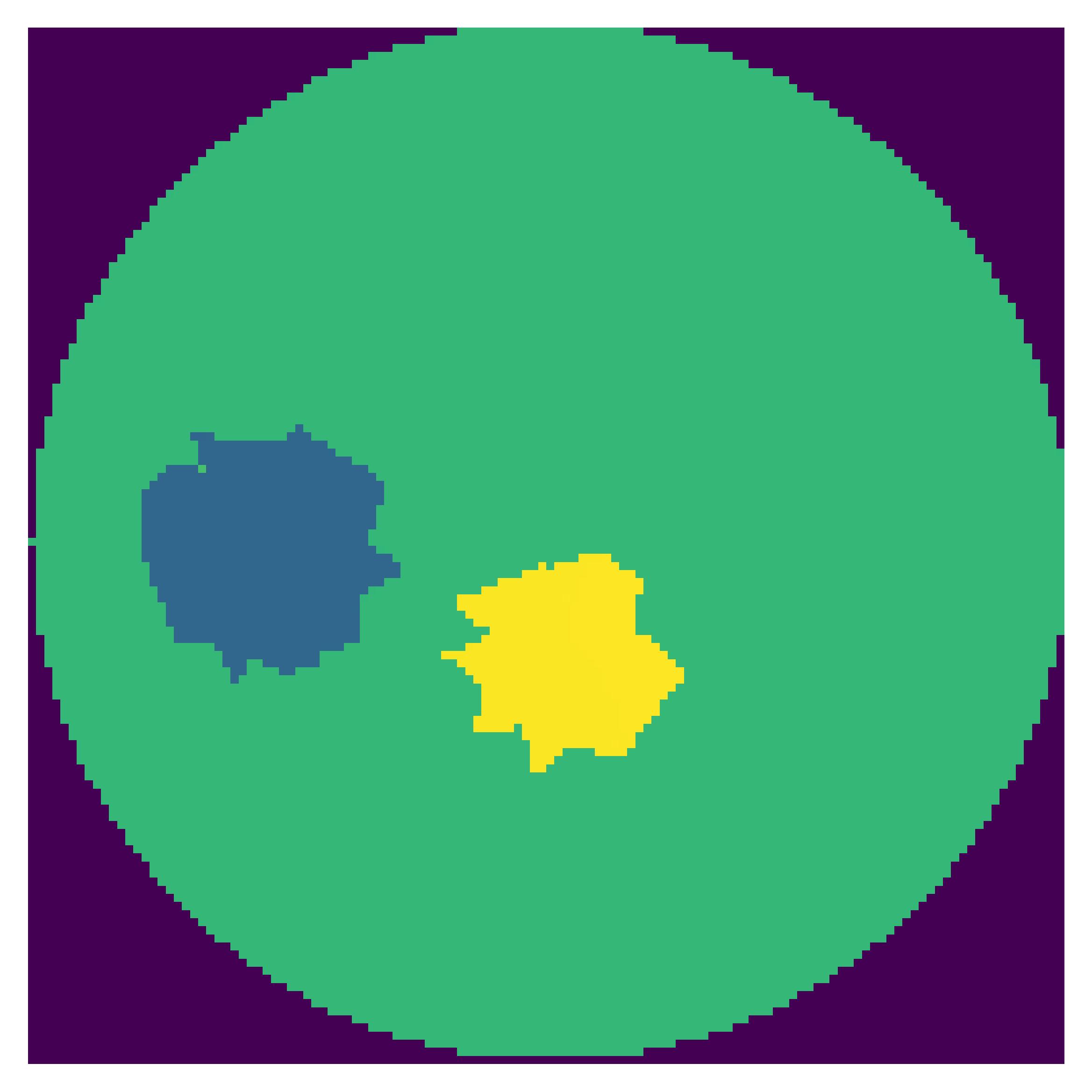}\vspace{5pt} 
    \includegraphics[width=1in]{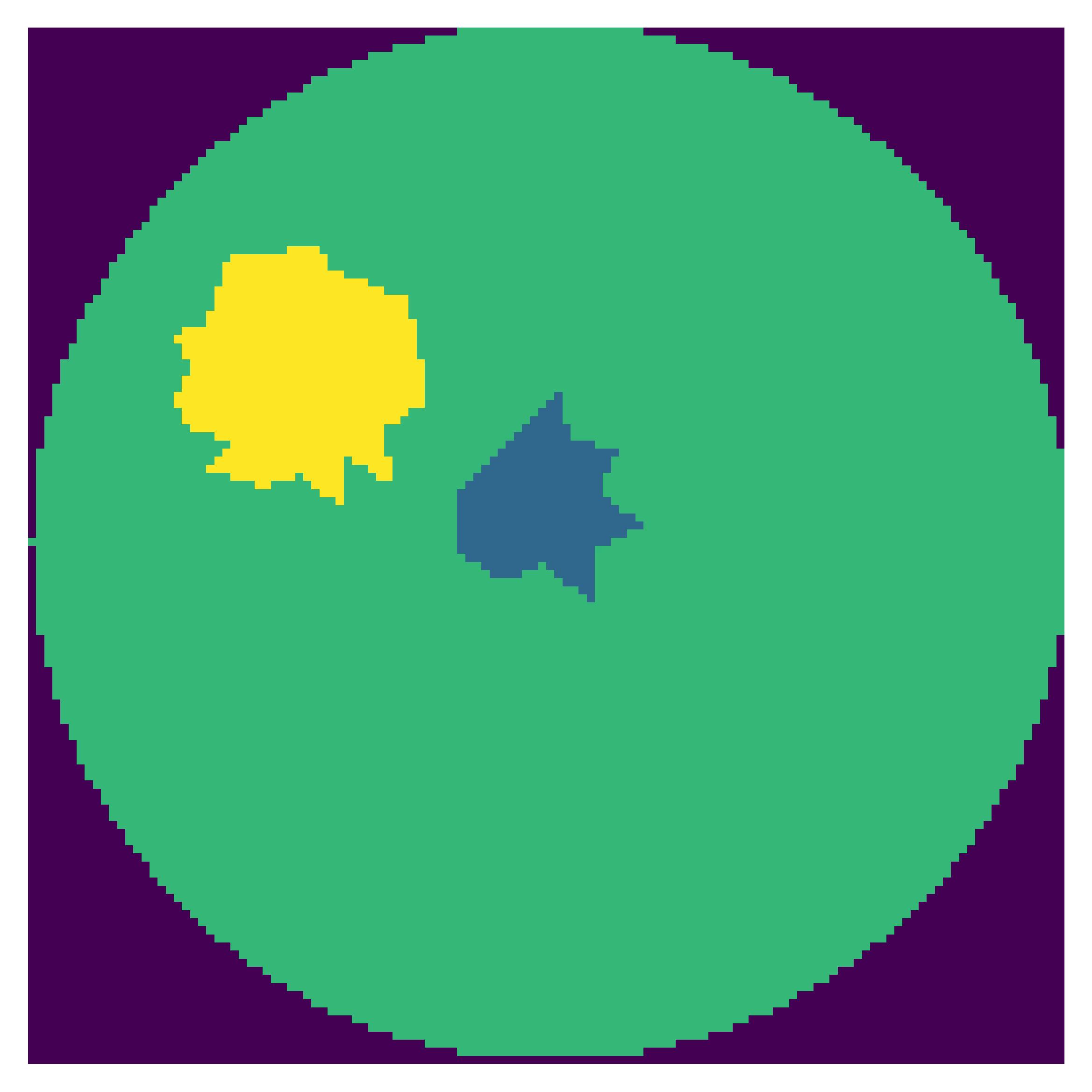}\vspace{5pt}
    \end{minipage}
}
\caption{\textbf{Generalization performance at the same noise level}, (a) original image of $128\times 128$ pixels, reconstructed image from (b) CVAE, and (c) CNF, as well as (d) $\mathrm{CSD}^*$.}
\label{fig:gener_40db_2}
\end{figure}

When comparing the performances of these models trained directly on 40dB with 2 anomalies dataset, shown in Table \ref{tab:gener_40db2}, we observe that CNF outperforms the others, with little change in $\mathrm{CSD}^*$. However, the performance of CVAE dramatically decreases, showing an increase of 90.6\% in MSE. Moreover, since CNF trained on 40dB with 4 anomalies performs well, this may be the reason for its good generalization on 40dB dataset with 2 anomalies, but $\mathrm{CSD}^*$ still performs better in terms of DR metrics.
As illustrated in Figure \ref{fig:gener_40db_2}, it is apparent that CVAE is unable to generate even the general shape of anomalies. CNF can generate the general outline, but its generalization effect is not very good for smaller-sized anomalies. Conversely, $\mathrm{CSD}^*$ performs well, with clear edges and shapes that are very similar to those of real images.

\begin{figure}[htbp]
\centering
\subfigure[GT]{
    \begin{minipage}[b]{0.22\linewidth} 
    \centering
    \includegraphics[width=1in]{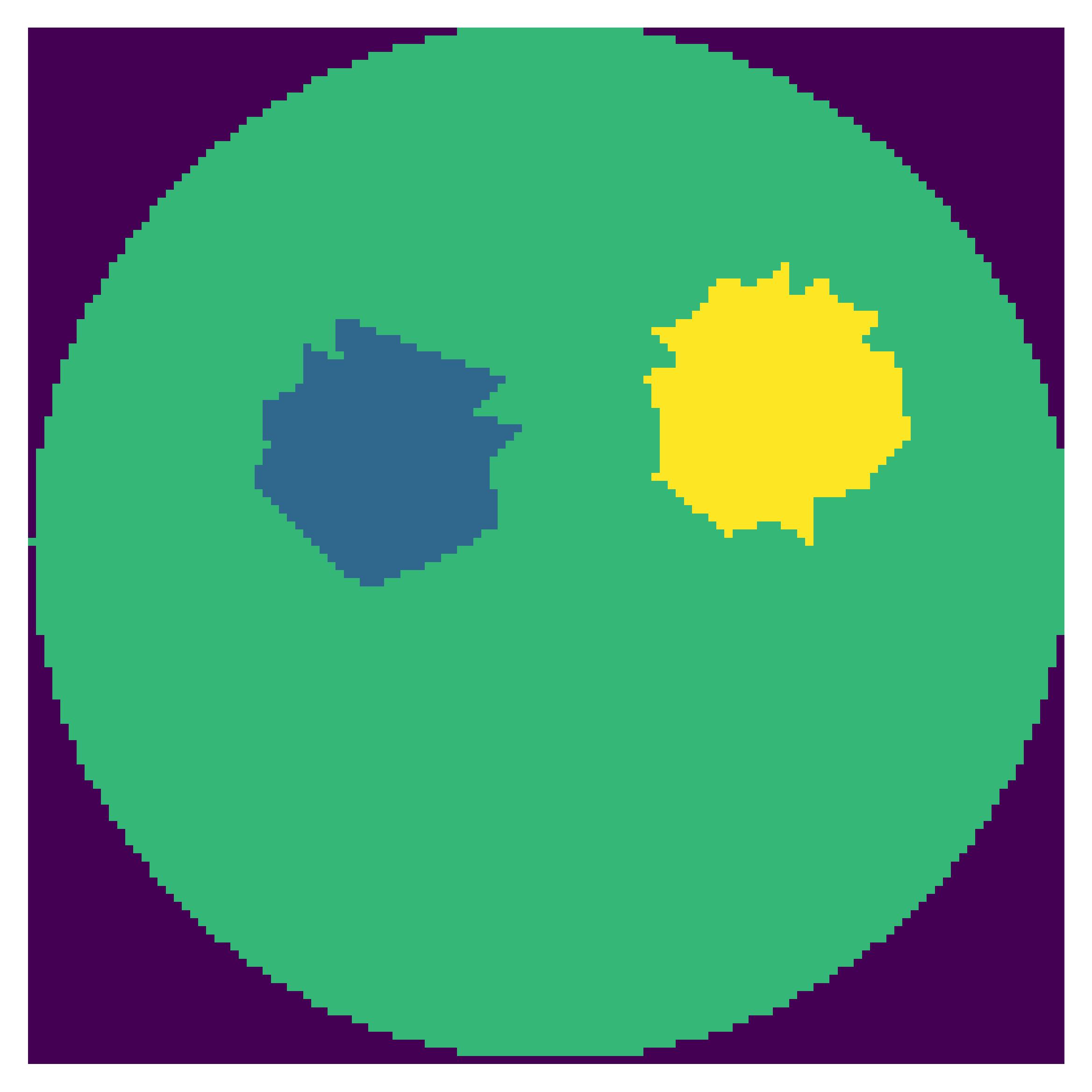}\vspace{5pt} 
    \includegraphics[width=1in]{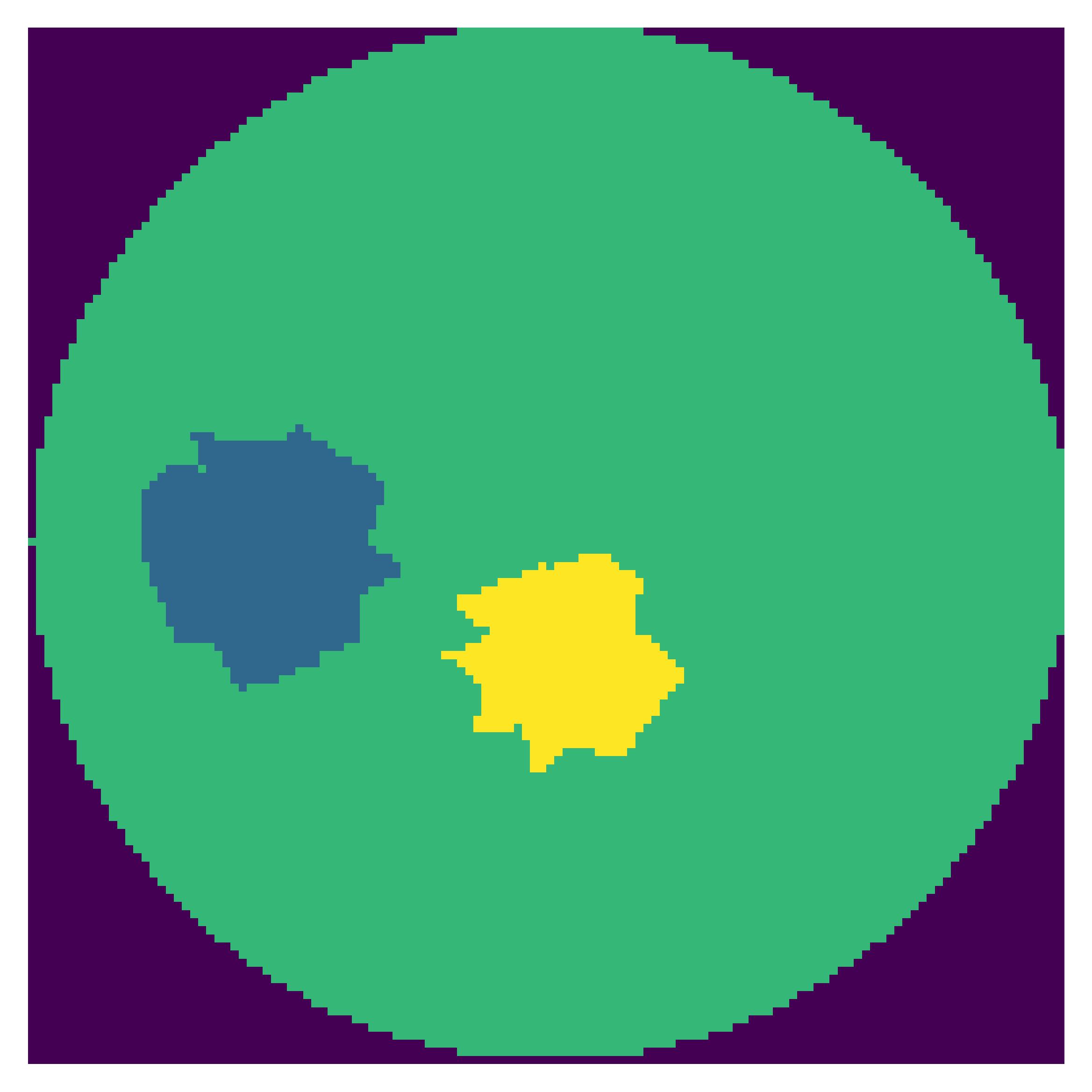}\vspace{5pt}
    \end{minipage}
}
\hspace{-12mm} 
\subfigure[CVAE]{
    \begin{minipage}[b]{0.22\linewidth}
    \centering
    \includegraphics[width=1in]{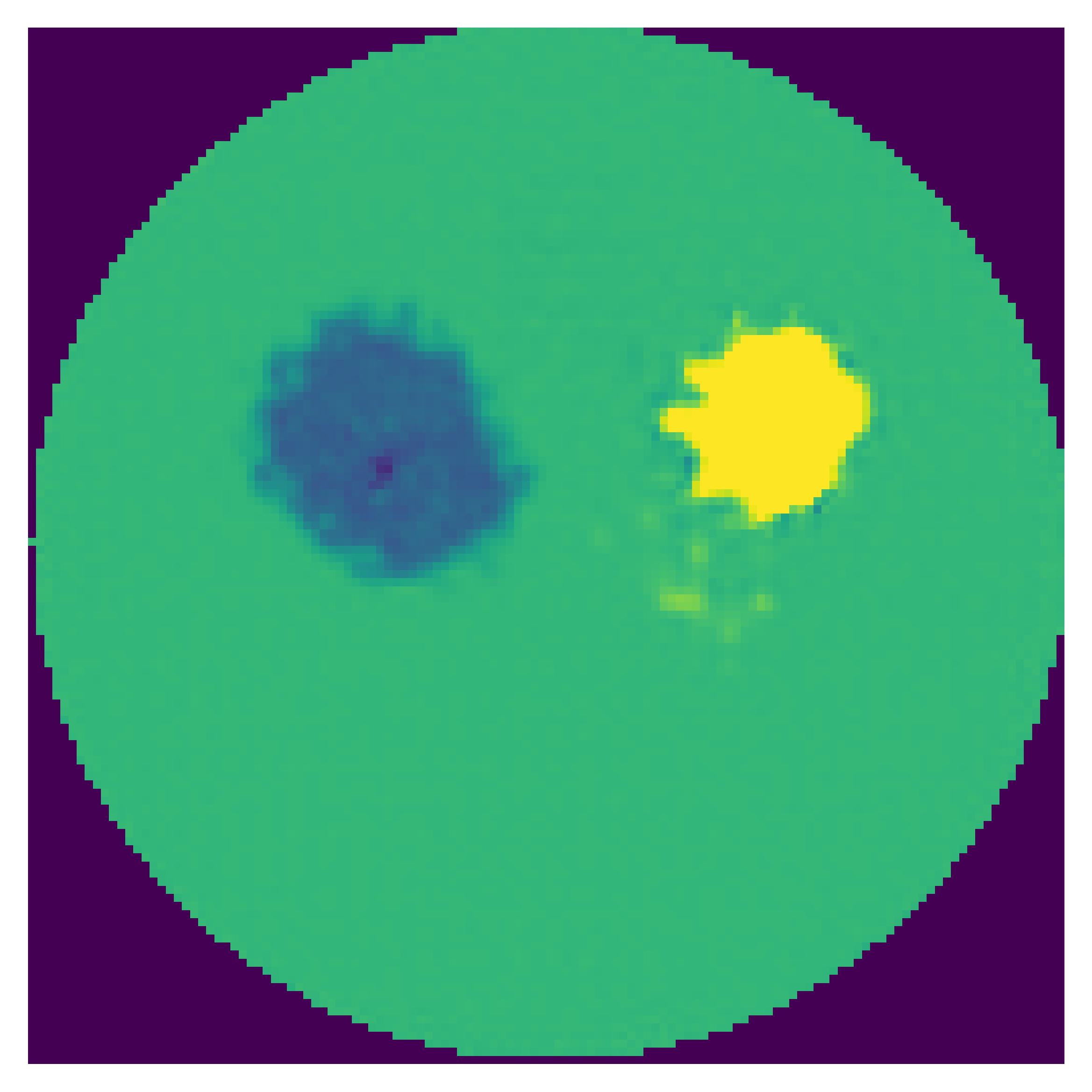}\vspace{5pt} 
    \includegraphics[width=1in]{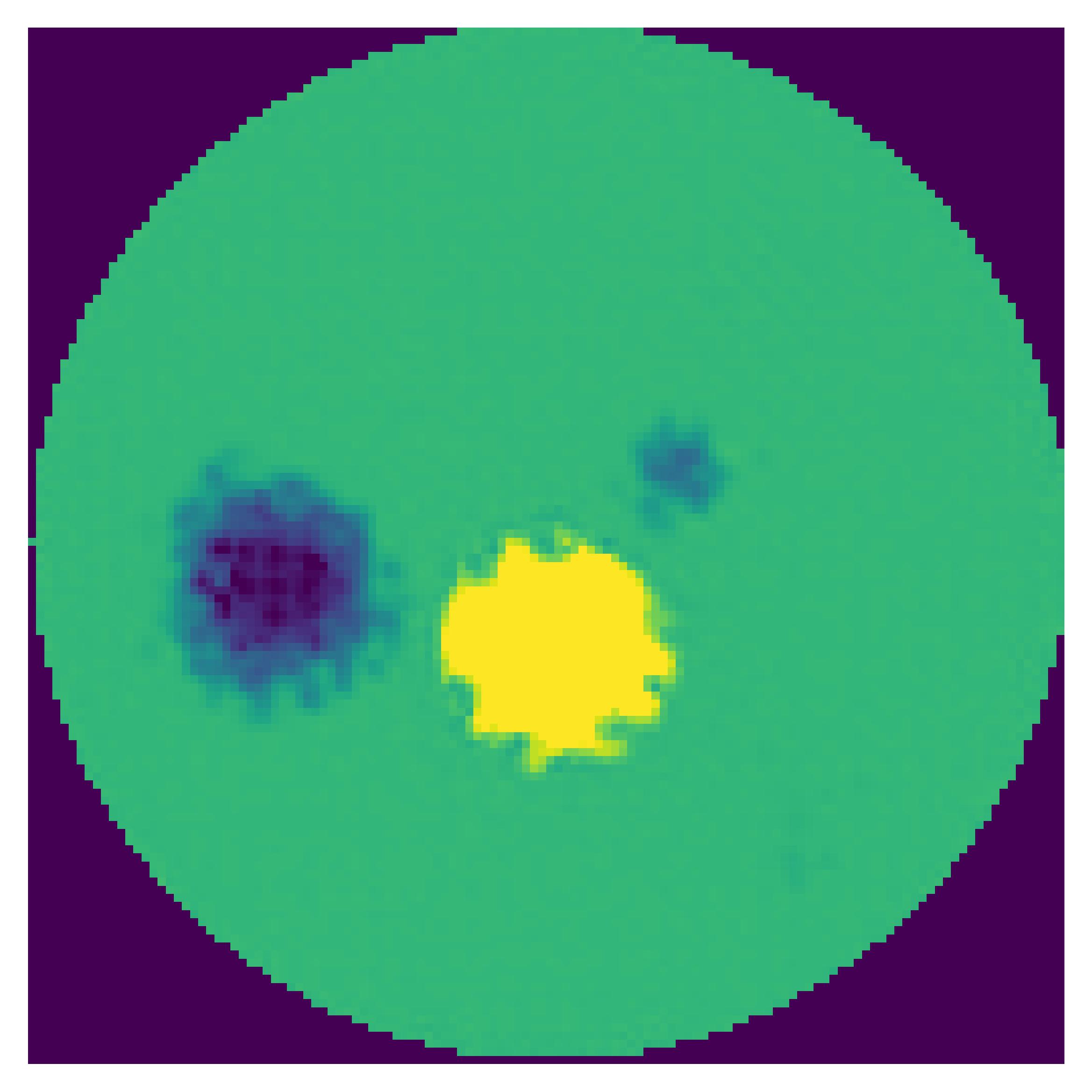}\vspace{5pt}
    \end{minipage}
}
\hspace{-12mm} 
\subfigure[CNF]{
    \begin{minipage}[b]{0.22\linewidth}
    \centering
    \includegraphics[width=1in]{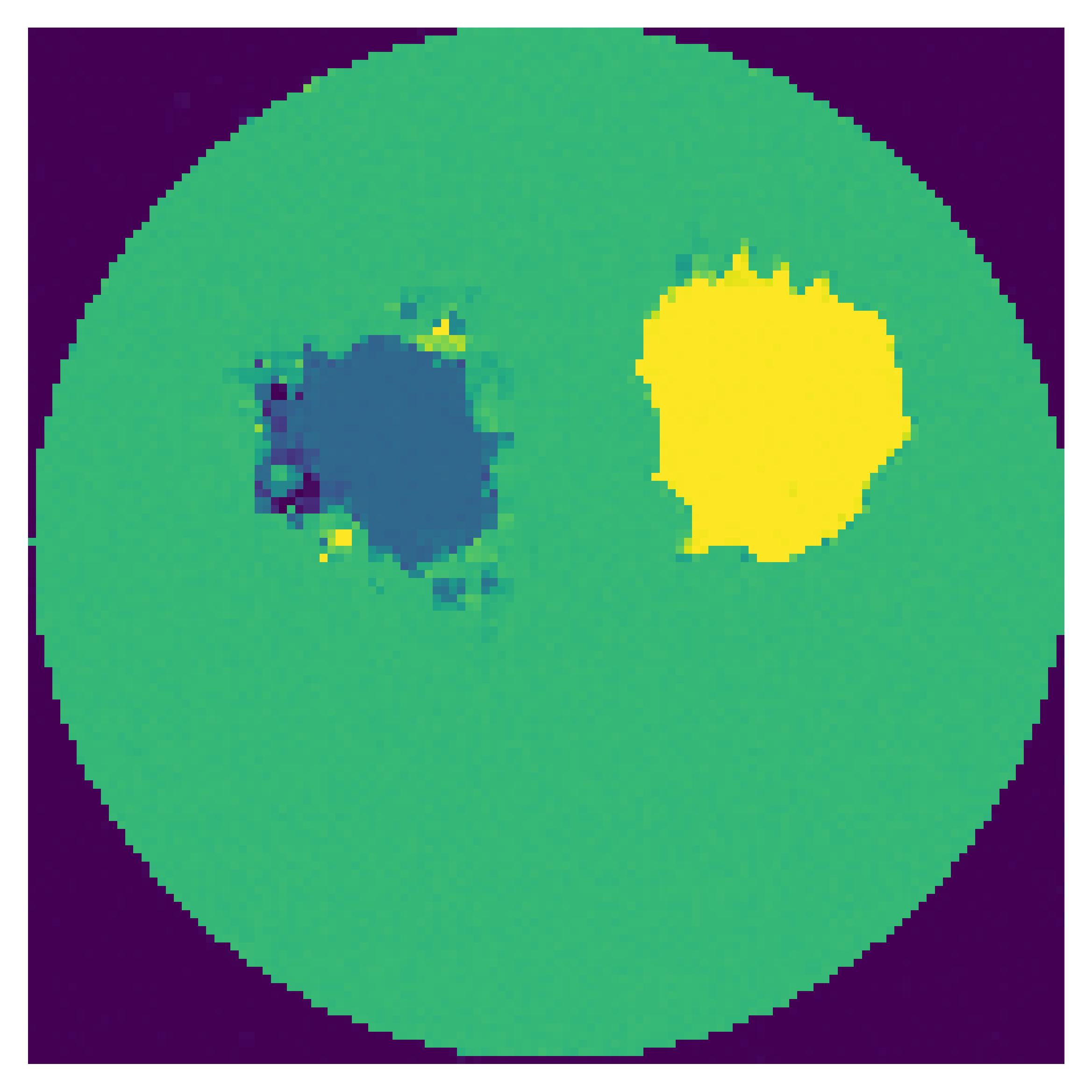}\vspace{5pt} 
    \includegraphics[width=1in]{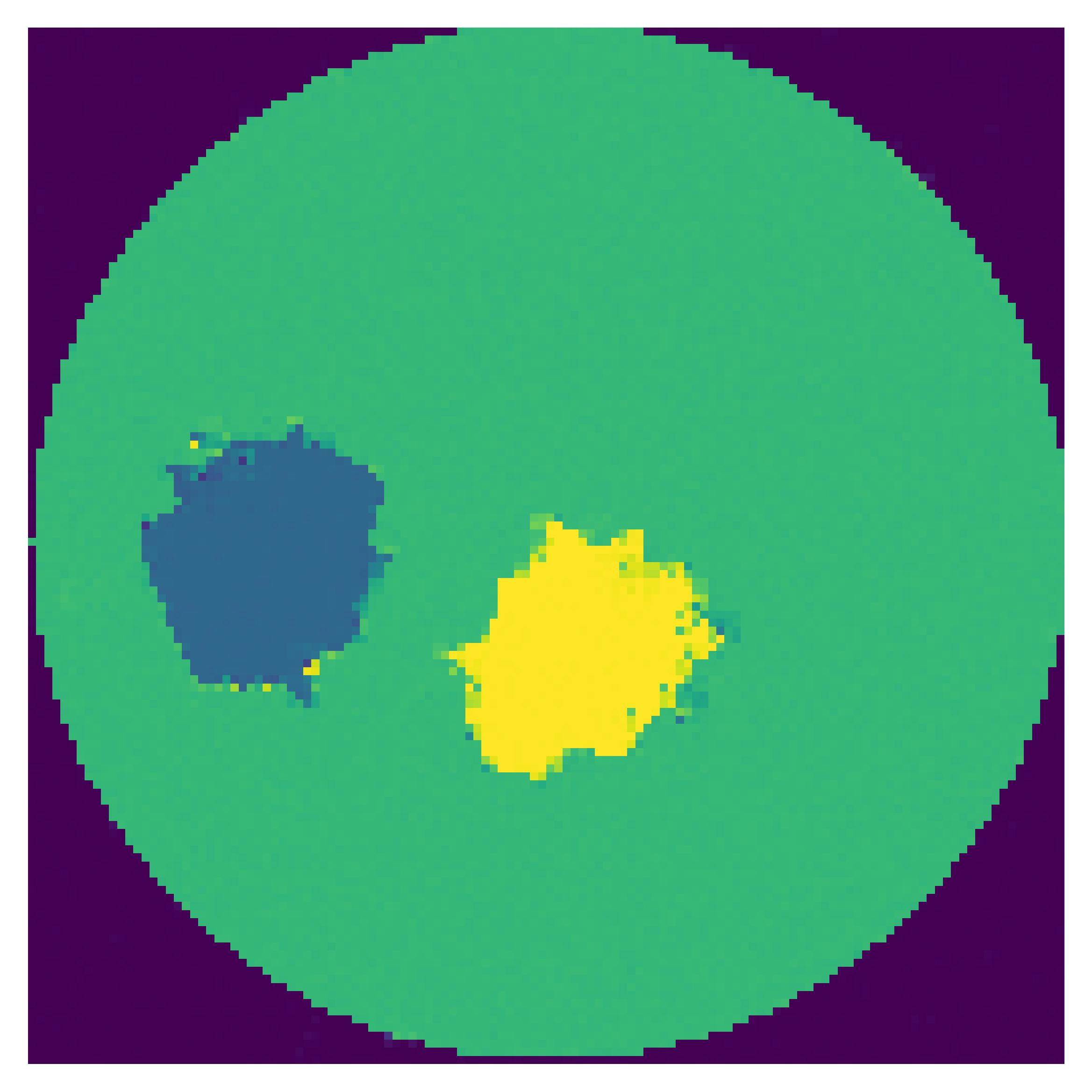}\vspace{5pt}
    \end{minipage}
}
\hspace{-12mm} 
\subfigure[$\mathrm{CSD}^*$]{
    \begin{minipage}[b]{0.22\linewidth}
    \centering
    \includegraphics[width=1in]{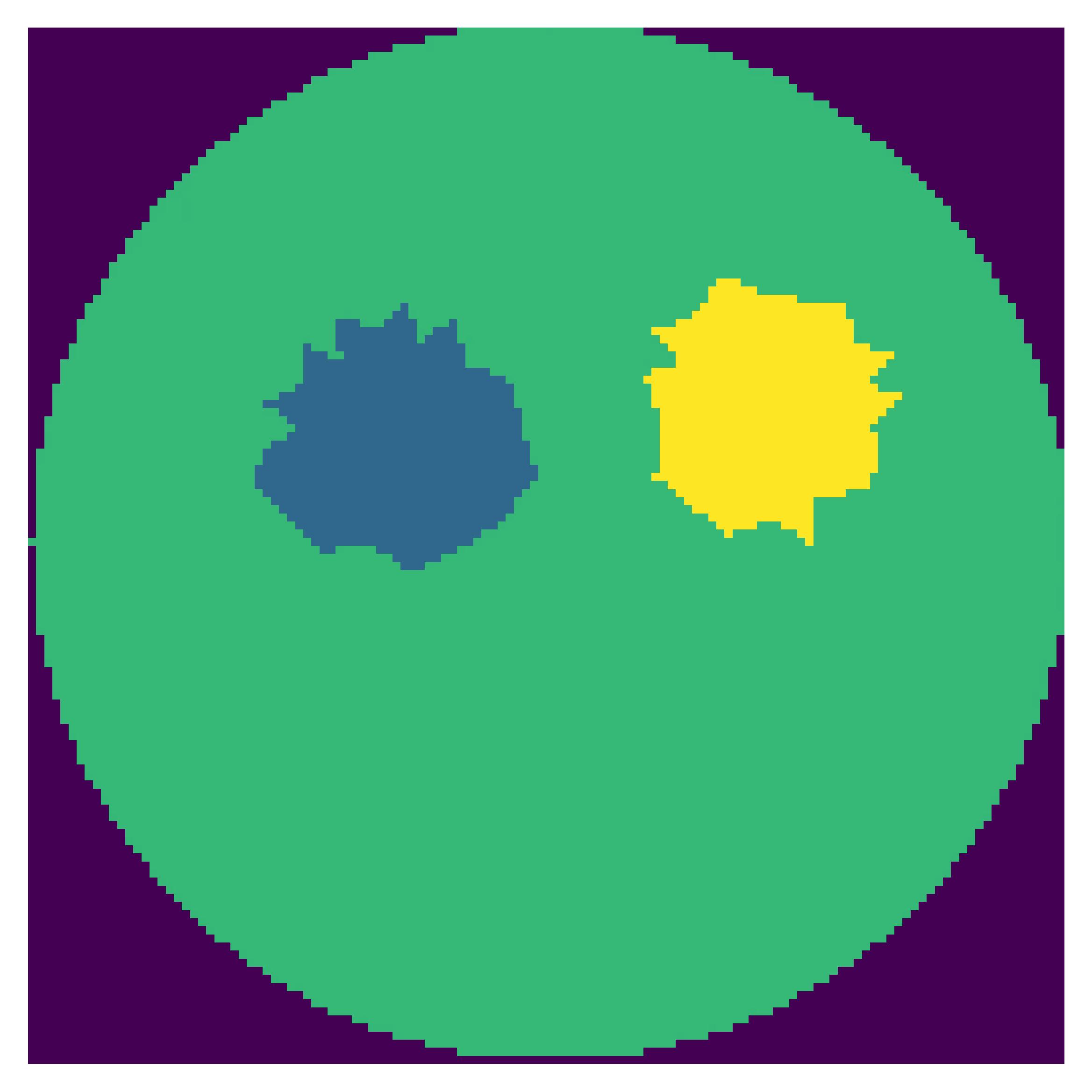}\vspace{5pt} 
    \includegraphics[width=1in]{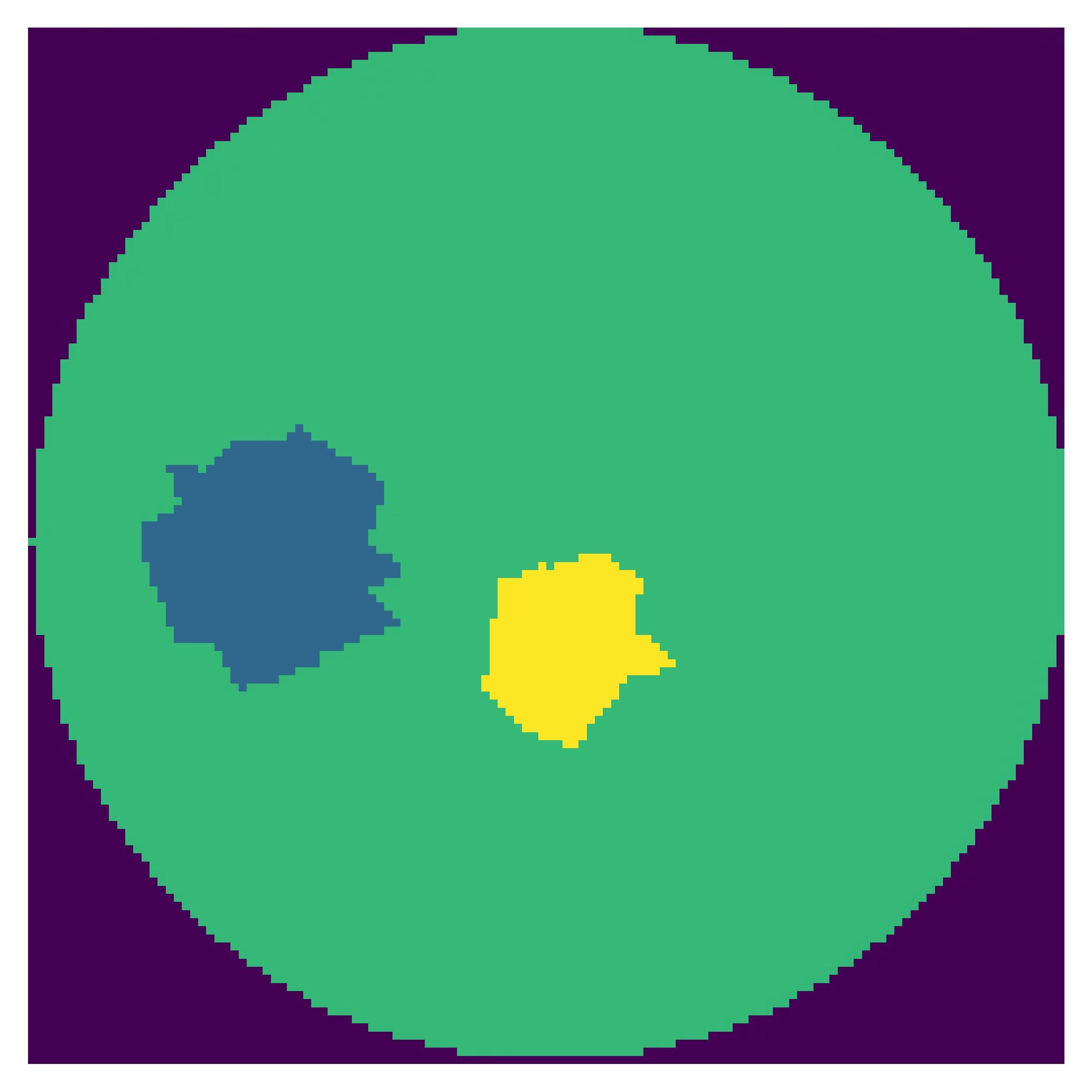}\vspace{5pt}
    \end{minipage}
}
\caption{\textbf{Generalization performance at the higher noise level(25dB)}, (a) original image of $128\times 128$ pixels, reconstructed image from (b) CVAE, and (c) CNF, as well as (d) $\mathrm{CSD}^*$. }
\label{fig:gener_25db_2}
\end{figure}

\begin{table*}[ht]
\resizebox{\linewidth}{!}{
\begin{threeparttable}
    \centering
    \caption{Generalization performance at the higher noise level(25dB).}
    \label{tab:gener_25db2}
    \begin{tabular}{lcccccc}
    \toprule
        \textbf{Method} & \textbf{MSE} & \textbf{PSNR} & \textbf{SSIM} & \textbf{RE} & \textbf{AE} & \textbf{DR} \\ \midrule
        CVAE &0.0149$\pm$0.006 &24.459$\pm$1.628 &0.870$\pm$0.030 &0.048$\pm$0.014 &0.038$\pm$0.011
        &1.313$\pm$0.198  \\[0.8ex]
        CNF &0.0074$\pm$0.0023 &\textbf{25.067$\pm$1.388} &0.903$\pm$0.018 &0.024$\pm$0.005 &0.019$\pm$0.004 &0.596$\pm$0.094  \\[0.8ex]
        $\mathrm{CSD}^*$ &\textbf{0.0064$\pm$0.0024} &24.747$\pm$2.809 &\textbf{0.918$\pm$0.024} &\textbf{0.021$\pm$0.007} &\textbf{0.016$\pm$0.005} &\textbf{0.903$\pm$0.153}  \\
    \bottomrule
    \end{tabular}   
\end{threeparttable}}
\end{table*}

As the noise level increases, while the performance of $\mathrm{CSD}^*$ on generalization is similar to its direct performance on 25dB with 2 anomalies, CNF and CVAE show some extent of performance degradation (see Table \ref{tab:gener_25db2}), with CVAE experiencing a particularly significant decline, as indicated by a 122.4\% increase in MSE. Overall, $\mathrm{CSD}^*$ has the best generalization performance at the high noise level, with the exception of slightly higher PSNR values of CNF.
In terms of image quality shown in Figure \ref{fig:gener_25db_2}, CVAE performs worst, with inconsistent shapes, unclear boundaries, and incorrect color representations of anomalies. CNF can only roughly localize anomalies and cannot even generate similar shapes to the ground truth, while the $\mathrm{CSD}^*$ model can partially generate samples that are close to the ground truth.

Overall, $\mathrm{CSD}^*$ shows better generalization stability, which, according to~\cite{guo2023physics}, may be due to the more the input is processed by physics, the better the generalization will be. In other words, the preprocessing in the Gauss-Newton method involves more physics and thus reduces the network’s burden.

\section{Conclusion and outlook} \label{sec:conclusion}
In conclusion, this study investigated the application of conditional variational autoencoder (CVAE), conditional normalizing flow (CNF), and conditional score-based diffusion model ($\mathrm{CSD}^*$) for EIT image reconstruction, generalization, and model efficiency. 
From the reconstruction perspective, CVAE excels in high-level noise with fewer anomalies, while $\mathrm{CSD}^*$ demonstrates better performance with more anomalies at the same noise level. Additionally, CNF performs better in low-level noise regardless of the number of anomalies.
None of the generalization results for CVAE are very satisfactory, while CNF performs well in low-level noise, and $\mathrm{CSD}^*$ excels in high-level noise.
In terms of efficiency, CVAE outperforms the other models due to its small architecture with fewer parameters. In contrast, CNF is time-consuming for its larger parameter size, while $\mathrm{CSD}^*$ is relatively efficient as it only utilizes the score-based diffusion model as a post-processing operator and avoids repetitive training when datasets change.
We also discovered that no single method consistently outperforms the others across all noise settings.

A main perspective for future work is to scale the proposed methodology to large images, for example,  
by learning statistically independent latent representations through hierarchical DGMs to increase the robustness of the reconstruction model,
by replacing CNN with graph neural networks that 
providing more flexibility in handling element's size, number, and topology arising in EIT problem~\cite{herzberg2021graph},
by incorporating the physics model into the DGMs to improve the convergence rate as well as interpretability~\cite{guo2023physics}.
Additionally, future work could also explore the application of the proposed methodology to other imaging problems, particularly those involving highly non-regular models, such as optical diffraction tomography and electromagnetic imaging.

\vskip 0.3in
\acks{We declare that we have no financial and personal relationships with other people or organizations that can inappropriately influence our work, and there is no professional or other personal interest of any nature or kind in any product, service, and/or company that could be construed as influencing the position presented in, or the review of, the manuscript entitled.

The work is supported by the NSF of China (12101614) and the NSF of Hunan (2021JJ40715). We are grateful to the High-Performance Computing Center of Central South University for assistance with the computations.}


\vskip 0.2in
\bibliography{sample}

\begin{thebibliography}{35}
\providecommand{\natexlab}[1]{#1}
\providecommand{\url}[1]{\texttt{#1}}
\expandafter\ifx\csname urlstyle\endcsname\relax
  \providecommand{\doi}[1]{doi: #1}\else
  \providecommand{\doi}{doi: \begingroup \urlstyle{rm}\Url}\fi

\bibitem[Adler and Holder(2021)]{adler2021electrical}
Andy Adler and David Holder.
\newblock \emph{{Electrical Impedance Tomography: Methods, History and Applications}}.
\newblock CRC Press, 2021.

\bibitem[Ardizzone et~al.(2019)Ardizzone, L{\"u}th, Kruse, Rother, and K{\"o}the]{ardizzone2019guided}
Lynton Ardizzone, Carsten L{\"u}th, Jakob Kruse, Carsten Rother, and Ullrich K{\"o}the.
\newblock Guided image generation with conditional invertible neural networks.
\newblock \emph{arXiv preprint arXiv:1907.02392}, 2019.

\bibitem[Bar and Sochen(2021)]{bar2021strong}
Leah Bar and Nir Sochen.
\newblock Strong solutions for pde-based tomography by unsupervised learning.
\newblock \emph{SIAM Journal on Imaging Sciences}, 14\penalty0 (1):\penalty0 128--155, 2021.

\bibitem[Bohra et~al.(2022)Bohra, Pham, Dong, and Unser]{bohra2022bayesian}
Pakshal Bohra, Thanh-an Pham, Jonathan Dong, and Michael Unser.
\newblock Bayesian inversion for nonlinear imaging models using deep generative priors.
\newblock \emph{IEEE Transactions on Computational Imaging}, 8:\penalty0 1237--1249, 2022.

\bibitem[Chung et~al.(2021)Chung, Huh, Kim, Park, and Ye]{9495275}
Hyungjin Chung, Jaeyoung Huh, Geon Kim, Yong~Keun Park, and Jong~Chul Ye.
\newblock Missing cone artifact removal in odt using unsupervised deep learning in the projection domain.
\newblock \emph{IEEE Transactions on Computational Imaging}, 7:\penalty0 747--758, 2021.
\newblock \doi{10.1109/TCI.2021.3098937}.

\bibitem[Chung et~al.(2022)Chung, Sim, and Ye]{chung2022come}
Hyungjin Chung, Byeongsu Sim, and Jong~Chul Ye.
\newblock Come-closer-diffuse-faster: Accelerating conditional diffusion models for inverse problems through stochastic contraction.
\newblock In \emph{Proceedings of the IEEE/CVF Conference on Computer Vision and Pattern Recognition}, pages 12413--12422, 2022.

\bibitem[Chung et~al.(2023)Chung, Kim, Mccann, Klasky, and Ye]{chung2023diffusion}
Hyungjin Chung, Jeongsol Kim, Michael~Thompson Mccann, Marc~Louis Klasky, and Jong~Chul Ye.
\newblock Diffusion posterior sampling for general noisy inverse problems.
\newblock In \emph{The Eleventh International Conference on Learning Representations}, 2023.
\newblock URL \url{https://openreview.net/forum?id=OnD9zGAGT0k}.

\bibitem[Colibazzi et~al.(2022)Colibazzi, Lazzaro, Morigi, and Samor{\'e}]{colibazzi2022learning}
Francesco Colibazzi, Damiana Lazzaro, Serena Morigi, and Andrea Samor{\'e}.
\newblock Learning nonlinear electrical impedance tomography.
\newblock \emph{Journal of Scientific Computing}, 90\penalty0 (1):\penalty0 1--23, 2022.

\bibitem[Denker et~al.(2021)Denker, Schmidt, Leuschner, and Maass]{denker2021conditional}
Alexander Denker, Maximilian Schmidt, Johannes Leuschner, and Peter Maass.
\newblock Conditional invertible neural networks for medical imaging.
\newblock \emph{Journal of Imaging}, 7\penalty0 (11):\penalty0 243, 2021.

\bibitem[Dinh et~al.(2015)Dinh, Krueger, and Bengio]{DBLP:journals/corr/DinhKB14}
Laurent Dinh, David Krueger, and Yoshua Bengio.
\newblock {NICE:} non-linear independent components estimation.
\newblock In Yoshua Bengio and Yann LeCun, editors, \emph{3rd International Conference on Learning Representations, {ICLR} 2015, San Diego, CA, USA, May 7-9, 2015, Workshop Track Proceedings}, 2015.
\newblock URL \url{http://arxiv.org/abs/1410.8516}.

\bibitem[Dinh et~al.(2017)Dinh, Sohl-Dickstein, and Bengio]{dinh2017density}
Laurent Dinh, Jascha Sohl-Dickstein, and Samy Bengio.
\newblock Density estimation using real {NVP}.
\newblock In \emph{International Conference on Learning Representations}, 2017.
\newblock URL \url{https://openreview.net/forum?id=HkpbnH9lx}.

\bibitem[Gehre et~al.(2012)Gehre, Kluth, Lipponen, Jin, Sepp{\"a}nen, Kaipio, and Maass]{gehre2012sparsity}
Matthias Gehre, Tobias Kluth, Antti Lipponen, Bangti Jin, Aku Sepp{\"a}nen, Jari~P Kaipio, and Peter Maass.
\newblock Sparsity reconstruction in electrical impedance tomography: an experimental evaluation.
\newblock \emph{Journal of Computational and Applied Mathematics}, 236\penalty0 (8):\penalty0 2126--2136, 2012.

\bibitem[Goodfellow et~al.(2020)Goodfellow, Pouget-Abadie, Mirza, Xu, Warde-Farley, Ozair, Courville, and Bengio]{goodfellow2020generative}
Ian Goodfellow, Jean Pouget-Abadie, Mehdi Mirza, Bing Xu, David Warde-Farley, Sherjil Ozair, Aaron Courville, and Yoshua Bengio.
\newblock Generative adversarial networks.
\newblock \emph{Communications of the ACM}, 63\penalty0 (11):\penalty0 139--144, 2020.

\bibitem[Grenander and Miller(1994)]{grenander1994representations}
Ulf Grenander and Michael~I Miller.
\newblock Representations of knowledge in complex systems.
\newblock \emph{Journal of the Royal Statistical Society: Series B (Methodological)}, 56\penalty0 (4):\penalty0 549--581, 1994.

\bibitem[Guo and Jiang(2021)]{guo2021construct}
Ruchi Guo and Jiahua Jiang.
\newblock Construct deep neural networks based on direct sampling methods for solving electrical impedance tomography.
\newblock \emph{SIAM Journal on Scientific Computing}, 43\penalty0 (3):\penalty0 B678--B711, 2021.

\bibitem[Guo et~al.(2023)Guo, Huang, Li, Zhang, and Eldar]{guo2023physics}
Rui Guo, Tianyao Huang, Maokun Li, Haiyang Zhang, and Yonina~C Eldar.
\newblock Physics-embedded machine learning for electromagnetic data imaging: Examining three types of data-driven imaging methods.
\newblock \emph{IEEE Signal Processing Magazine}, 40\penalty0 (2):\penalty0 18--31, 2023.

\bibitem[Herzberg et~al.(2021)Herzberg, Rowe, Hauptmann, and Hamilton]{herzberg2021graph}
William Herzberg, Daniel~B Rowe, Andreas Hauptmann, and Sarah~J Hamilton.
\newblock Graph convolutional networks for model-based learning in nonlinear inverse problems.
\newblock \emph{IEEE Transactions on Computational Imaging}, 7:\penalty0 1341--1353, 2021.

\bibitem[Kawar et~al.(2022)Kawar, Song, Ermon, and Elad]{kawar2022jpeg}
Bahjat Kawar, Jiaming Song, Stefano Ermon, and Michael Elad.
\newblock {JPEG} artifact correction using denoising diffusion restoration models.
\newblock In \emph{NeurIPS 2022 Workshop on Score-Based Methods}, 2022.
\newblock URL \url{https://openreview.net/forum?id=O3WJOt79289}.

\bibitem[Kingma and Welling(2014)]{kingma2014autoencoding}
Diederik~P. Kingma and Max Welling.
\newblock Auto-encoding variational bayes.
\newblock In \emph{International Conference on Learning Representations}, 2014.
\newblock URL \url{https://openreview.net/forum?id=33X9fd2-9FyZd}.

\bibitem[Li(2021)]{li2021differentiable}
Dongzhuo Li.
\newblock Differentiable gaussianization layers for inverse problems regularized by deep generative models.
\newblock \emph{arXiv e-prints}, pages arXiv--2112, 2021.

\bibitem[Liu et~al.(2018)Liu, Yang, Xu, Xia, Dai, Ji, You, Dong, Shi, and Fu]{liu2018pyeit}
Benyuan Liu, Bin Yang, Canhua Xu, Junying Xia, Meng Dai, Zhenyu Ji, Fusheng You, Xiuzhen Dong, Xuetao Shi, and Feng Fu.
\newblock pyeit: A python based framework for electrical impedance tomography.
\newblock \emph{SoftwareX}, 7:\penalty0 304--308, 2018.

\bibitem[Parisi(1981)]{parisi1981correlation}
Giorgio Parisi.
\newblock Correlation functions and computer simulations.
\newblock \emph{Nuclear Physics B}, 180\penalty0 (3):\penalty0 378--384, 1981.

\bibitem[Saharia et~al.(2022)Saharia, Ho, Chan, Salimans, Fleet, and Norouzi]{saharia2022image}
Chitwan Saharia, Jonathan Ho, William Chan, Tim Salimans, David~J Fleet, and Mohammad Norouzi.
\newblock Image super-resolution via iterative refinement.
\newblock \emph{IEEE Transactions on Pattern Analysis and Machine Intelligence}, 2022.

\bibitem[Seo et~al.(2019)Seo, Kim, Jargal, Lee, and Harrach]{seo2019learning}
Jin~Keun Seo, Kang~Cheol Kim, Ariungerel Jargal, Kyounghun Lee, and Bastian Harrach.
\newblock A learning-based method for solving ill-posed nonlinear inverse problems: a simulation study of lung eit.
\newblock \emph{SIAM journal on Imaging Sciences}, 12\penalty0 (3):\penalty0 1275--1295, 2019.

\bibitem[Singh et~al.(2022)Singh, Jandial, Chopra, Ramesh, Krishnamurthy, and Balasubramanian]{singh2022conditioning}
Vedant Singh, Surgan Jandial, Ayush Chopra, Siddharth Ramesh, Balaji Krishnamurthy, and Vineeth~N Balasubramanian.
\newblock On conditioning the input noise for controlled image generation with diffusion models.
\newblock \emph{arXiv preprint arXiv:2205.03859}, 2022.

\bibitem[Somersalo and Kaipio(2004)]{somersalo2004statistical}
Erkki Somersalo and Jari Kaipio.
\newblock Statistical and computational inverse problems.
\newblock \emph{Applied Mathematical Sciences}, 160, 2004.

\bibitem[Song et~al.(2023)Song, Vahdat, Mardani, and Kautz]{song2023pseudoinverseguided}
Jiaming Song, Arash Vahdat, Morteza Mardani, and Jan Kautz.
\newblock Pseudoinverse-guided diffusion models for inverse problems.
\newblock In \emph{International Conference on Learning Representations}, 2023.
\newblock URL \url{https://openreview.net/forum?id=9_gsMA8MRKQ}.

\bibitem[Song et~al.(2021{\natexlab{a}})Song, Durkan, Murray, and Ermon]{song2021maximum}
Yang Song, Conor Durkan, Iain Murray, and Stefano Ermon.
\newblock Maximum likelihood training of score-based diffusion models.
\newblock In A.~Beygelzimer, Y.~Dauphin, P.~Liang, and J.~Wortman Vaughan, editors, \emph{Advances in Neural Information Processing Systems}, 2021{\natexlab{a}}.
\newblock URL \url{https://openreview.net/forum?id=AklttWFnxS9}.

\bibitem[Song et~al.(2021{\natexlab{b}})Song, Sohl-Dickstein, Kingma, Kumar, Ermon, and Poole]{song2021scorebased}
Yang Song, Jascha Sohl-Dickstein, Diederik~P Kingma, Abhishek Kumar, Stefano Ermon, and Ben Poole.
\newblock Score-based generative modeling through stochastic differential equations.
\newblock In \emph{International Conference on Learning Representations}, 2021{\natexlab{b}}.
\newblock URL \url{https://openreview.net/forum?id=PxTIG12RRHS}.

\bibitem[Song et~al.(2022)Song, Shen, Xing, and Ermon]{song2022solving}
Yang Song, Liyue Shen, Lei Xing, and Stefano Ermon.
\newblock Solving inverse problems in medical imaging with score-based generative models.
\newblock In \emph{International Conference on Learning Representations}, 2022.
\newblock URL \url{https://openreview.net/forum?id=vaRCHVj0uGI}.

\bibitem[Vincent(2011)]{vincent2011connection}
Pascal Vincent.
\newblock A connection between score matching and denoising autoencoders.
\newblock \emph{Neural computation}, 23\penalty0 (7):\penalty0 1661--1674, 2011.

\bibitem[Wang et~al.(2004{\natexlab{a}})Wang, Wang, and Yin]{1315989}
Huaxiang Wang, Chao Wang, and Wuliang Yin.
\newblock A pre-iteration method for the inverse problem in electrical impedance tomography.
\newblock \emph{IEEE Transactions on Instrumentation and Measurement}, 53\penalty0 (4):\penalty0 1093--1096, 2004{\natexlab{a}}.
\newblock \doi{10.1109/TIM.2004.831180}.

\bibitem[Wang et~al.(2004{\natexlab{b}})Wang, Bovik, Sheikh, and Simoncelli]{wang2004image}
Zhou Wang, Alan~C Bovik, Hamid~R Sheikh, and Eero~P Simoncelli.
\newblock Image quality assessment: from error visibility to structural similarity.
\newblock \emph{IEEE transactions on image processing}, 13\penalty0 (4):\penalty0 600--612, 2004{\natexlab{b}}.

\bibitem[Winkler et~al.(2020)Winkler, Worrall, Hoogeboom, and Welling]{winkler2020learning}
Christina Winkler, Daniel Worrall, Emiel Hoogeboom, and Max Welling.
\newblock Learning likelihoods with conditional normalizing flows, 2020.
\newblock URL \url{https://openreview.net/forum?id=rJg3zxBYwH}.

\bibitem[Zhang et~al.(2020)Zhang, Guo, Li, Yang, Xu, and Abubakar]{zhang2020supervised}
Ke~Zhang, Rui Guo, Maokun Li, Fan Yang, Shenheng Xu, and Aria Abubakar.
\newblock Supervised descent learning for thoracic electrical impedance tomography.
\newblock \emph{IEEE Transactions on Biomedical Engineering}, 68\penalty0 (4):\penalty0 1360--1369, 2020.

\end{thebibliography}

\end{document}